\documentclass[]{pasj01}
\usepackage{lscape}
\draft
\Received{}
\Accepted{}
\usepackage{url}
\usepackage{graphicx}
\begin{document}
\title{
Diversity of ${\gamma}$-ray and Radio Variabilities of Bright Blazars and Implications for ${\gamma}$-ray Emission Location}

\author{Hai-Ming Zhang\altaffilmark{1,2}, Zhen-Jie Wang\altaffilmark{1}, Jin Zhang\altaffilmark{3\dag}, Ting-Feng Yi \altaffilmark{4}, Liang Chen\altaffilmark{5}, Rui-Jing Lu\altaffilmark{1}, En-Wei Liang\altaffilmark{1}}
\altaffiltext{1}{Guangxi Key Laboratory for Relativistic Astrophysics, Department of Physics, Guangxi University, Nanning 530004, China}
\altaffiltext{2}{School of Astronomy and Space Science, Nanjing University, Nanjing 210023, China}
\altaffiltext{3}{Key Laboratory of Space Astronomy and Technology, National Astronomical Observatories, Chinese Academy of Sciences, Beijing 100012, China}
\altaffiltext{4}{Department of Physics, Yunnan Normal University, Kunming 650500, China}
\altaffiltext{5}{Key Laboratory for Research in Galaxies and Cosmology, Shanghai Astronomical Observatory, Chinese Academy of Sciences, 80 Nandan Road, Shanghai 200030, China}
\email{jinzhang@bao.ac.cn}
\maketitle

\begin{abstract}
Violent multi-wavelength variabilities are observed in $\gamma$-ray-selected blazars. We present an analysis of long-term light curves for eight bright blazars to explore the co-variation pattern in the $\gamma$-ray and radio bands. We extract their $\gamma$-ray light curves and spectra with data observed by the \emph{Fermi} Large Area Telescope (LAT) since 2008. We find diverse co-variation patterns between the $\gamma$-ray and radio (at 43 GHz) fluxes in these sources. The $\gamma$-ray and radio fluxes of 3C 454.3 and PKS 1633+382 are correlated without any time-lag, suggesting that they are from the same radiation region. Similar correlation is also observed in 3C 273 and PKS 1222+216, but the radio flux is lag behind the $\gamma$-ray flux approximately $\sim$160 days and $\sim$290 days, respectively. This likely suggests that their $\gamma$-ray emission regions are located at the upstream of their radio cores at 43 GHz. The $\gamma$-ray and radio fluxes of the other four blazars are not correlated, implying that the $\gamma$-ray and radio emission may be from different regions in their jets. The $\gamma$-ray light curves of the eight blazars can be decomposed into some long timescale variability components and fast spike flares. We propose that they may be attributed to the central engine activity and the magnetic reconnection process or turbulence in the local emission region, respectively.
\end{abstract}

\section{Introduction}

Blazars are a special class of active galactic nuclei (AGNs) and host a relativistic jet pointing toward us (Urry \& Padovani 1995). They are divided into BL Lacertae objects (BL Lacs) and flat-spectrum radio quasars (FSRQs) according to the spectral features in the optical band. Blazars are characterized with the broadband non-thermal emission and the obvious variabilities in multiwavelength. So far, most confirmed extragalactic $\gamma$-ray emission sources are blazars (Ackermann et al. 2015), and they are the targets to research the extragalactic background light (Aharonian et al. 2006) and the high-energy cosmic rays (Aartsen et al. 2018). Many theoretical models were proposed to explain the energy dissipation and particle acceleration in relativistic jets of blazars, however, the crucial issue is the location of the $\gamma$-ray emission along the jet; inside or outside the broad-line region (BLR).

For the $\gamma$-ray radiation region inside the BLR scenario, the inverse Compton (IC) scattering of the BLR photons by relativistic electrons would dominate the $\gamma$-ray emission (Sikora et al. 1994; Ghisellini et al. 2010; Zhang et al. 2013, 2014, 2015). It is supported by the observations of spectral breaks in the 2--10 GeV range, which can be well reproduced by the absorption of $\gamma$-rays via photon-photon pair production on the He {\sc\,ii} Lyman recombination continuum and lines (Poutanen \& Stern 2010). The correlation of the increased emission line flux with millimeter core ejections and $\gamma$-ray, optical, and ultraviolet flares also implies that the BLR extends beyond the $\gamma$-emitting region (Le\'{o}n-Tavares et al. 2013; Isler et al. 2013). However, the Klein-Nashina (KN) effect and the absorption of $\gamma$-rays via photon-photon pair production (Liu \& Bai 2006) make it inappropriate to use the BLR photons as the seed photons providing the $\gamma$-ray emission above 10 GeV. Instead, the thermal radiation from torus might dominate the IC scattering process to produce the $\gamma$-ray emission for these TeV FSRQs (e.g., Sikora et al. 1994; Tavecchio \& Mazin 2009; Kang et al. 2014; Kang 2017). Furthermore, the correlation between the $\gamma$-ray and millimeter signals (e.g., Wehrle et al. 2012) also favors the torus as the external radiation fields. The characteristic variability timescales of blazars may provide some implications to discriminate the two kinds of external radiation fields (e.g., Sikora et al. 2009; Nalewajko et al. 2014), intraday/hours for the BLR case and days/weeks for the torus scenario, respectively.

The radio core might represent a transition region from synchrotron self-absorbed regions into optically thin regions (Blandford \& K\"{o}nigl 1979), and is usually located at parsec (pc) scale from the central engine. Based on the observations of the very long baseline interferometry (VLBI) and the $\gamma$-ray detectors (Energetic Gamma Ray Experiment Telescope, EGRET and \emph{Fermi} Large Area Telescope, \emph{Fermi}/LAT), a clear connection  between the occurrence of a $\gamma$-ray flare and the ejection of a new superluminal component from the radio core is observed for many sources (Boccardi et al. 2017, and references therein). It is suggested that the $\gamma$-ray emission and the millimeter-wave emission are co-spatial, and thus the $\gamma$-ray emission may be produced at pc-scale from the central engine (e.g., Marscher et al. 2008, 2010; Schinzel et al. 2012; Karamanavis et al. 2016). The plasma may attain its terminal speed and the equipartition between magnetic energy and particle energy at the radio core (Marscher et al. 2010; G\'{o}mez et al. 2016; see also Boccardi et al. 2017). Hence, the mm-VLBI observations may be a good probe to study the location of $\gamma$-ray emission region.

The corresponding issue to the $\gamma$-ray emission location is the energy dissipation and particle acceleration mechanisms in the relativistic jets of blazars. In the sub-pc-scale from the central engine, the jet that is dominated by the Poynting flux is collimated due to the confinement provided by the magnetic field and/or by the external medium, and is accelerated to relativistic speeds by magnetic pressure gradients (Boccardi et al. 2017 for a review). The particles in jet may be accelerated by the magnetic reconnection process (e.g., Sikora et al. 2005; Giannios et al. 2009). In the pc-scale a kinetic-flux dominated jet forms (Boccardi et al. 2017). Alternatively, the reconfinement shocks produced by the interaction of the jet with the external medium in the pc to several-pc scale may be also the dissipation and particle acceleration mechanisms (e.g., Nalewajko 2012; Nalewajko et al. 2014; Boccardi et al. 2017).

In this paper, we comprehensively analyze the long-term light curves observed by \emph{Fermi}/LAT for eight blazars with the multiply exponential functions to study their variability behaviors, and combine the long-term observational data at 43 GHz to estimate the locations of the $\gamma$-ray emission. The reduction of the \emph{Fermi}/LAT data is given in Section 2. Data analysis methods for the light curves are described in Section 3. Results and discussion about the results are presented in Section 4 and Section 5, respectively. A summary is given in Section 6.

\section{Sample Selection and \emph{Fermi}/LAT Data Analysis}

There are 38 \emph{Fermi}/LAT sources in the Very Long Baseline Array (VLBA)-BU Blazar Monitoring Program\footnote{\url{http://www.bu.edu/blazars/VLBAproject.html}}. We download their third \emph{Fermi}/LAT catalog (3FGL) Aperture Photometry Light-curves\footnote{\url{https://fermi.gsfc.nasa.gov/ssc/data/access/lat/4yr_catalog/ap_lcs.php?ra=12-13}} with 30-day time resolution, and then these blazars that have at least one outburst with the flux variation more than one order of magnitude in the \emph{Fermi}/LAT energy band are selected. Finally eight bright blazars are included in our analysis, one BL Lac and seven FSRQs. The \emph{Fermi}/LAT is a pair conversion $\gamma$-ray telescope and is sensitive to photon energies greater than 20 MeV with a field of view of about 2.4 sr. It surveys the whole sky every 3 hr (Atwood et al. 2009). The Pass 8 data of these sources were collected from the Fermi Science Support Center\footnote{\url{http://fermi.gsfc.nasa.gov/ssc}} with a temporal coverage from 2008 August 6 (Modified Julian Day, MJD 54684) to 2017 June 20 (MJD 57924). Since the sources are monthly observed with the VLBA at 43 GHz and also for guaranteeing the confidence level of the long-term light curves, the time-bin of 30-day is used to analyze the \emph{Fermi}/LAT data.

The data analysis was performed with the standard analysis tool \textit{gtlike/pyLikelihood}, which is part of the \emph{Fermi} Science Tools software package (version v10r0p5). Photon events with energies from 100 MeV to 300 GeV were considered. In order to reduce the contamination from the Earth limb $\gamma$-rays, our analysis was restricted to a maximum zenith angle of 100$^{\circ}$, i.e., the events with zenith angles $>100^{\circ}$ were excluded. The P8R2\_SOURCE\_V6 set of instrument response functions was used. The photons were extracted from a circular region of interest (ROI) with a radius of 10$^{\circ}$, centered at the targets. All point sources in the third \emph{Fermi}/LAT source catalog (Acero et al. 2015) located in the ROI and an additional surrounding 10$^{\circ}$ wide annulus were modeled in the fits. In the model file, the spectral parameters for sources lying within the ROI were kept free and for sources lying within the annulus were fixed. The isotropic background, including the sum of residual instrumental background and extragalactic diffuse $\gamma$-ray background, was fitted with a model derived from the isotropic background at high Galactic latitude, i.e., ``iso-P8R2-SOURCE-V6-v06.txt", and the Galactic diffuse GeV emission was modeled with ``gll-iem-v06.fits"\footnote{\url{http://fermi.gsfc.nasa.gov/ssc/data/access/lat/BackgroundModels.html}}.

The significance of the $\gamma$-ray signal from the sources was evaluated using the maximum-likelihood test statistic (TS), which is defined by TS=2($\ln\mathcal{L}_{1}-\ln\mathcal{L}_{0})$ (Mattox et al. 1996), where $\mathcal{L}_{0}$ is the likelihood of background (null hypothesis) and $\mathcal{L}_{1}$ is the likelihood of the hypothesis for including the point source. The spectral analysis in the energy range of 0.1--300 GeV for the sources was performed using the \textit{unbinned likelihood analysis}. For these sources the spectral shape can be well reproduced by a power law (PL) function, i.e., $dN(E)/dE = N_{0}(E/E_{0})^{-\Gamma_{\gamma}}$. Hence, we used the standard unbinned maximum-likelihood fit technique and a PL spectral function to analyze each time-bin. All the light curves in this paper were obtained using the derived fluxes with PL fits. The $\gamma$-ray light curves of these blazars in time-bins of 30-day are shown in Figures 1--8. If TS$<$9, an upper limit is presented, where TS=9 approximatively corresponds to $\sim3\sigma$ detection (Mattox et al. 1996).

\section{Data Analysis Methods}

\subsection{Variation Timescale Analysis}

The global light curves at the $\gamma$-ray band of these blazars are different and complex. In order to show the temporal evolution and search for the typical timescale of variability in the $\gamma$-ray band we fit the light curves of the eight blazars by a sum of exponentials (see also Britto et al. 2016; Prince et al. 2017; Shukla et al. 2018), which can grant the decay and rise time for the different peaks in the light curves. The fitting function is defined as
\begin{equation} \label{1}
F(t) = \sum_{i=1}^{\kappa}{2F_0}_i[\exp(\frac{t_i-t}{{T_{\rm r}}_i})+\exp(\frac{t-t_i}{{T_{\rm d}}_i})]^{-1}+F_{\rm C},
\end{equation}
where $\kappa$ indicates the total number of flares, ${F_0}_i$ is the flux at time $t_i$ approximately representing the flare amplitude, ${T_{\rm r}}_i$ and ${T_{\rm d}}_i$ are the rise and decay time of the $i$th flare, $F_{\rm C}$ is the constant flux (or constant baseline) of the light curves. We fit the global light curves with the function of Equation (1) and take each peak as a single exponential component. Note that some light curves are characterized by the complexly temporal profiles with short timescales while others likely show the broad outbursts superimposing the small flares. For the latter case, we add a long timescale variability component besides the fast spike flares to fit the light curves. Our strategy and procedure to fit the light curves are as follows.

First, we pick up each peak of a light curve as a potential component. Second, we fit the highest and its two adjacent peaks with three-component models. The component of the highest peak then can be decomposed by the $\chi^2$-minimization technique. Third, we subtract the highest component from the light curve and repeat the second step until all the data points are involved in our fits. Fourth, we create a global light curve model by adding all components into Euqation (1), except for those components whose relative errors of their amplitudes are larger than 50\%. $t_i$, ${T_{\rm r}}_i$, and ${T_{\rm d}}_i$ of each component are fixed at the values derived in the second and third steps, but their amplitudes are setting as free parameters. We then make global fits to obtain the amplitude of each component and underlying flux $F_{\rm C}$. Fifth, we calculate the residual of the global fit and smooth the residual curve to find out the possible broad components. Then, we add the possible broad components to the global light curve model and refine the global fit again. Sixth, we remove the potential components whose relative errors of their amplitudes are larger than 50\% from our global light curve model. We refine our fit in order to avoid over-fit to a light curve by the $\chi^2$ minimization technique and adopt a criterion of $\chi^2_{\rm r}<1.1$ for our fits (e.g., Protassov et al. 2002).

On the basis of the flare-finding with the multiply exponential-fitting method, we obtain the rise time ($T_{\rm r}$) and the total radiation energy ($E_{\gamma}$) of each flare, where $E_{\gamma}$ is derived by the integrating emission in the duration of each exponential component in the 0.1--300 GeV energy band. We use a broken power-law function (BPLF) to fit the cumulative distributions of $T_{\rm r}$ and $E_{\gamma}$. The BPLF is defined as (e.g., Liang et al. 2007; Wang et al. 2015)
\begin{equation}
N(X)=N_{0}\left[\left(\frac{X}{X_{\rm b}}\right)^{\alpha_{1}\omega}+\left(\frac{X}{X_{\rm b}}\right)^{\alpha_{2}\omega}\right]^{-\frac{1}{\omega}},
\end{equation}
where $\alpha_{1}$ ($\alpha_{2}$) is the slope before (after) the break $X_{\rm b}$ and $\omega$ represents the peak sharpness of the cumulative distribution.

Considering the large redshift of some sources, we make the redshift correction for the timescale of $T_{\rm r}$. Due to the selection bias (light curves in time-bins of 30 days), the cumulative distributions of $T_{\rm r}/(1+z)$ and $E_{\gamma}$ show a flat part at the short-time (and low-energy) regime, and thus we only consider the fitting result above the break of cumulative distribution (see also Cliver et al. 2012; Wang et al. 2015). The derived slopes of the cumulative distributions for $T_{\rm r}/(1+z)$ and $E_{\gamma}$ are named as $\alpha_{\rm T}$ and $\alpha_{\rm E}$, respectively.

\subsection{PSD and DCF Analysis for $\gamma$-ray and Radio Emission}

We use the power spectral density (PSD; see Ivezi\'{c} et al. (2014) for details) to study how the variability amplitude is distributed among the timescales in the $\gamma$-ray and 43 GHz. The \emph{Fermi}/LAT light curves of these sources cover 108 months and are derived in time-bins of 30 days and the VLBA observations at 43 GHz in the VLBA-BU-BLAZAR Program are also performed about once per month, hence the corresponding PSDs approximatively range in $10^{-3}-10^{-2}$ day$^{-1}$. Generally the PSD curves can be described by a power-law (or a broken power-law) function, i.e., $\propto f^{-\beta}$, where $f$ is the temporal frequency. If $\beta=0$, it means that the variability power is equal on each timescale and corresponds to a white noise light curve. Usually $1\leq\beta\leq2$, it indicates that the flux variations on longer timescales dominate the variations on shorter timescales and corresponds to a red noise light curve. The PSDs of the light curves in 43 GHz and $\gamma$-ray band for the eight blazars are also fitted with the power-law function using the \emph{stingray} package (Huppenkothen et al. 2019).

The long-term monitoring data with the VLBA observations at 43 GHz for the eight blazars are uneven, which are taken from the VLBA-BU Blazar Monitoring Program. In order to investigate the correlation of the variability between $\gamma$-ray and radio at 43 GHz, we use the discrete cross-correlation function (DCF, Edelson \& Krolik 1988) to analyze the long-term light curves in the two energy bands since the radio data are uneven. The DCF is defined as
\begin{equation}
\rm UDCF_{ij} = \frac{(y_{i}-\bar{y})(x_{j}-\bar{x})}{[(\sigma_{y}^2-e_{y}^2)(\sigma_{x}^2-e_{x}^2)]^{1/2}},
\end{equation}
\begin{equation}
\rm DCF(\tau)=\frac{1}{M}\rm UDCF_{ij},
\end{equation}
where $\sigma_{x}$ (or $\sigma_{y}$) and $e_{x}$ (or $e_{y}$) are the variance and error for time series x (or y), $\Delta t_{ij}=t_{j}-t_{i}$ is the associated time-lag. The DCF at time-lag of $\tau$ is then computed by binning and averaging $\rm UDCF_{ij}$ over \emph{M} pairs of points that are within $\tau-\delta t/2 \leq \triangle t_{ij} < \tau+\delta t/2$, where $\delta t$ is the bin-size of the DCF calculation. Due to the uneven radio data, $\delta t=40$ days is taken to ensure at least one data point in each bin-size. The period of observational data covers about 3300 days, considering the limit of the light curve length, and thus the DCF results between -800 days to 800 days are presented for the clear presentation of the results.  Note that a few time-bins in 3C273, 3C 454.3, CTA 102, and PKS 1222+216 are shown as the upper limits with TS$<$9, which are not included in the DCF analysis. And the non-detection of the correlation for some sources should not be due to the contamination of the noises.

We use the Monte Carlo method developed by Max-Moerbeck et al. (2014) to estimate the confidence levels of correlation coefficients for the DCF analysis results. The strategy and procedure are as follows (see Figure \ref{simulation_LC} as an example). First, the PSD model is defined as a broken power-law, i.e., the slope of zero is set at low frequencies while at the high frequencies the slope is taken the best-fit value of the PSD of light curve. Second, we simulate a light curve according to the defined PSD model and make it to be 1000 times longer than the real one using the codes from Emmanoulopoulos et. al (2013). Note that the break frequency of the PSD model is at the lowest frequency ($\sim 1/3300$) sampled by the real data. Third, we adjust the amplitude of light curve to make its probability density function (PDF) same as the one of observed data (details to see  Emmanoulopoulos et. al 2013). Fourth, we cut the simulated light curve into 1000 segments, each has the same time interval and span as the real data. Fifth, we perform the same procedure for the light curve in another band, and obtain another 1000 segments. Sixth, we make DCF analysis for the 1000 pairs of simulated light curves. And then the distribution of the random cross-correlation coefficient at each segment can be used to assess the confidence level of correlation coefficient for the real light curves.

\section{Results}

The data analysis results are displayed in Figures 1--8. Each figure corresponds to one blazar and contains 7 or 8 panels. The derived $\gamma$-ray light curves in time-bins of 30 days with the fitting lines by Equation (1) together with the long-term light curves at 43 GHz observed with the VLBA are presented in the top panel (a) (and panel (a1) for 3C 273 and PKS 1222+216). The DCF results of light curves between $\gamma$-ray and 43 GHz and the PSDs with the power-law fitting results in the two energy bands are shown in the middle panels (b), (c), and (d), respectively. $\Gamma_{\gamma}$ as a function of $L_{\gamma}$, where $\Gamma_{\gamma}$ is the photon spectral index and $L_{\gamma}$ is the integral luminosity of the \emph{Fermi}/LAT energy band (0.1--300 GeV) for each time-bin of the $\gamma$-ray light curves, the rise time ($T_{\rm r}/(1+z)$) distributions, and the cumulative distributions of $T_{\rm r}/(1+z)$ and $E_{\gamma}$ are given in the bottom panels (e), (f), and (g), respectively. In the following, we will describe the results for each source respectively and then some general characters of these blazars are presented. Note that in this paper we focus on the variability behaviors in the long-term light curves of these blazars, not the fast variabilities in short timescales of hours or sub-day.

\begin{figure}
 \centering
   \includegraphics[angle=0,scale=0.43]{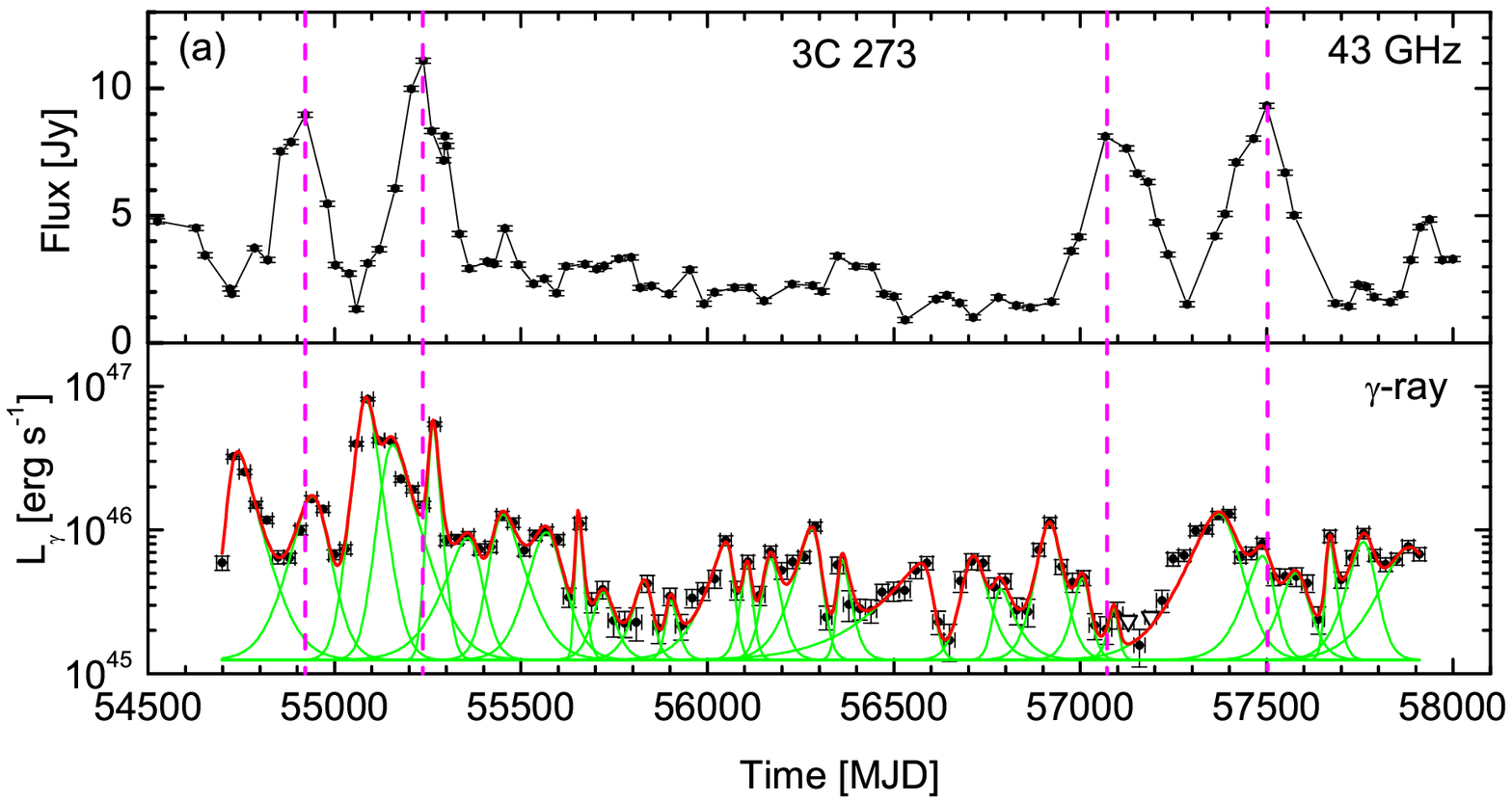}
   \includegraphics[angle=0,scale=0.43]{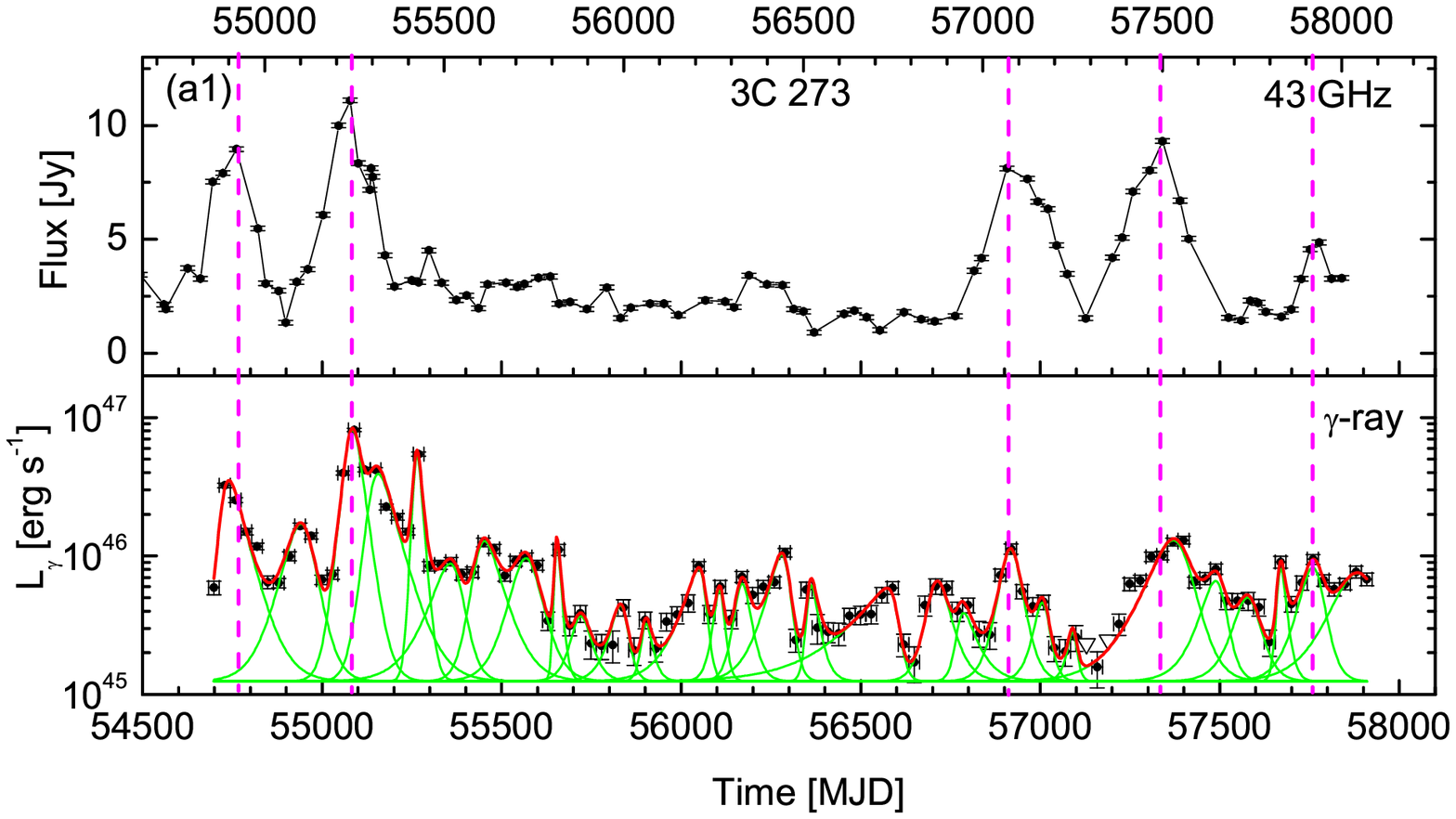}
   \includegraphics[angle=0,scale=0.26]{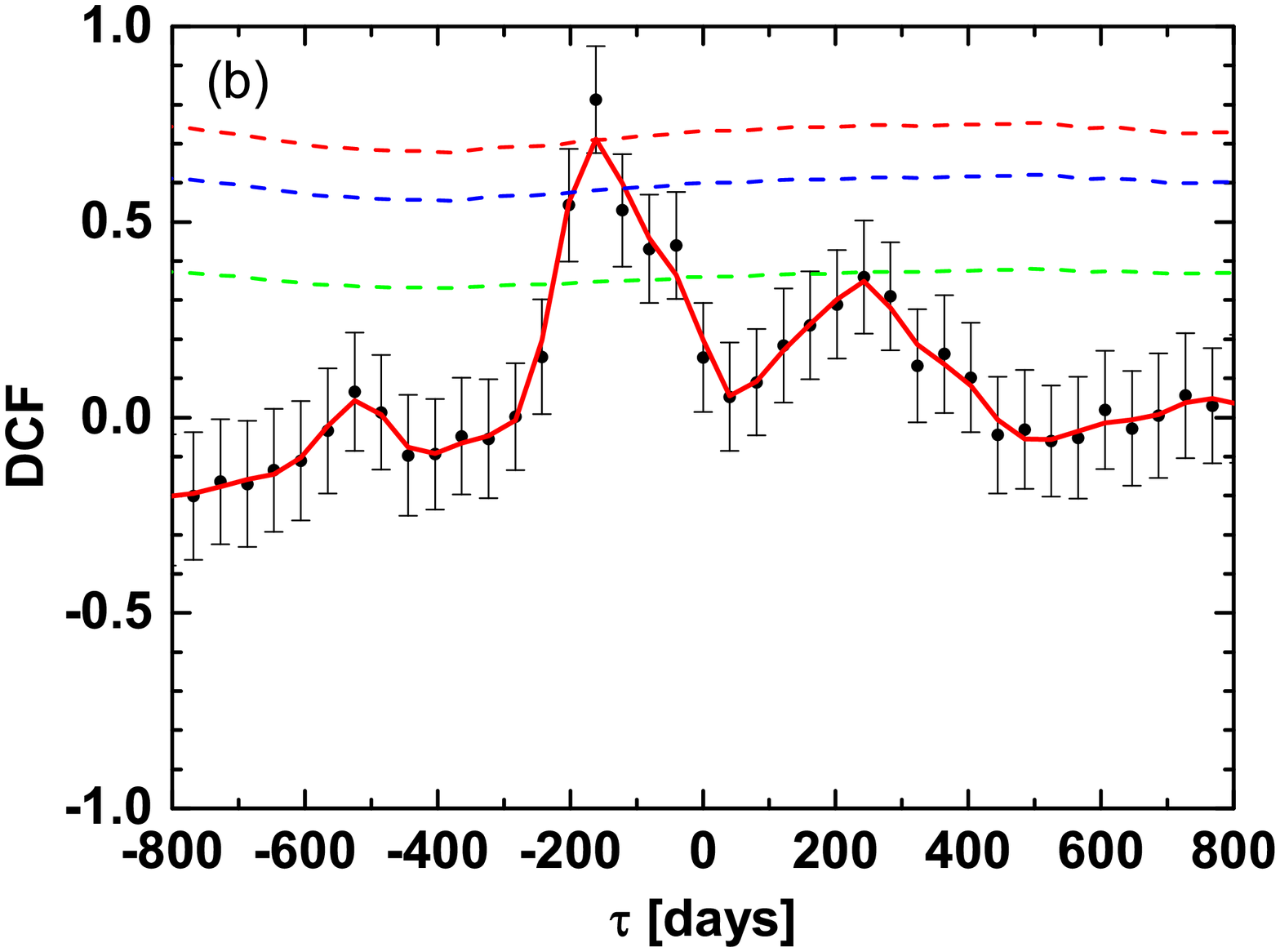}
   \includegraphics[angle=0,scale=0.24]{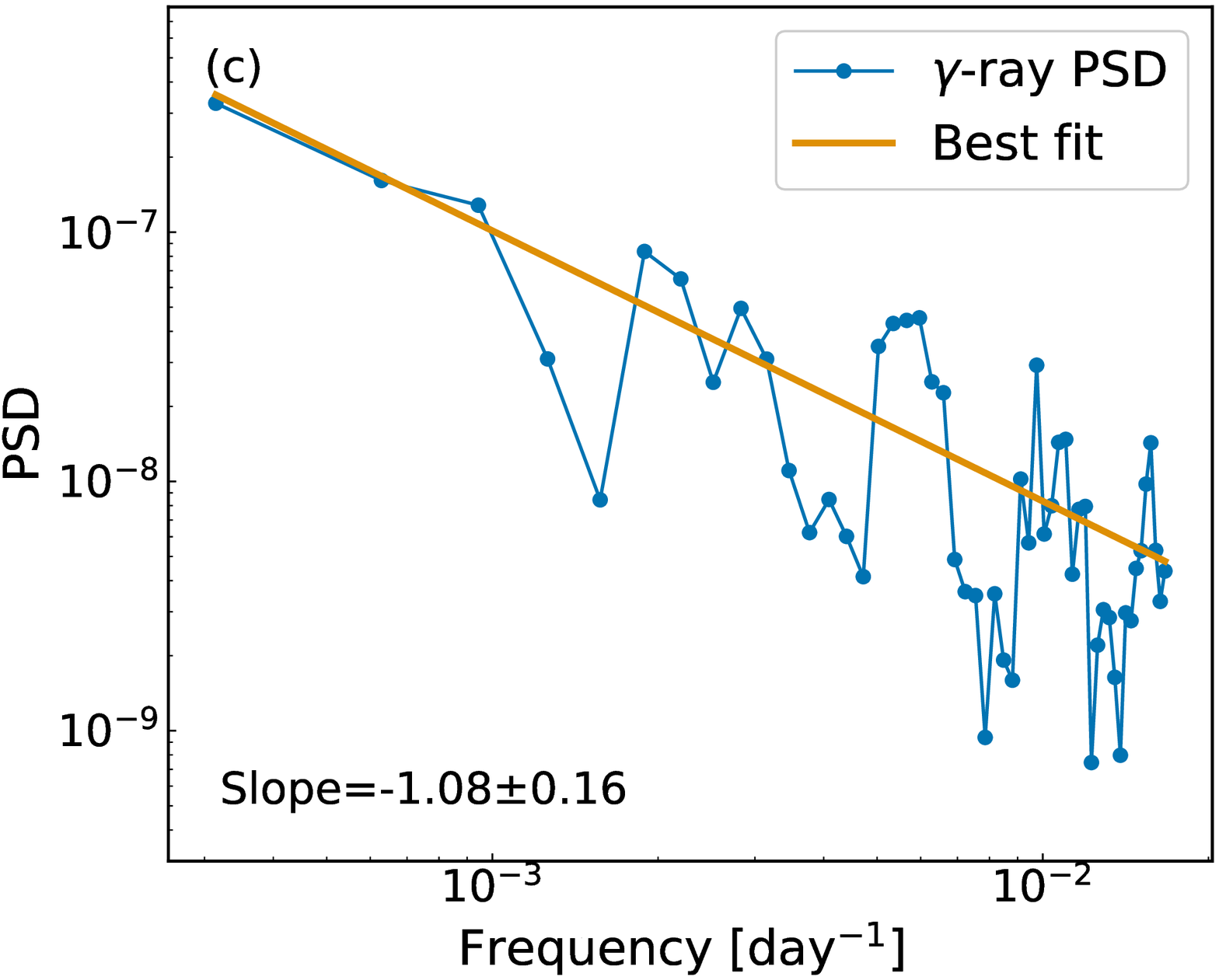}
   \includegraphics[angle=0,scale=0.24]{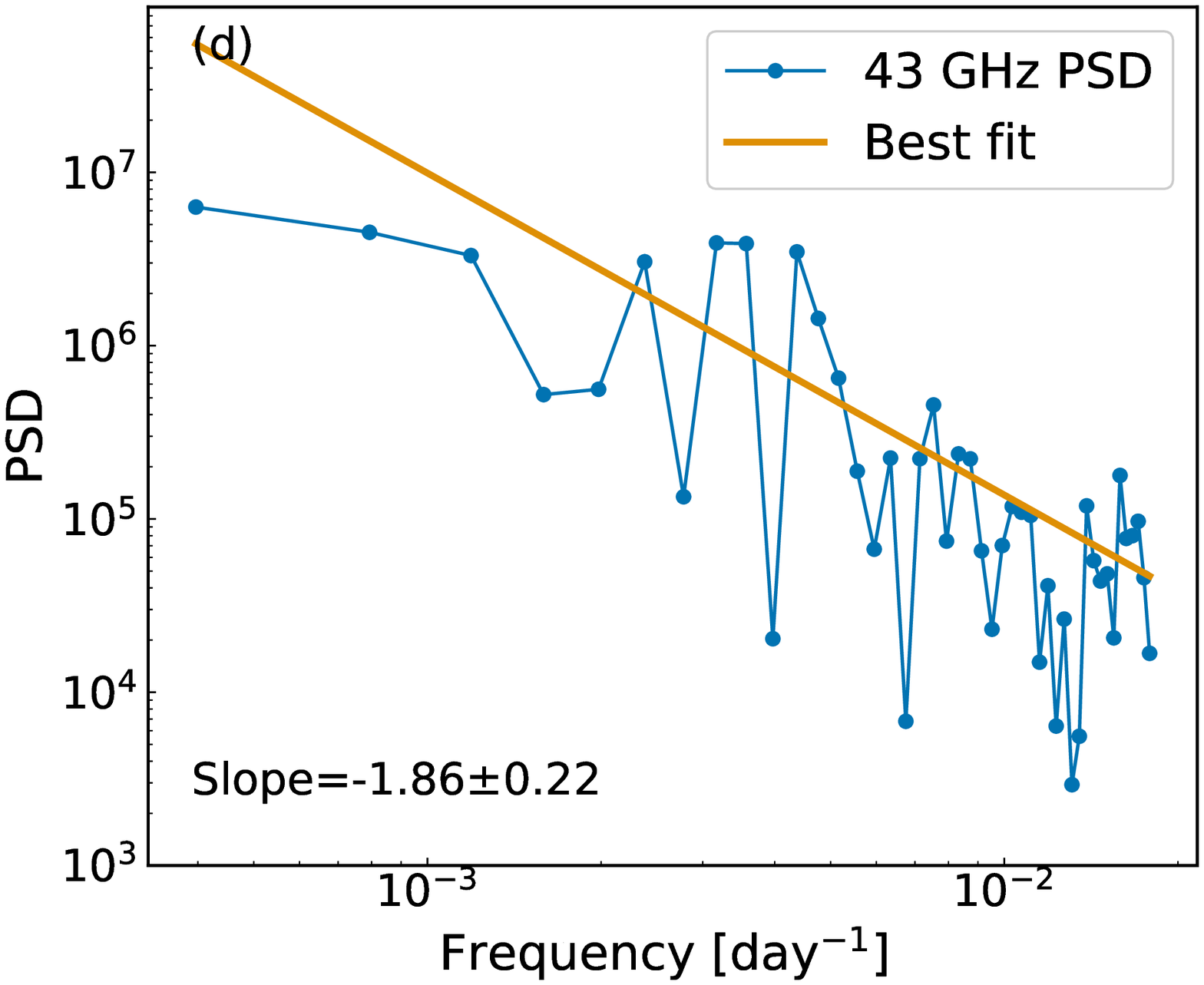}
   \includegraphics[angle=0,scale=0.25]{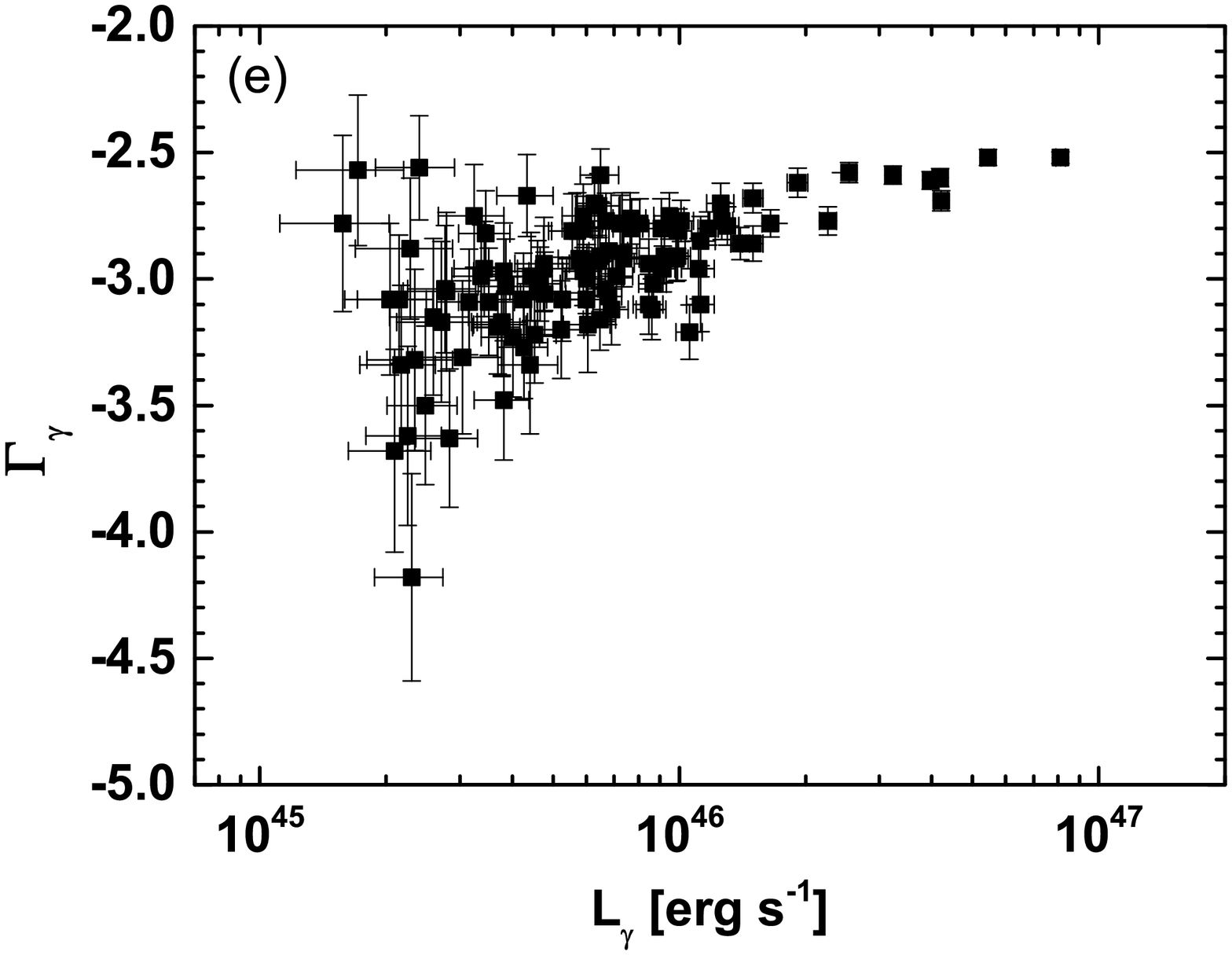}
   \includegraphics[angle=0,scale=0.30]{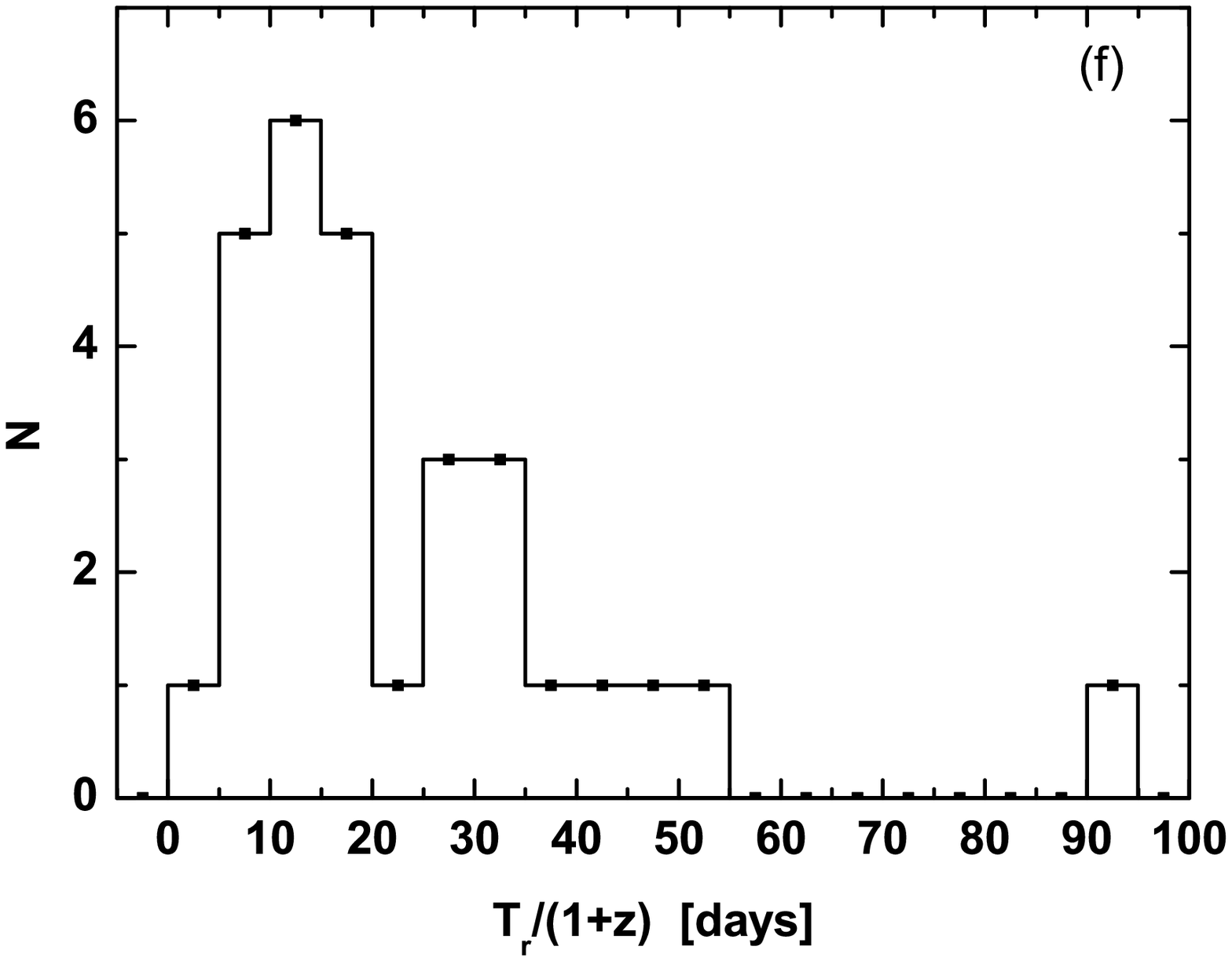}
   \includegraphics[angle=0,scale=0.30]{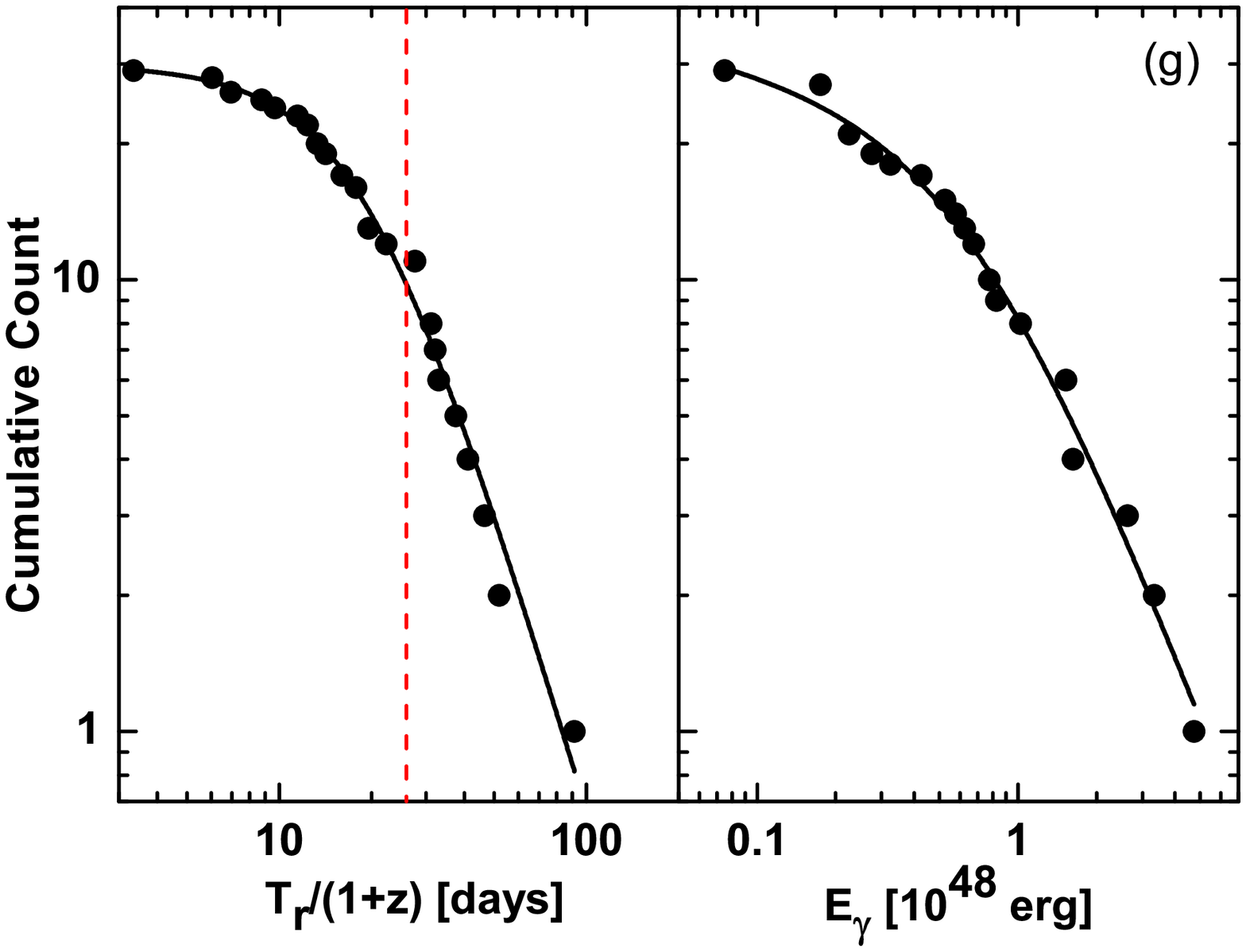}
\caption{Analysis of long-term light curves for 3C 273. Panel (a): Light curve observed by the \emph{Fermi}/LAT from 2008 August 6 (MJD 54684) to 2017 September 20 (MJD 57924) in time-bins of 30 days together with light curve in 43 GHz observed with the VLBA. The opened triangles in the $\gamma$-ray light curve represent the upper limits with TS$<$9. The green lines indicate the fitting results by Equation (1) and the red lines are the sum. The magenta vertical dashed lines indicate some radio flare peaks. Panel (a1): Same as Panel (a), but the time-axes of the light curves in the two energy bands are re-scaled according the DCF result (Panel (b)) between $\gamma$-ray and radio bands, i.e., 160 days. The coloured dash lines in Panel (b) show the confidence levels of random cross correlations, i.e., 1$\sigma$ (green line), 90\% (blue line), and 2$\sigma$ (red line), which are obtained by the Monte Carlo simulations (Max-Moerbeck et al. 2014). Panels (c) and (d): PSDs of the light curves in the $\gamma$-ray and 43 GHz band. The dark yellow lines display the best-fit results with the power-law function using the \emph{stingray} package. The derived slopes are also presented in each panel. Panel (e): $\Gamma_{\gamma}$ as a function of $L_{\gamma}$. Panel (f): Distribution of the rise time ($T_{\rm r}/(1+z)$, in the rest frame) for each exponential component in the $\gamma$-ray light curve (Panel (a)). Panel (g): Cumulative distributions of rise time ($T_{\rm r}/(1+z)$) and energy ($E_{\gamma}$) for each exponential component in the $\gamma$-ray light curve (Panel (a)). The red vertical dash line is representative of $30/(1+z)$ days. The black solid lines are the best fitting lines with Equation (2) for the cumulative distributions.}\label{LC_273}
\end{figure}

\begin{figure}
 \centering
   \includegraphics[angle=0,scale=0.7]{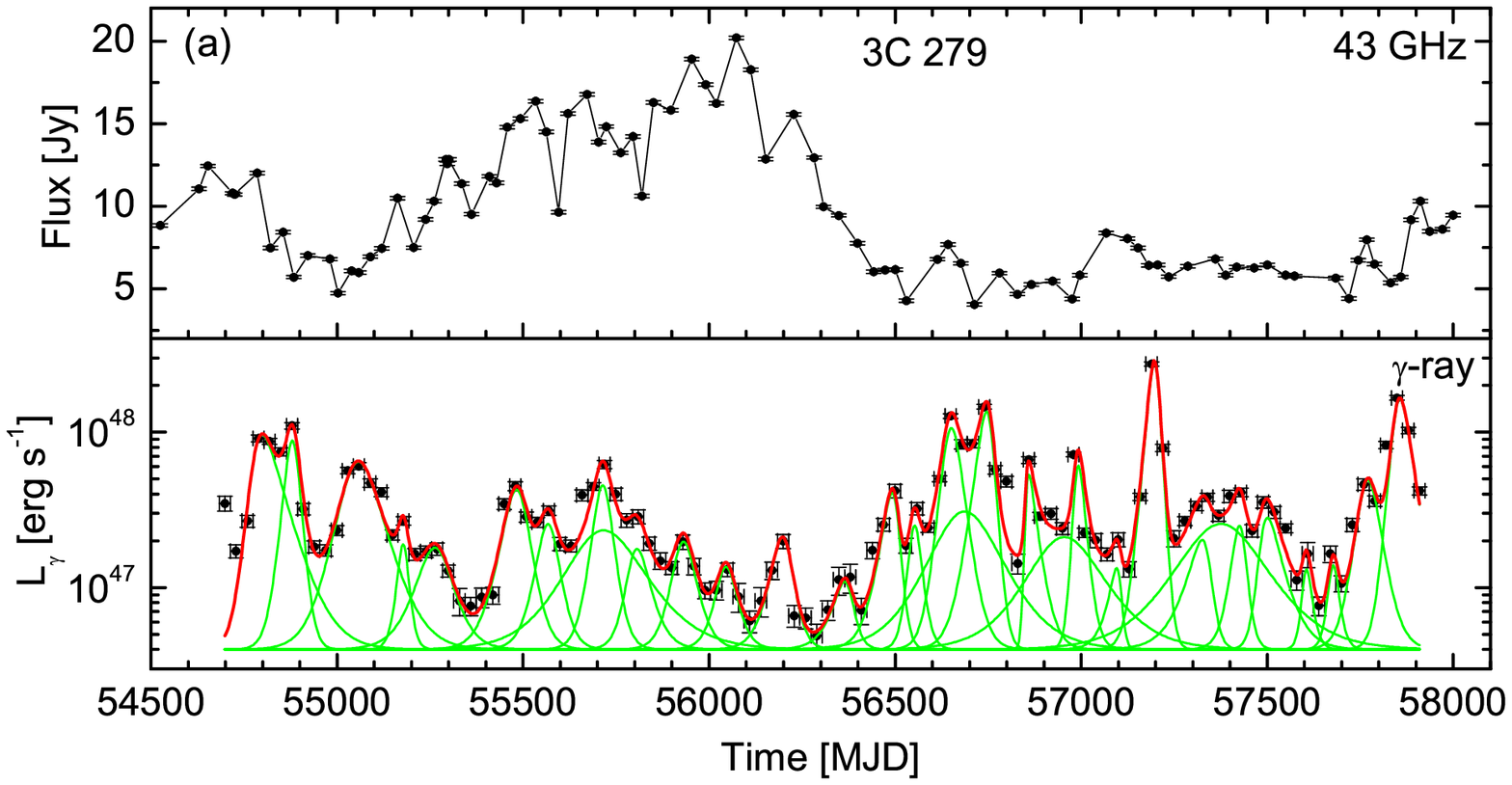}\\
   \includegraphics[angle=0,scale=0.26]{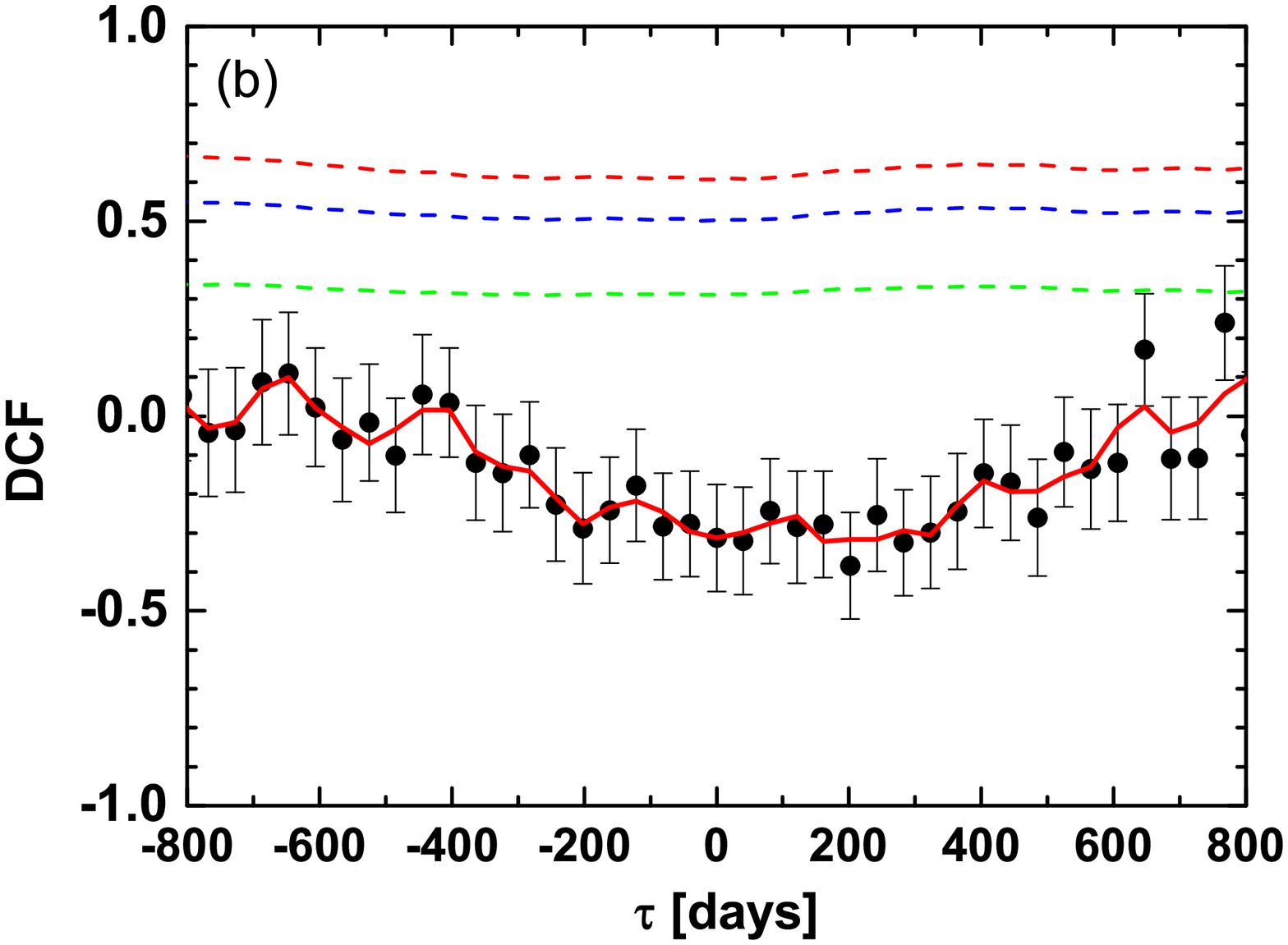}
   \includegraphics[angle=0,scale=0.24]{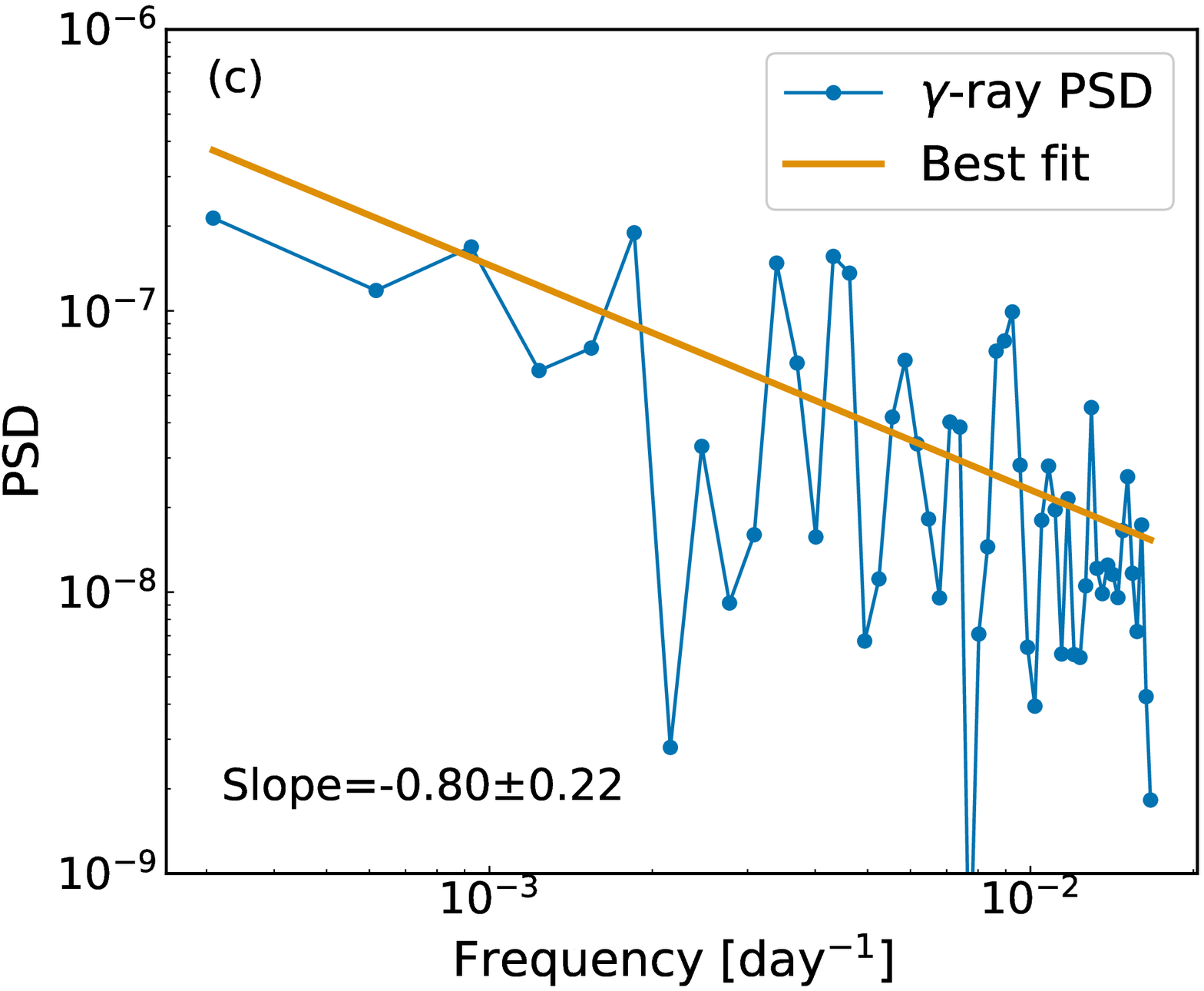}
   \includegraphics[angle=0,scale=0.24]{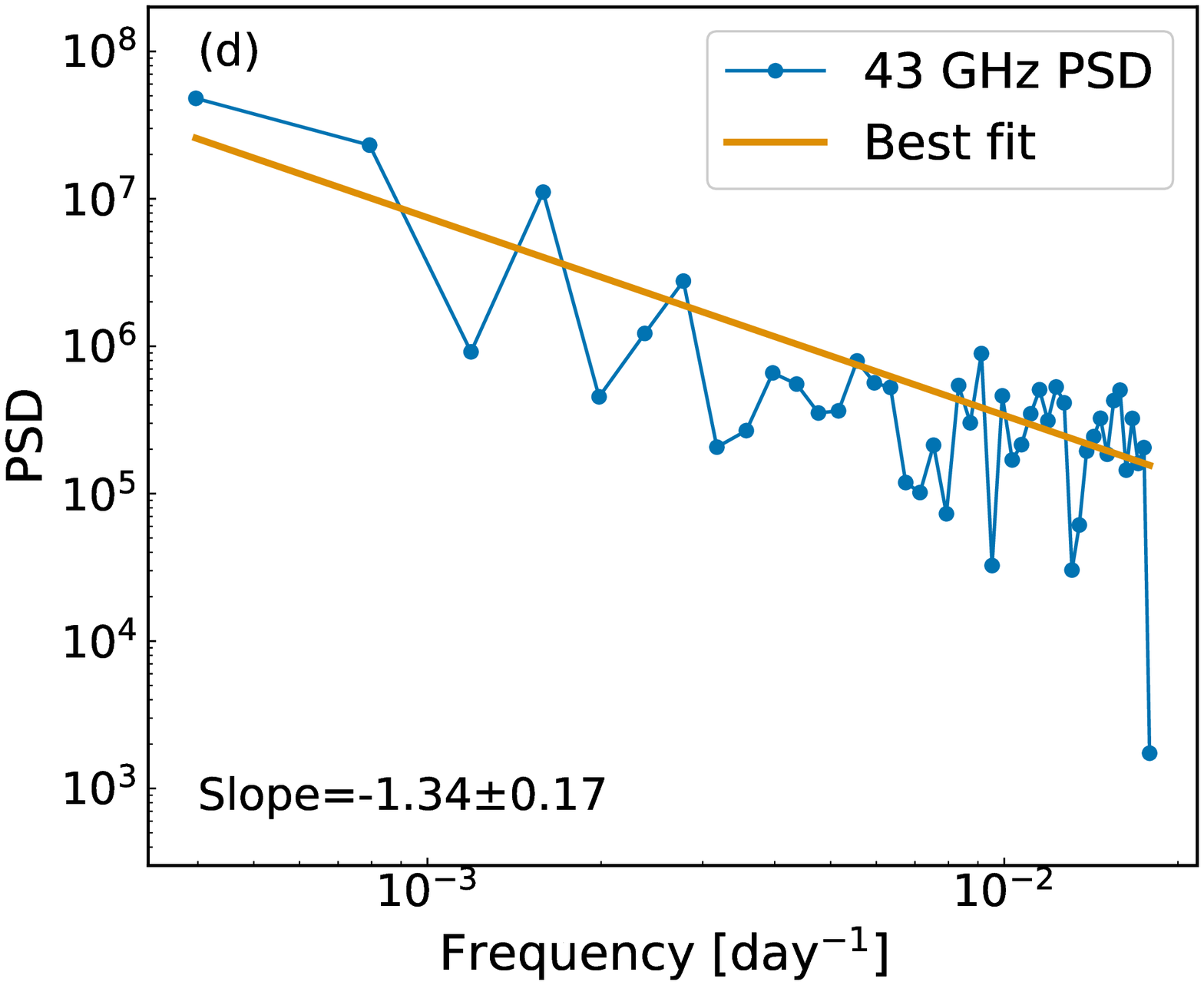}
   \includegraphics[angle=0,scale=0.25]{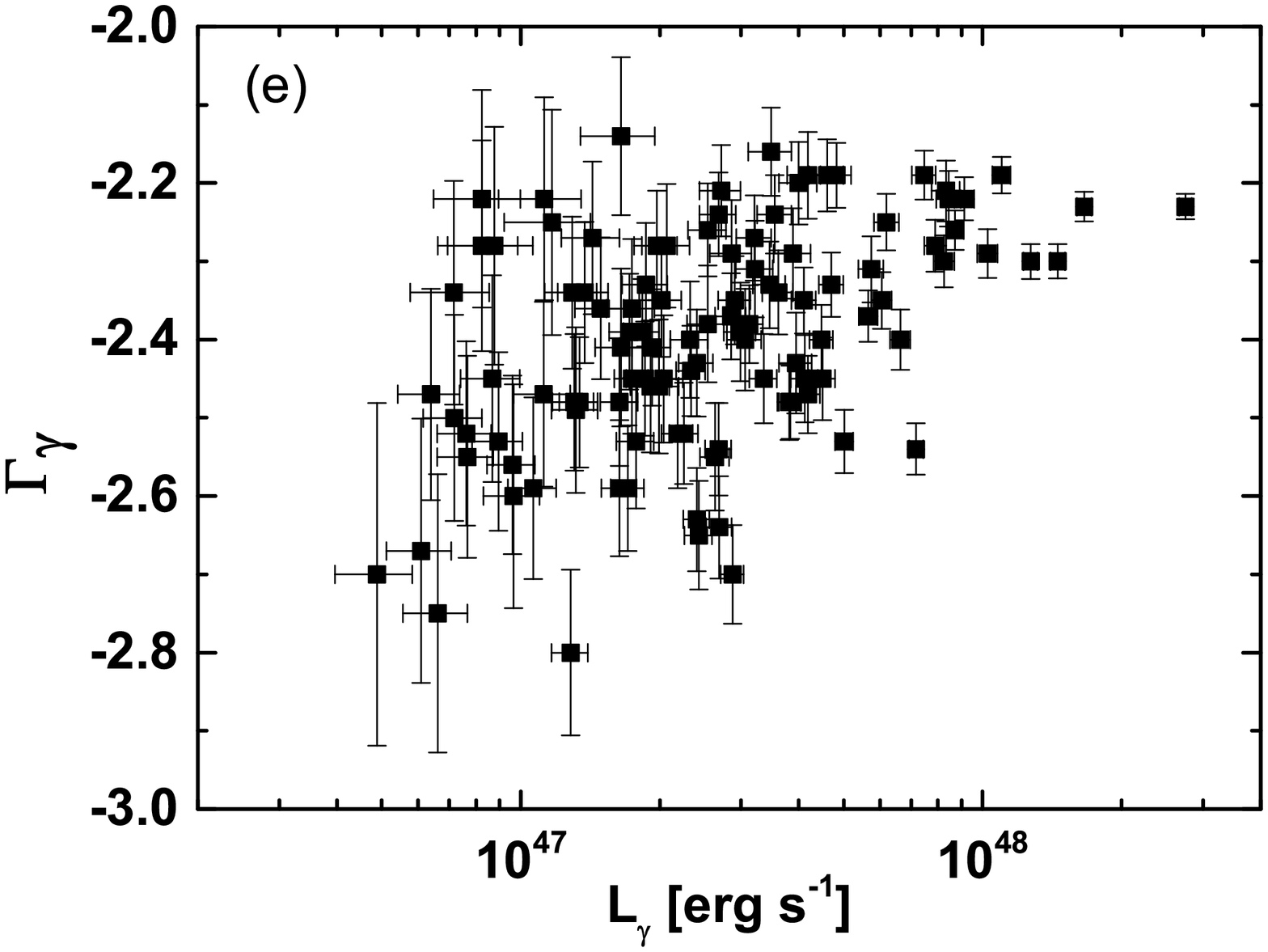}
   \includegraphics[angle=0,scale=0.30]{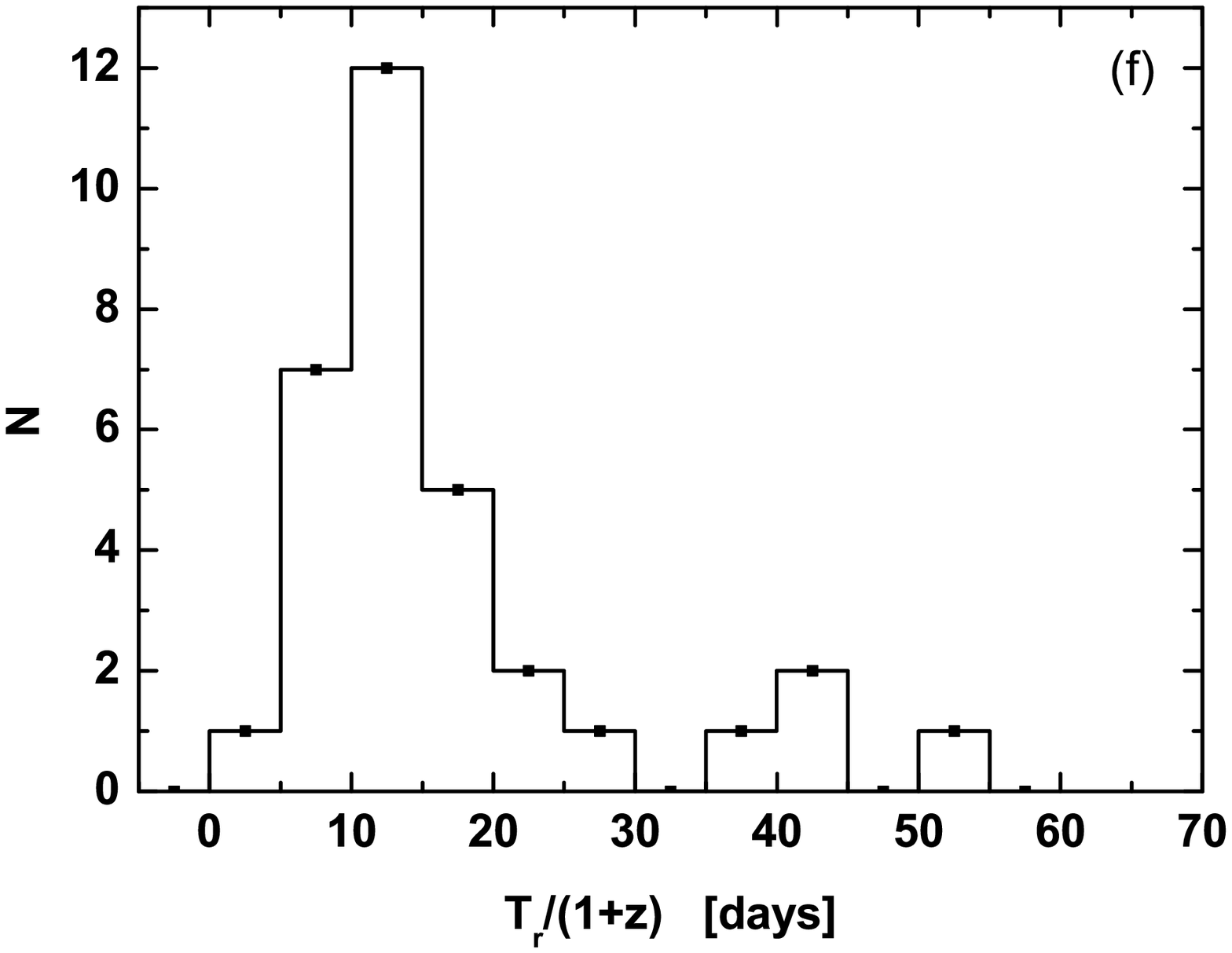}
   \includegraphics[angle=0,scale=0.30]{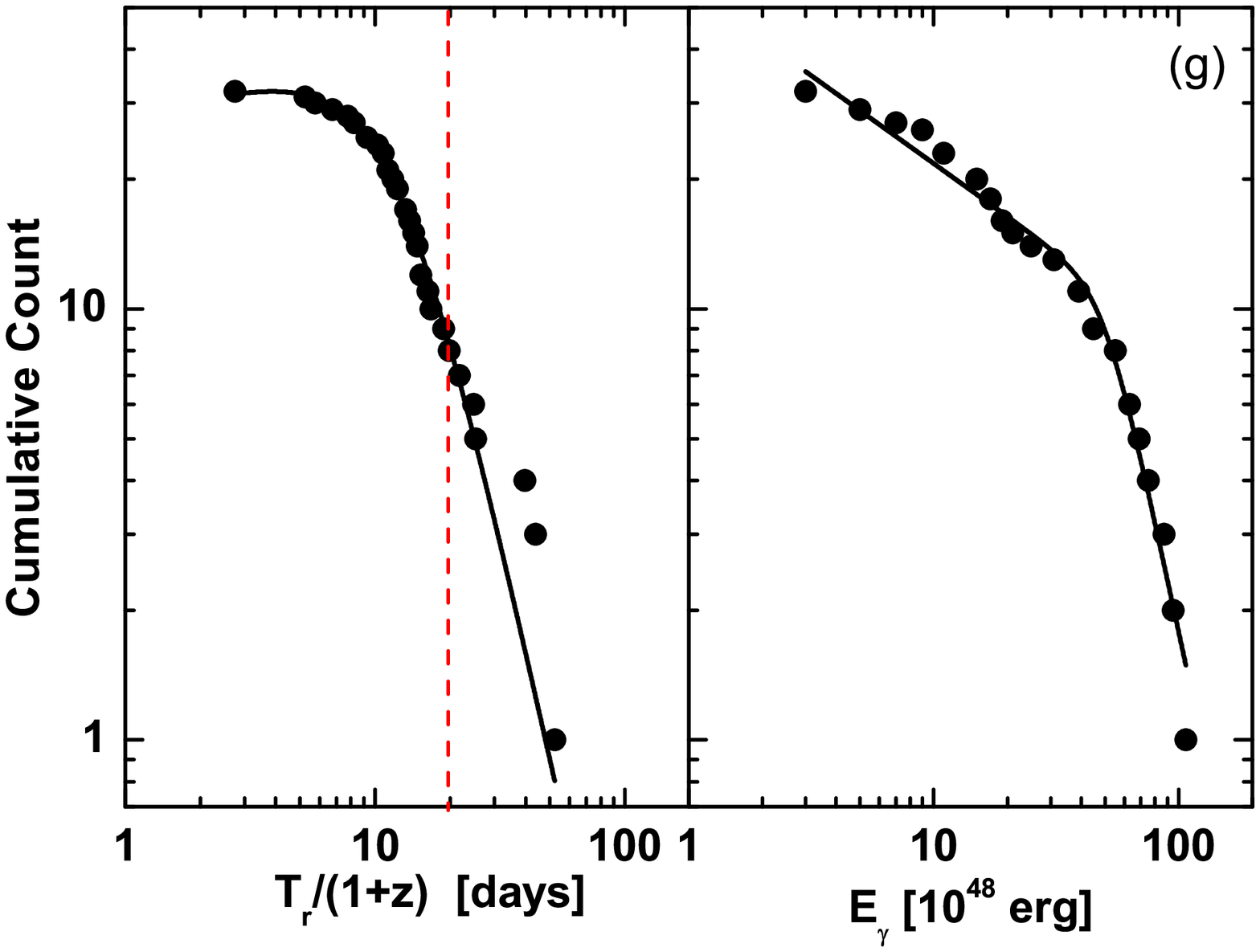}
\caption{Same as Figure 1, but for 3C 279.}\label{LC_279}
\end{figure}

\begin{itemize}

\item \emph{3C 273.} As illustrated in Figure \ref{LC_273}(a), the highest flux (on $\sim$MJD 55090) in the $\gamma$-ray light curve corresponds to a very low state at 43 GHz, but there is one obvious radio flare respectively before and after this $\gamma$-ray flare. The same feature is also observed for the $\gamma$-ray flare around $\sim$MJD 57370. According to the DCF result of variability between $\gamma$-ray and 43 GHz, the flux variations in the $\gamma$-ray band are correlated with and leading the radio flux variations about $\sim$160 days at a $2\sigma$ confidence level. We re-scale the time-axes of the light curves in the two energy bands and find that the five main radio flares are almost accompanied by the $\gamma$-ray flares as shown in Figure \ref{LC_273}(a1). The derived slopes of PSDs in the two energy bands are consistent with the red-noise process. $L_{\gamma}$ ranges from $1.6\times10^{45}$ to $8.1\times10^{46}$ erg s$^{-1}$ while $\Gamma_{\gamma}$ is in the range from $-4.18$ to $-2.52$. $\Gamma_{\gamma}$ is correlated with $L_{\gamma}$ with a Pearson correlation coefficient of $r=0.58$ and chance probability of $p=9.0\times10^{-11}$, displaying the behavior of ``harder when brighter". The $T_{\rm r}/(1+z)$ distribution clusters at 5--20 days. The slopes of the cumulative distributions for $T_{\rm r}/(1+z)$ and $E_{\gamma}$, which are obtained by fitting the cumulative distributions with Equation (2), are $\alpha_{\rm T}=2.27\pm0.21$ and $\alpha_{\rm E}=1.51\pm0.19$, respectively.

\item \emph{3C 279.} There is a large outburst superimposing several small flares in the radio light curve, which is not simultaneously observed in the $\gamma$-ray band, as displayed in Figure \ref{LC_279}(a). At the same time, the several big $\gamma$-ray flares (e.g., $\sim$MJD 56740 and MJD 57190) are also not accompanied by the radio flares. It seems like that the totally different variability behaviors are presented in the two energy bands, which is also consistent with the DCF result; no correlation of flux variation over $1\sigma$ confidence level is observed between the two energy bands. The derived PSD slopes in the two energy bands are also different; that in the radio band ($-1.34\pm0.17$) corresponds to the red-noise process while that in the $\gamma$-ray band ($-0.80\pm0.22$) is not. $L_{\gamma}$ ranges from $4.9\times10^{46}$ to $2.8\times10^{48}$ erg s$^{-1}$ while $\Gamma_{\gamma}$ is in the range from $-2.80$ to $-2.14$. However, only a correlation tendency is observed between $\Gamma_{\gamma}$ and $L_{\gamma}$ with $r=0.47$ and $p=2.7\times10^{-7}$. $T_{\rm r}/(1+z)$ clusters at 5--20 days. The derived slopes of the cumulative distributions for $T_{\rm r}/(1+z)$ and $E_{\gamma}$ are $\alpha_{\rm T}=2.62\pm0.22$ and $\alpha_{\rm E}=2.64\pm0.34$, respectively.

\item \emph{3C 454.3} The correlations of variability between $\gamma$-ray and radio bands for some individual flares have been widely reported for this source. We find that the long-term variability behaviors in the two energy bands are also very similar, however more spike flares are presented in the $\gamma$-ray light curve than in 43 GHz. The DCF result indicates that the flux variations of the long-term light curves in the two energy bands are correlated without time-lag at a $2\sigma$ confidence level. The PSD slopes in the two energy bands are also roughly consistent and correspond to the red-noise process. The largest luminosity variation among the eight blazars is observed in this source, almost changing three orders of magnitude, from $6.1\times10^{46}$ to $4.1\times10^{49}$ erg s$^{-1}$. $\Gamma_{\gamma}$ is strongly correlated with $L_{\gamma}$ with $r=0.70$ and $p=2.2\times10^{-16}$, showing the behavior of ``harder when brighter". $T_{\rm r}/(1+z)$ clusters at 5--20 days. The slopes of the cumulative distributions for $T_{\rm r}/(1+z)$ and $E_{\gamma}$ are $\alpha_{\rm T}=3.35\pm0.78$ and $\alpha_{\rm E}=1.23\pm0.14$, which are roughly compatible with those reported in Zhang et al. (2018). They analyzed the long-term \emph{Fermi}/LAT light curve in time-bins of 1-day and obtained $\alpha_{\rm T}=2.28\pm0.05$ and $\alpha_{\rm E}=1.46\pm0.02$.

\item \emph{CTA 102.} As displayed in Figure \ref{LC_102}(a), there are several big outbursts with smooth flux changing in the long-term light curve at 43 GHz while only spike flares are presented in the $\gamma$-ray band. There is an obvious flare at 43 GHz around MJD 55000 when the $\gamma$-ray emission is in a low state. The maximum luminosity variation of this source is also almost three orders of magnitude, from $1.2\times10^{47}$ to $6.0\times10^{49}$ erg s$^{-1}$, however no such big flux change is observed at 43 GHz. The DCF result also indicates that no correlation of flux variation over $1\sigma$ confidence level is observed between the two energy bands, but the PSD slopes in both bands are consistent with the red-noise process. The largest change of $\Gamma_{\gamma}$ among the eight sources is presented in CTA 102 from $-3.95$ to $-2.03$. $\Gamma_{\gamma}$ is also strongly correlated with $L_{\gamma}$ with $r=0.66$ and $p=7.9\times10^{-14}$. $T_{\rm r}/(1+z)$ narrowly clusters at 10--15 days. Note that the $T_{\rm r}/(1+z)$ cumulative distribution of CTA 102 does not show a very flat part in the short time regime like other seven blazars, and the slope ($4.47\pm1.19$) above the break is much larger than others, hence we take the slope below the break of the $T_{\rm r}/(1+z)$ cumulative distribution for this source, i.e., $\alpha_{\rm T}=1.03\pm0.38$ and $\alpha_{\rm E}=0.87\pm0.10$.

\item \emph{PKS 1222+216.} It seems that the global light curves in the two energy bands are very similar, three large outbursts superimposing some small flares. The DCF result also indicates that the flux variations between $\gamma$-ray and 43 GHz are statistically correlated at a $2\sigma$ confidence level, but it has a very broad peak, from $\sim$ $-360$ days to $-40$ days. The time-lag of the highest fluxes in the two energy bands (on MJD 55330 and MJD 55620) is $\sim$290 days, and then we re-scale the time-axes of the light curves in the two energy bands according to this time-lag. As displayed in Figure \ref{LC_1222}(a1), the flux variations in the two energy bands seem totally accordant in this case. The PSD slopes in the two energy bands are consistent within the errors and correspond to the red-noise process. $\Gamma_{\gamma}$ is correlated with $L_{\gamma}$ with $r=0.51$ and $p=1.7\times10^{-8}$. $T_{\rm r}/(1+z)$ narrowly clusters at 5--15 days, and we obtain $\alpha_{\rm T}=1.64\pm0.11$ and $\alpha_{\rm E}=1.49\pm0.45$.

\item \emph{PKS 1510--089.} There are many big flares in the both energy bands, however, the four main flares at 43 GHz correspond to different $\gamma$-ray emission states, as illustrated in Figure \ref{LC_1510}(a). The DCF result indicates that no significant correlation of flux variations over 90\% confidence level is observed between $\gamma$-ray and 43 GHz in the long-term light curves. The PSD slope of the radio band ($-1.18\pm0.18$) corresponds to the red-noise process while that in the $\gamma$-ray band ($-0.94\pm0.23$) is roughly consistent with the red-noise process. And that only a correlation tendency is observed between $\Gamma_{\gamma}$ and $L_{\gamma}$ with $r=0.47$ and $p=2.5\times10^{-7}$. $T_{\rm r}/(1+z)$ clusters at 5--20 days, and we obtain $\alpha_{\rm T}=3.13\pm0.36$ and $\alpha_{\rm E}=1.26\pm0.25$.

\item \emph{PKS 1633+382.} The flux variations between $\gamma$-ray and 43 GHz are correlated at a $2\sigma$ confidence level, as illustrated by the long-term light curves and the DCF result. The derived PSD slopes in the two energy bands are consistent within the errors and correspond to the red-noise process. The DCF curve of this source displays a broad bump approximately peaking at 0, likely indicating that the flux variations between $\gamma$-ray and 43 GHz are coincident without time-lag. $\Gamma_{\gamma}$ is correlated with $L_{\gamma}$ with $r=0.53$ and $p=4.8\times10^{-9}$. $T_{\rm r}/(1+z)$ narrowly clusters at 5--10 days, and we obtain $\alpha_{\rm T}=2.51\pm0.23$ and $\alpha_{\rm E}=1.47\pm0.15$.

\item \emph{S5 0716+714.} It is the only one BL Lac in our sample and also a TeV source. We can find that the obviously different variability behaviors are demonstrated in the two energy bands; it seems that several big outbursts superimposing small flares are displayed in the $\gamma$-ray light curve while only the spike flares are shown at 43 GHz. As illustrated by the DCF result, no correlation of flux variation over 90\% confidence level between the two energy bands is observed. It shows the flattest PSD at 43 GHz among the eight blazars with slope of $-0.72\pm0.25$. The PSD slope ($-1.05\pm0.32$) in $\gamma$-ray band is marginally consistent with the red-noise process. The strongest correlation between $\Gamma_{\gamma}$ and $L_{\gamma}$ is observed among the eight blazars with $r=0.80$ and $p\sim0$, so the obvious behavior of ``harder when brighter" is presented in this BL Lac. $T_{\rm r}/(1+z)$ clusters at 5--20 days, and we obtain $\alpha_{\rm T}=1.71\pm0.18$ and $\alpha_{\rm E}=2.59\pm0.21$.

\end{itemize}

Half the sources in our sample display that the flux variations between $\gamma$-ray and radio bands are correlated; two sources (3C 273 and PKS 1222+216) show a time-lag while no time-lag is observed in the other two sources (3C 454.3 and PKS 1633+382). No correlation of flux variation between the two energy bands is observed in the rest four blazars. Note that there are four TeV blazars in our sample, except for PKS 1222+216, the other three TeV blazars (3C 279, PKS 1510--089, and S5 0716+714) do not show the flux correlation in the long-term light curves between radio and $\gamma$-ray bands.

As displayed in Figure \ref{slope_PSD}, the PSD slopes of radio band are steeper than that of $\gamma$-ray band on average, suggesting a higher dominance of red-noise in radio band than in $\gamma$-ray band. It means that there are more rapid flux variations in $\gamma$-ray band than at 43 GHz for these blazars. It seems likely that the TeV blazars have the flatter PSDs than other blazars. It may indicate that the TeV blazars have more power at high frequencies of PSDs, or in short timescale variations than other blazars. As reported by Chidiac et al. (2016), the different slopes of PSDs might indicate the different dominant radiation processes responsible for the observed variations.

It is found that the TeV BL Lac S5 0716+714 displays the strongest correlation between $\Gamma_{\gamma}$ and $L_{\gamma}$, however, only the tendency of the $\Gamma_{\gamma}$--$L_{\gamma}$ correlation is presented in the two TeV FSRQs, 3C 279 and PKS 1510--089. The rest five FSRQs show the significant correlation between $\Gamma_{\gamma}$ and $L_{\gamma}$. Note that the GeV emission of the TeV BL Lac S5 0716+714 has the different radiation origin from that of FSRQs; it is produced by the synchrotron self-Compton (SSC; Zhang et al. 2012) process while the GeV emission of FSRQs is from the external Compton scattering (Zhang et al. 2013, 2014, 2015).

\section{Discussion}

\subsection{Relations between $\gamma$-ray Emission Region and Radio Core}

The radio core is observed at the location where the optical depth of synchrotron self-absorption is unit, and thus the core position offset at two frequencies can be used for estimating the physical parameters of jets (Lobanov 1998). The distance in pc of the observed radio core from the jet apex is given by (Lobanov 1998; Pushkarev et al. 2012)
\begin{equation}
r_{\rm core}(\nu)=\frac{\Omega_{\rm r\nu}}{\nu\sin\theta}\approx\frac{\Omega_{\rm r\nu}(1+\beta^2_{\rm app})^{1/2}}{\nu},
\end{equation}
where $\nu$ is the observed frequency in GHz, $\Omega_{\rm r\nu}$ is the core shift measure defined in Lobanov (1998), $\theta$ is the viewing angle, $\beta_{\rm app}$ is the apparent jet speed. The distance between $\gamma$-ray emission region and radio core can be calculated by (Pushkarev et al. 2010; Lisakov et al. 2017)
\begin{equation}
\Delta r=\frac{\delta\Gamma c\Delta t^{\rm obs}}{1+z}=\frac{\beta_{\rm app}c\Delta t^{\rm obs}}{(1+z)\sin\theta},
\end{equation}
where $\Gamma$ is the bulk Lorentz factor, $\delta$ is the Doppler factor, $c$ is the speed of light, and $\Delta t^{\rm obs}$ is the time-lag of flux variations between $\gamma$-ray and radio emission. The values of $\Omega_{\rm r\nu}$ and $\beta_{\rm app}$ together with the derived magnetic field strength ($B_1$) at 1 pc for the eight blazars are taken from Pushkarev et al. (2012) and given in Table 2. The distances ($r_{\rm core,43}$) between the radio core at 43 GHz and the jet apex of these blazars are estimated with Equation (5) and shown in Table 2. The corresponding magnetic field strengths of the radio core at 43 GHz is $B_{\rm core,43}=B_1(\frac{1~pc}{r_{\rm core,43}})$, which are also presented in Table 2.

As illustrated by the $T_{\rm r}$ distributions of the eight blazars, they peak at $\sim$15 days, which are consistent with the predicted timescale of flux changes if the $\gamma$-ray emission zone is located at the pc-scale distances from central engine (Sikora et al. 2009). The radiative region size can be estimated by $R\leq c\Delta t\delta/(1+z)$. If taking $\Delta t/(1+z)=15$ days and $\delta=15$ (the distribution median of a blazar sample, Zhang et al. 2014), we obtain $R\sim230$ light-days, i.e., $R\sim6\times10^{17}$ cm. The distance of the dissipation region from the black hole is assumed to be $10R$, which is larger than the BLR radii ($R_{\rm BLR}$) for 3C 273, 3C 279, 3C 454.3, and PKS 1510--089 ($R_{\rm BLR}$ are from Zhang et al. 2014). This likely implies that the $\gamma$-ray variabilities originate in the locations outside the BLR, maybe in the pc-scale and nearby the radio core.

The VLBI observations at 43 GHz have revealed the emergence of a new, superluminal component from the radio core around the time of the $\gamma$-ray flares for these blazars, likely indicating that the $\gamma$-ray flares are related with the activities of the central engine and the radiation region is nearby the radio core.

\begin{itemize}

\item \emph{3C 273.} Lisakov et al. (2017) suggested that the strong $\gamma$-ray flare around MJD 55090 is followed by the radio outburst and also accompanied by a newborn component. Using the time-lag between radio and $\gamma$-ray light curves, they estimated the location of the $\gamma$-ray emission zone, which is close to the jet apex and located at 2--7 pc upstream from the 43 GHz core. By investigating the variability activity in multiwavelength for the period of MJD 54600--56000, Chidiac et al. (2016) reported that the $\gamma$-ray emission is correlated with and leading the radio emission, and they suggested that the $\gamma$-ray emission region is located at a distance of $1.2\pm0.9$ pc from the jet apex. However, Rani et al. (2013) suggested that the $\gamma$-ray emission region is located within the BLR ($<$1.6 pc) and the multiple flares indicate the presence of multiple shock scenario at the base of the jet, which is also consistent with the emergences of four components during that period (Jorstad et al. 2012). There is also probably a new superluminal knot that passed the 43 GHz core (Marscher et al. 2012), which produced the $\gamma$-ray flare around MJD 54730.

\item \emph{3C 279.} Rani et al. (2018) reported that six bright $\gamma$-ray flares superimposing on a long-term outburst ($\sim$MJD 56600--56850) are observed, and two components are ejected during this extreme flaring activity; one (MJD 56611--56678) is accompanied by the first three flares indicating the 43 GHz core as the site of energy dissipation. another one (MJD 56790--56836) is ejected after the last three flares suggesting the energy dissipation located at upstream from the 43 GHz core nearby the black hole. However, by studying the long-term multiwavelength polarization observations after the $\gamma$-ray flare on MJD 56646, Kang et al. (2015) concluded that the $\gamma$-ray flare region may be different from the polarized radio emission region. Using the multiwavelength data over a time period of 6 yr (MJD 54500--56900), Pati\~{n}o-\'{A}lvarez et al. (2018) further reported that there are different variability behaviors in different activity periods, so they suggested that the dominant emission mechanism and the location of $\gamma$-ray emission may change over time.

\item \emph{3C 454.3.} Its $\gamma$-ray outbursts are accompanied by the activities in radio band and the variations of pc-scale jet structure, which have been widely reported. On the basis of the observations with the \emph{Fermi}/LAT and VLBA at 43.2 GHz in the period of January 2008 to March 2010 ($\sim$MJD 54466--55286), Schinzel et al. (2012) proposed that the $\gamma$-ray emission may originate from the pc-scale jet extending to 23 pc. Jorstad et al. (2013) reported that the three prominent $\gamma$-ray flares ($\sim$MJD 55167, 55294, 55520) are co-spatial with the lower frequency events, and two components with superluminal apparent speeds in the pc-scale jet are associated with the first and third flares, respectively. By studying the $\gamma$-ray burst from November 2010 to January 2011 (MJD 55480--55570) together with the simultaneous observations at millimeter and far-IR, the strong correlations of flux variability among multifrequency and the VLBA observations that a bright superluminal knot was blended with the core during the $\gamma$-ray outburst favor a $\gamma$-ray emission location within the radio core at 43 GHz (Wehrle et al. 2012).

\item \emph{CTA 102.} The $\gamma$-ray outburst detected by the \emph{Fermi}/LAT in 2012 September--October ($\sim$MJD 56171--56231) is coincident with flares at all the other frequencies and is related to the passage of a new superluminal knot (N4) through the 43 GHz core, and it was proposed that the $\gamma$-ray emission region is located at a short distance downstream of the radio core, more than 12 pc from the black hole (Casadio et al. 2015). The $\gamma$-ray flare in April 2013 (MJD 56387--56394) may be also due to the passage of component N4 through knot C1 (Casadio et al. 2015). Li et al. (2018) reported that the $\gamma$-ray emission spanning between 2015 June 1 and 2016 October 1 (MJD 57174--57662) is produced at the distance of 5.7--16.7 pc from the central engine and in the upstream of the 43 GHz core, and a helical jet is used to interpret the long-term variability in flux density, polarization degree and other parameters.

\item \emph{PKS 1222+216.} It was reported that a new superluminal knot passed the 43 GHz core during MJD 55300--55350 (Marscher et al. 2012). Recently, Lee et al. (2019) suggested that a jet component likely was ejected at the end of October 2014, which is coincident with the GeV $\gamma$-ray flare in the mid-November 2014 ($\sim$MJD 56976).

\item \emph{PKS 1510--089.} Marscher et al. (2010) reported that a new ejected superluminal knot is responsible for the eight major $\gamma$-ray flares occurred during MJD 54832--54982. The knot propagates through a helical magnetic field where the jet flow is accelerated and the $\gamma$-ray flares are erupted at a variety of locations as disturbances pass down the jet. By investigating the connection between the radio and $\gamma$-ray activities during 2007--2010, Orienti et al. (2011) suggested that some $\gamma$-ray flares are produced in the same region as the radio emission. The connection between four radio (43 GHz) and $\gamma$-ray flares with $\gamma$-ray emission leading about 54 days during January 2011 to September 2013 ($\sim$MJD 55500--56500) was reported and explained by a shock-in-jet model (Beaklini et al. 2017). A high state in the GeV $\gamma$-ray band during 2012 January 1 to April 7 (MJD 55927--56024) may be concurrent with the ejection of a new component from the 43 GHz core (Aleksi\'{c} et al. 2014). A high state in the TeV band is detected by the MAGIC on May 2015 (MJD 57150--57170), during which the source is also in a high state in the GeV band as show in Figure \ref{LC_1510}(a), and likely it was accompanied by the emission of a new jet component in radio band (Ahnen et al. 2017).

\item \emph{PKS 1633+382.} A new superluminal knot may pass the 43 GHz core during MJD 55050--55100 (Marscher et al. 2012), which is coincident with a $\gamma$-ray flare as displayed in Figure \ref{LC_1633}(a). By studying the multifrequency VLBI observations and the \emph{Fermi}/LAT observations during the period from March 2012 to August 2015 (MJD 56000--57200), it was found that two superluminal knots should be ejected in MJD $56185\pm30$ and MJD $56520\pm30$ from the radio core when the largest $\gamma$-ray fluxes were observed (Algaba et al. 2018b). And a significant correlation is found between radio and $\gamma$-ray fluxes with $\gamma$-ray leading radio emission of $70\pm40$ days, showing that they arise from different regions, located at $1\pm13$ pc and $40\pm13$ pc from the black hole, respectively (Algaba et al. 2018a).

\item \emph{S5 0716+714.} With the high-frequency VLBI kinematical study during September 2008 to October 2010 ($\sim$MJD 54720--55500), Rain et al. (2015) suggested that a moving disturbance from the base of the jet produces the $\gamma$-ray variations in the upstream of the 43 GHz core, close to the central engine (see also Rain et al. 2014), and the disturbance propagates along a helical path, so the Doppler factor variations are also responsible for the observed flux variations. Recently, Ahnen et al. (2018) suggested that its TeV outbursts in January 2015 originate from the entrance and exit of a superluminal knot in and out a recollimation shock in the inner jet.

\end{itemize}

These observations provide evidence for that the $\gamma$-ray emission is connected with the radio core. We should note that the variability investigations in multiwavelength for individual sources mainly focus on the period covering the large $\gamma$-ray outbursts, however, it is hard to say which $\gamma$-ray flare corresponds to which radio flare considering the long-term light curves for some sources. For example PKS 1510--089 as illustrated in Figure \ref{LC_1510}(a), the result of the DCF analysis between $\gamma$-ray and radio bands indicates that the long-term flux variations in the $\gamma$-ray band are not correlated with that at 43 GHz.

3C 454.3 and PKS 1633+382 have the similar emission behaviors in the $\gamma$-ray and radio bands according to the DCF and PSD analysis of their long-term light curves, and thus we propose that the emission regions of their $\gamma$-ray flares may be located at pc-scale. The lack of a significant time delay between $\gamma$-ray and radio bands likely indicates that the $\gamma$-ray emission regions are transparent at 43 GHz, however, considering the 30-day time-bin of the light curves and the 40-day bin-size of DCF, their $\gamma$-ray emission regions may be nearby the radio core, are not exactly the radio core. As listed in Table 2, the estimated distance of the 43 GHz core from the jet apex is 7.28 pc for 3C 454.3 and 14.53 pc for PKS 1633+382, respectively. The derived magnetic field strength by fitting the broadband spectral energy distributions (SEDs) is larger than that of the 43 GHz core ($B_{\rm core,43}$), indicating that their $\gamma$-ray emission regions may be nearby the 43 GHz core but be located at the upstream of the 43 GHz core.

For 3C 273 and PKS 1222+216, the long-term light curves between $\gamma$-ray and radio bands are also correlated, but there is a significant time-lag with $\gamma$-ray leading radio. Using Equation (6), we calculate the distance between $\gamma$-ray emission region and 43 GHz core. It is $\sim$ 27 pc with $\Delta t^{\rm obs}$=160 days for 3C 273. Since no observation data of core shift is available in the literature for 3C 273, we cannot assess the location of its 43 GHz core. It is found that 27 pc is larger than the $r_{\rm core,43}$ values of other blazars. The $\Delta r$ value of PKS 1222+216 ranges from 5.57 pc to 50.12 pc for $\Delta t^{\rm obs}$=40--360 days, which will also be larger than its $r_{\rm core,43}$ value for a long time-lag. As suggested by Lisakov et al. (2017), besides the optical depth effect, the larger size of the radio-emitting region than that of $\gamma$-rays can also give rise to a time delay. Both effects may work at this source, and thus a very broad peak is presented in its DCF curve. The absence of a cutoff in the GeV--TeV spectra of PKS 1222+216 indicates that the $\gamma$-ray emission lies outside the BLR (Aleksi\'{c} et al. 2011). Considering the long time-lags of flux variation between $\gamma$-ray and radio bands, we propose that the $\gamma$-ray radiation regions of 3C 273 and PKS 1222+216 may locate at the upstream of the 43 GHz core nearby their BLRs.

For the rest four blazars (S5 0716+714, 3C 279, CTA 102, and PKS 1510--089), no significant correlation of flux variation over 90\% confidence level between $\gamma$-ray and radio band is observed, and thus maybe there are multiple $\gamma$-ray emission regions, which may or may not be simultaneously active, located along the relativistic jet from the central engine (e.g., Brown 2013; Dotson et al. 2015). Pati\~{n}o-\'{A}lvarez et al. (2018) also suggested that the location of the $\gamma$-ray emission zone of 3C 279 may change depending on the activity state of the central engine. As given in Table 2, the derived magnetic field strengths of the four blazars by fitting their broadband SEDs are larger than their $B_{\rm core,43}$, hence we propose that their $\gamma$-ray emission regions should be located at the upstream of the 43 GHz core.

\subsection{Variability Mechanisms of $\gamma$-ray Emission}

$\alpha_{\rm T}$ vs. $\alpha_{\rm E}$ of the eight blazars are plotted in Figure \ref{slope_TE} and no correlation between them is observed. Wang et al. (2015) reported that the statistical distribution characteristics of X-ray flares in the Sun, Sgr A* (see also Li et al. 2015), and M87 are similar and can be explained using a three-dimensional fractal-diffusive self-organized criticality (SOC) model (see also for Mrk 421, Yan et al. 2018), i.e., $\alpha_{\rm T}=(S+1)/2$ and $\alpha_{\rm E}=3(S+1)/(S+5)$ and $S=$1, 2, and 3 are the Euclidean dimensions (Aschwanden 2012). And they suggested that the radiation regions (or the relativistic jets) are magnetically dominated in these systems. The values of $\alpha_{\rm T}$ and $\alpha_{\rm E}$ for the $\gamma$-ray flares in the eight blazars are roughly consistent with the SOC model; 3C 273, 3C 454.3, PKS 1510--089; PKS 1633+382, together with 3C 279 may correspond to the three-dimensional case. PKS 1222+216 and S5 0716+714 are for the two-dimensional scenario while CTA 102 is the one-dimensional case.

Note that the \emph{Fermi}/LAT light curves are obtained in time-bins of 30 days for the high confidence in this paper, but they may lose some variability profiles with short timescales. For the brightest blazar 3C 454.3, its long-term \emph{Fermi}/LAT light curve in time-bins of 1-day has been analyzed by Zhang et al. (2018). They obtained $\alpha_{\rm T}=2.28\pm0.05$ and $\alpha_{\rm E}=1.46\pm0.02$, which are more consistent with the values of $\alpha_{\rm T}$ and $\alpha_{\rm E}$ for $S$=3 case than that in this paper, and reported that the nature of 3C 454.3 may be consistent with the SOC system. As illustrated in the long-term $\gamma$-ray light curves of these blazars, it seems that there are two variability components for some blazars. As displayed in Figures 1--8, each $T_{\rm r}/(1+z)$ distribution is roughly separated two parts, two peaks or one peak and some discrete distributions. The cumulative distributions of $T_{\rm r}/(1+z)$ also show the slight difference at the long-time regime. They may be related to the two variability components, respectively. Especially for 3C 279 and S5 0716+714, the steeper $\alpha_{\rm E}$ than other sources may be due to the different variability components in the light curves. It likely implies that the long-timescale variability components in the light curves of these blazars may have the different physical origin from the fast spike flares.

In this work we focus on the long-term variabilities of these blazars and thus the $\gamma$-ray light curves are derived in time-bins of 30 days for the high confidence, however, the fast minute-scale and hour-scale variabilities have been observed in many blazars by the \emph{Fermi}/LAT (Brown et al. 2013; Ackermann et al. 2016; Paliya et al. 2015; Shukla et al. 2018). The short timescale variabilities may be due to the local disturbance or be from an emission blob smaller than the jet cross section (Zacharias \& Schlickeiser 2013).

The existence of a long timescale variability component with overlapping fast spike flares may suggest that the causes of variabilities in blazars are diverse; there may be more than one physical mechanism that contributes the observed variability. The long timescale variability component (week or month timescale) is attributed to the central engine activity, or the emergence of new component, and is interpreted in the context of a shock model. The origin of the fast spike flares may be due to the magnetic reconnection process or turbulence in the local emission region. Hence the derived values of $\alpha_{\rm T}$ and $\alpha_{\rm E}$ by the light curves in time-bins of 30 days are roughly but not well consistent with the SOC system. Both mechanisms can accelerate the particles to higher energy and result in the harder spectra when the sources are brighter. The similar mechanisms are also used to explain the light curves of $\gamma$-ray bursts (e.g., Zhang \& Yan 2011; Gao et al. 2012).

Generally the $\gamma$-ray flux variations are more violent with shorter timescale than the radio variabilities, and not all of the $\gamma$-ray flares are accompanied by the radio flares, hence not all of the $\gamma$-ray flares are connected with the central activities. Casadio et al. (2015) reported that the powerful $\gamma$-ray emission of CTA 102 is associated with a change in direction of the jet, which became oriented more closely to the line of sight, and the radiation blob or the shock wave moves along a helical path (see also Larionov et al. 2016). The long-term variabilities in multifrequency of PKS 1633+382 are also interpreted as an inhomogeneous bent jet, and the different emitting regions account for the variations in the viewing angle and the Doppler factor (Raiteri et al. 2012). It was also suggested that the different intensity variations in different frequencies of 3C 273 are due to a change of the Doppler factor in the processing jet, which has a large effect on boosting at high frequencies (Beaklini \& Abraham 2014). The change in the Doppler-boosting factor may also give rise to the behavior of ``harder when brighter" (e.g., Liu et al. 2010; Fan et al. 2017). So some flux variations of blazars may be due to the geometric effect.

In addition, the weak correlation between $\Gamma_{\gamma}$ and $L_{\gamma}$ for the two TeV FSRQs 3C 279 and PKS 1510--089 among the eight blazars likely implies that the change of the external photon field may contribute to the part variations of the $\gamma$-ray flux.

\section{Summary}

We have compiled the $\gamma$-ray long-term light curves in time-bins of 30 days together with the long-term light curves at 43 GHz of eight blazars, for which there is at least one outburst with the flux variation more than one order of magnitude in the \emph{Fermi}/LAT energy band. The multiple exponential functions are used to fit the global $\gamma$-ray light curves in order to investigate the variability characters in the $\gamma$-ray band, and we use the DCF and PSD methods to analyze the long-term light curves in the two energy bands. The significant correlation of flux variations between $\gamma$-ray and 43 GHz is observed for 3C 454.3 and PKS 1633+382 without time-lag and for 3C 273 and PKS 1222+216 with $\gamma$-ray leading radio band. Adding to the similar PSD slopes in the $\gamma$-ray and radio bands, we suggested that the $\gamma$-ray emission regions for 3C 454.3 and PKS 1633+382 may be located nearby the radio core while that for 3C 273 and PKS 1222+216 may be at the upstream of the radio core. For the rest four blazars, there might be multiple $\gamma$-ray emission regions, which may or may not be simultaneously active, located along the relativistic jet from the central engine.

All the eight blazars display the behavior of ``harder when brighter" in the $\gamma$-ray band, especially for S5 0716+714. BL Lac S5 0716+714 indeed presents some different features from the seven FSRQs. This may be due to the intrinsically different jet properties between BL Lacs and FSRQs (e.g., Zhang et al. 2014). It was also found that there may be two variability components in the long-term $\gamma$-ray light curves of these blazars, long timescale variability components and fast spike flares. We thus proposed that there may be more than one physical mechanism that contributes the observed variability; the long timescale variability component is attributed to the central engine activity and is interpreted in the context of a shock model while the fast spike flares may be due to the magnetic reconnection process or turbulence in the local emission region.

\begin{ack}
This study makes use of radio data from the VLBA-BU Blazar Monitoring Program (VLBA-BU-BLAZAR; http://www.bu.edu/blazars/VLBAproject.html), funded by NASA through the Fermi and Swift Guest Investigator Programs. This work is supported by the National Natural Science Foundation of China (grants 11973050, 11573034, 11533003, 11851304, 11863007, and U1731239). En-Wei Liang acknowledges support from the special funding from the Guangxi Science Foundation for Guangxi distinguished professors (grant 2017AD22006 for Bagui Yingcai \& Bagui Xuezhe).

\end{ack}

\begin{landscape}
\begin{table}[h]
  \caption{Results of Light Curves in $\gamma$-ray Band. }
  \label{table1}
  \begin{tabular}{lcccccccccc} \hline\hline
  Source\footnotemark[a] &$z$&$L_{\gamma}^{\rm max}$&$L_{\gamma}^{\rm min}$&$\Gamma_{\gamma}^{\rm max}$&$\Gamma_{\gamma}^{\rm min}$&$\kappa$&$\alpha_{\rm T}$&$\alpha_{\rm E}$&$r$\footnotemark[b]&$p$\footnotemark[b]\\
   & &erg s$^{-1}$&erg s$^{-1}$& &&&&&\\
  \hline
  3C 273$^{\rm Q}$&0.158 &$(8.12\pm0.21)\times10^{46}$& $(1.58\pm0.46)\times10^{45}$ & $-2.52\pm0.02$ &  $-4.18\pm0.41$ &29&$2.27\pm0.21$&$1.51\pm0.19$&0.58&9.0$\times10^{-11}$\\
  3C 279$^{\rm Q}$&0.536 & $(2.75\pm0.08)\times10^{48}$ & $(4.89\pm0.93)\times10^{46}$ & $-2.14\pm0.10$ & $-2.80\pm0.11$&32 & $2.62\pm0.22$ & $2.64\pm0.34$&0.47&2.7$\times10^{-7}$\\
  3C 454.3$^{\rm Q}$ &0.859 & $(4.06\pm0.05)\times10^{49}$ & $(6.14\pm1.95)\times10^{46}$ & $-2.07\pm0.01$ & $-3.03\pm0.29$&23 &$3.35\pm0.78$& $1.23\pm0.14$&0.70&2.2$\times10^{-16}$\\
  CTA 102$^{\rm Q}$  &1.037 & $(6.00\pm0.12)\times10^{49}$ & $(1.17\pm0.45)\times10^{47}$ & $-2.03\pm0.01$ &$-3.95\pm0.45$&29 & $1.03\pm0.38$&$0.87\pm0.10$&0.66&7.9$\times10^{-14}$\\
  PKS 1222+216$^{\rm Q}$&0.434&$(1.86\pm0.07)\times10^{48}$ & $(8.47\pm3.57)\times10^{45}$ &$-2.07\pm0.01$ & $-2.93\pm0.18$&38 & $1.64\pm0.11$&$1.49\pm0.45$&0.51&1.7$\times10^{-8}$\\
  PKS 1510--089$^{\rm Q}$&0.360&$(1.16\pm0.03)\times10^{48}$&$(4.83\pm0.49)\times10^{46}$ & $-2.09\pm0.02$ & $-2.62\pm0.07$&27 & $3.13\pm0.36$ &$1.26\pm0.25$&0.47&2.5$\times10^{-7}$\\
  PKS 1633+382$^{\rm Q}$&1.813&$(1.91\pm0.07)\times10^{49}$ & $(6.41\pm1.61)\times10^{47}$ &$-2.28\pm0.06$&$-2.96\pm0.22$&36 & $2.51\pm0.23$ & $1.47\pm0.15$&0.53&4.8$\times10^{-9}$\\
  S5 0716+714$^{\rm B}$ &0.300 & $(2.51\pm0.19)\times10^{47}$ & $(3.77\pm0.95)\times10^{45}$ & $-1.80\pm0.03$ & $-2.61\pm0.16$&35 & $1.71\pm0.18$ & $2.59\pm0.21$&0.80&$\sim$0\\
  \hline
  \end{tabular}
  \begin{tabnote}
   \footnotemark[a] The superscripts denote the different types of blazars, ``Q" for FSRQ and ``B" for BL Lac.\\
   \footnotemark[b] The correlation coefficient ($r$) and chance probability ($p$) between $\Gamma_{\gamma}$ and $L_{\gamma}$ using the Pearson correlation analysis.\\
     \end{tabnote}
\end{table}
\end{landscape}

\clearpage
\begin{table*}[htdp]
\caption{Derived Parameters of Sources with the VLBI Observations.}
\label{tab:2}
\begin{center}
\begin{tabular}{lccccccccc}
\hline\hline
  Source&$z$&$\Omega_{\rm r\nu}$ &$B_1$& $\theta$& $\beta_{\rm app}$& $r_{\rm core,43}$  &$B_{\rm core,43}$&  $B_{\rm BLR}$\footnotemark[a]  &$B_{\rm IR}$\footnotemark[a]\\
   & &[pc GHz]&[G]& &&[pc]&[G]&[G]&[G]\\
  \hline
  3C 273&0.158&--&--&3.3&13.40&--&--&8.5$\pm$1.6$^{\rm Z15}$&1.13$\pm$0.11$^{\rm K14}$\\
  3C 279&0.536&5.88&0.42&2.4&20.57&2.82&0.15&5.9$\pm$0.3$^{\rm Z15}$&0.48$\pm$0.04$^{\rm K14}$\\
  3C 454.3&0.859&22.00&1.13&1.3&14.19&7.28&0.16&5.1$\pm$0.8$^{\rm Z15}$&0.85$\pm$0.20$^{\rm K14}$\\
  CTA 102&1.037&46.48&2.12&3.7&15.41&16.69&0.13&$\sim$4.1$^{\rm P18}$&$\sim$1.01$^{\rm G18}$\\
  PKS 1222+216&0.432&17.03&0.90&5.1&21.10&8.37&0.11&--&$\sim$0.35$^{\rm Ku}$\\
  PKS 1510--089&0.36&13.50&0.73&3.4&20.14&6.33&0.12&3.1$\pm$0.5$^{\rm Z15}$&$\sim$0.29$^{\rm A17}$\\
  PKS 1633+382&1.814&21.21&1.62&2.6&29.45&14.53&0.11&$\sim$1.5$^{\rm G10}$&--\\
  S5 0716+714&0.31&10.16&0.49&5.3&10.06&2.39&0.21&$<$0.7$^{\rm Z12}$&--\\
  \hline
\end{tabular}
\end{center}
\begin{tabnote}
  1. The values of $\Omega_{\rm r\nu}$, $B_1$, and $\beta_{\rm app}$, are taken from Pushkarev et al. (2012). Note that no data for 3C 273 are available in the literature. The values of $\theta$ are from Savolainen et al. (2010). \\
  2. \footnotemark[a]$B_{\rm BLR}$ and $B_{\rm IR}$ denote the derived magnetic field strength by fitting the broadband SEDs, i.e., the seed photons originate from the BLR and torus, respectively. For BL Lac S5 0716+714, the magnetic field strength is derived by considering the SSC process. The superscripts denote the references: ``Z15" is Zhang et al. (2015); ``K14" is Kang et al. (2014); ``P18" is Prince et al. (2018); ``G18" is Gasparyan et al. (2018); ``Ku" is Kushwaha et al. (2014); ``A17" is Ahnen et al. (2017); ``G10" is Ghisellini et al. (2010); ``Z12" is Zhang et al. (2012).\\
\end{tabnote}
\end{table*}

\clearpage

\clearpage

\clearpage
\begin{figure}
 \centering
   \includegraphics[angle=0,scale=0.7]{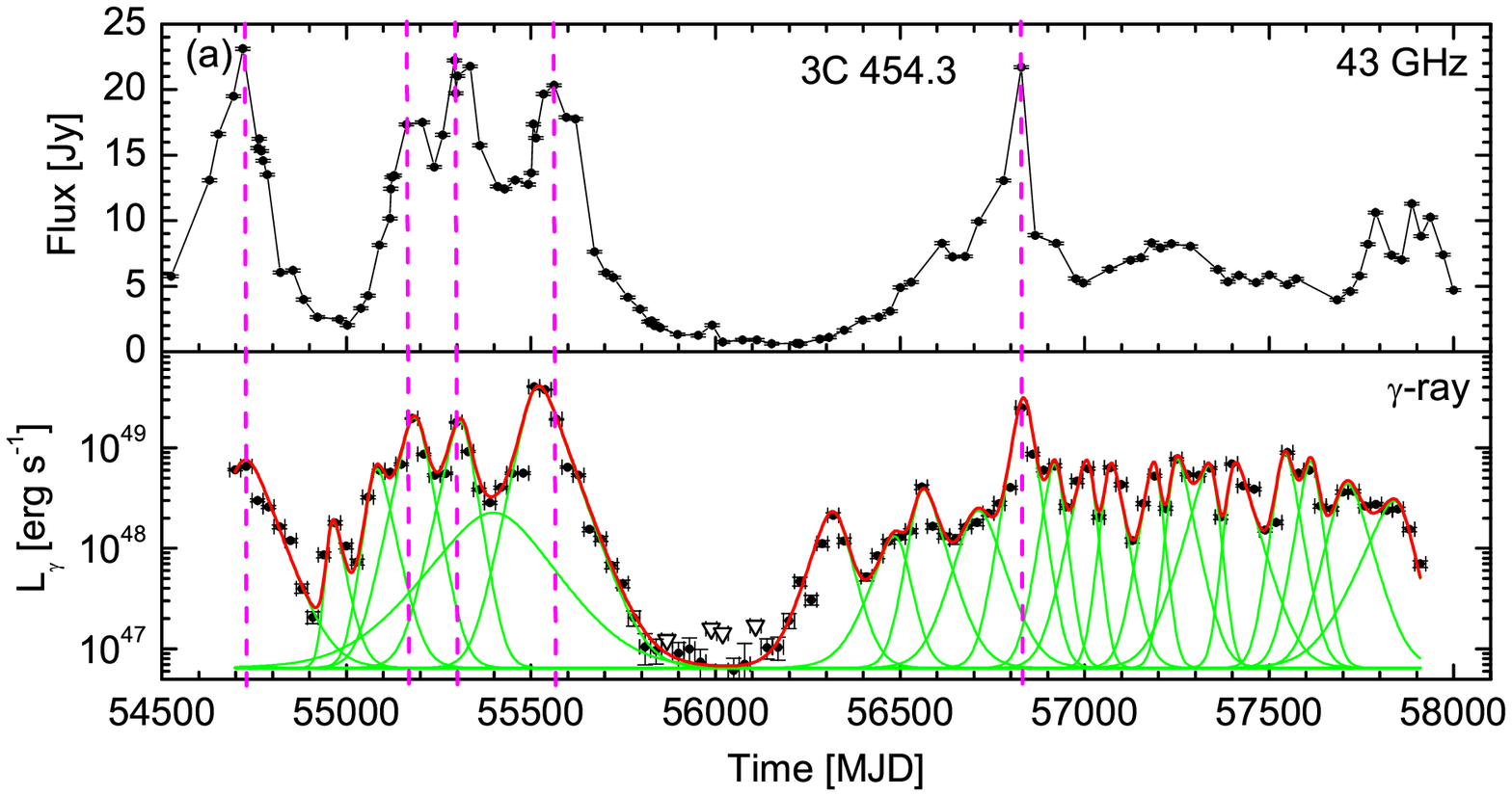}\\
   \includegraphics[angle=0,scale=0.26]{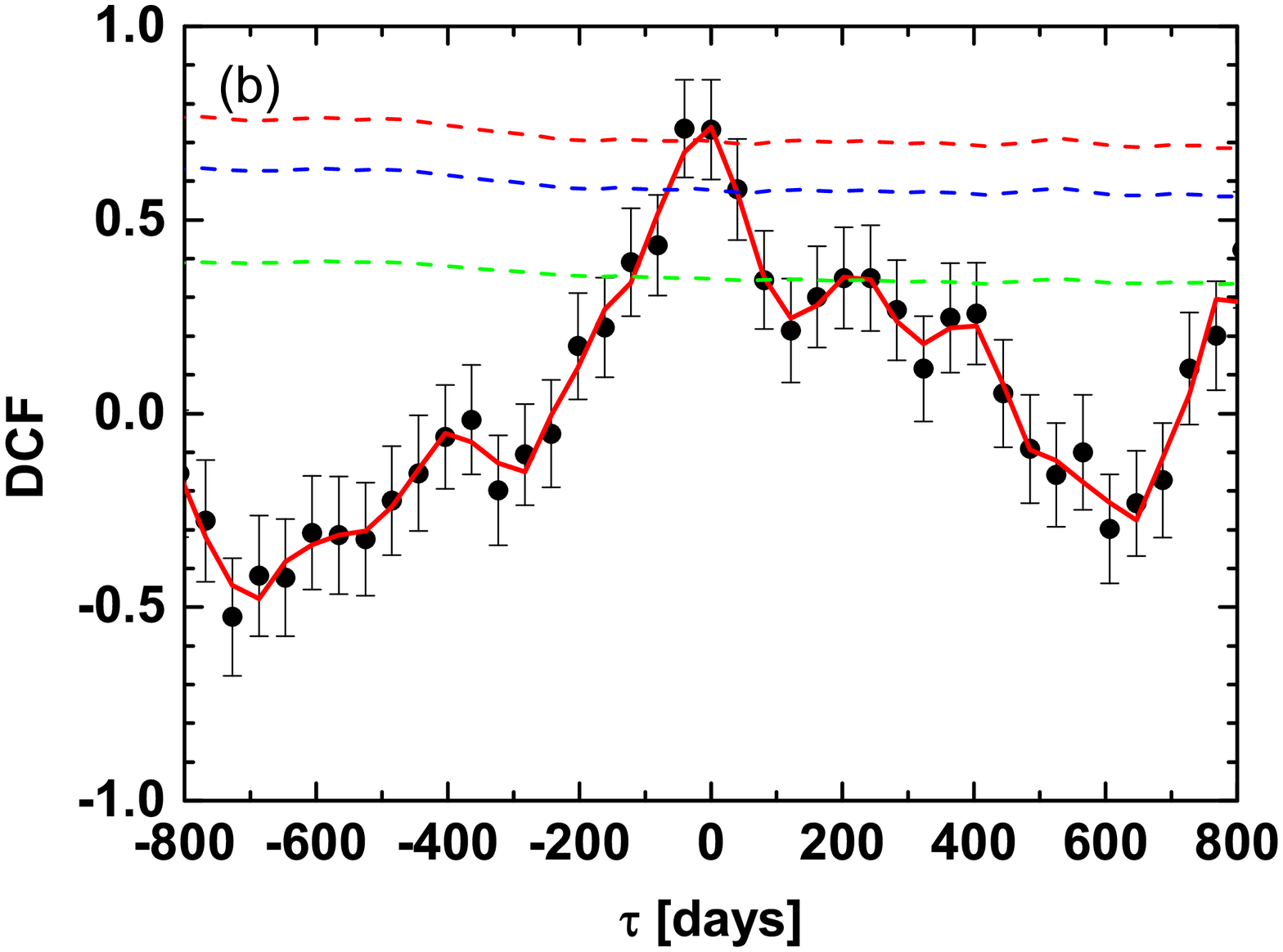}
   \includegraphics[angle=0,scale=0.24]{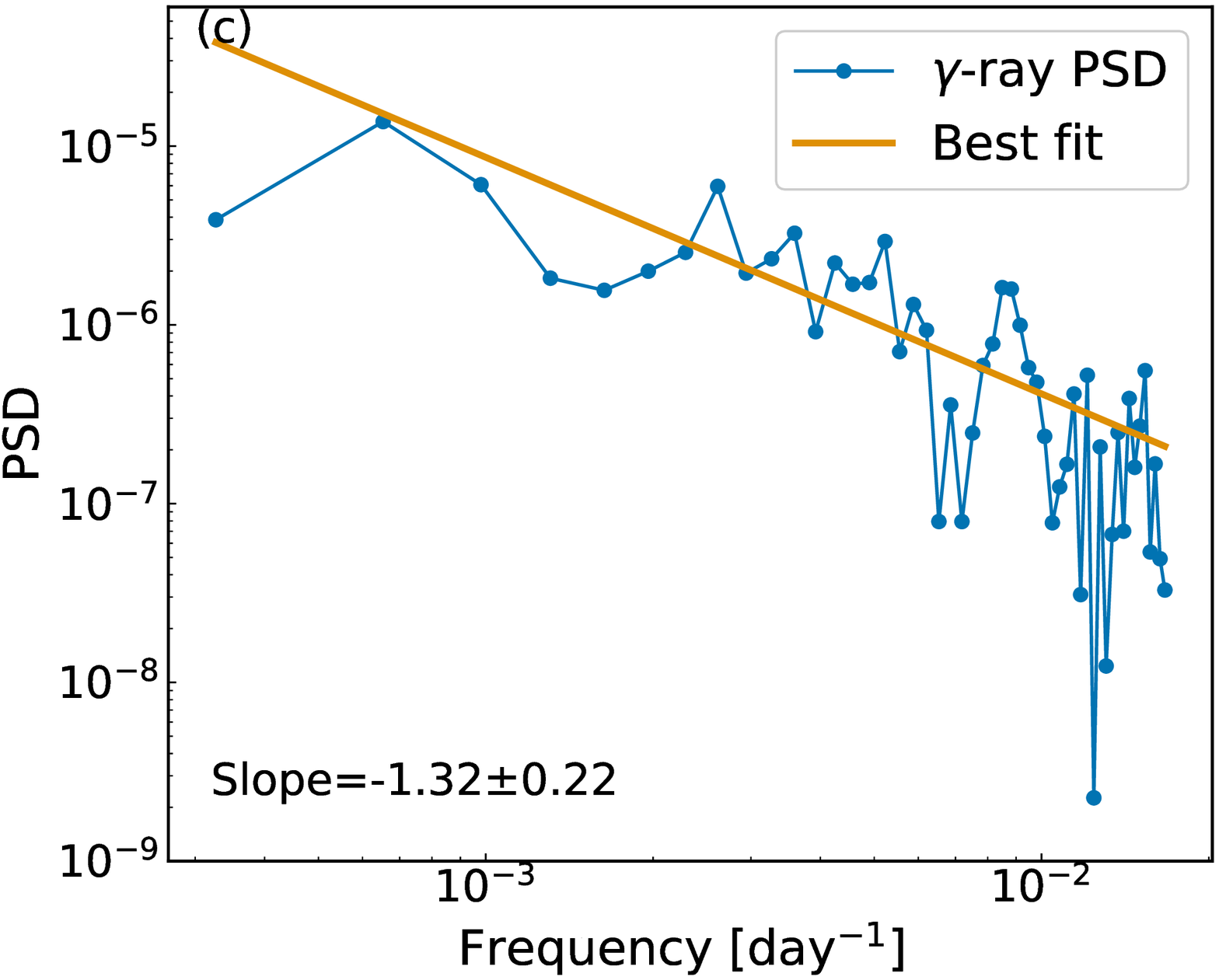}
   \includegraphics[angle=0,scale=0.24]{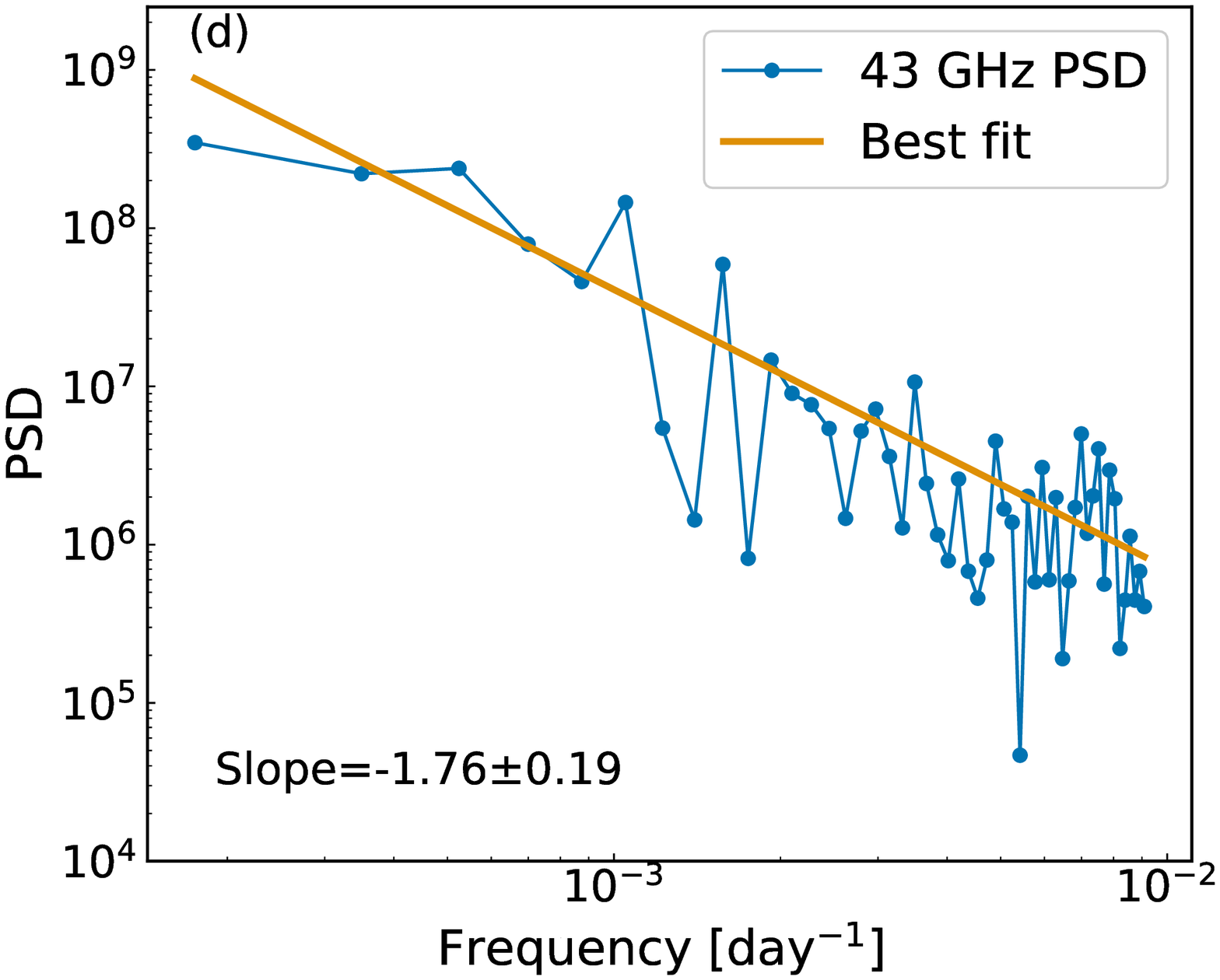}
   \includegraphics[angle=0,scale=0.25]{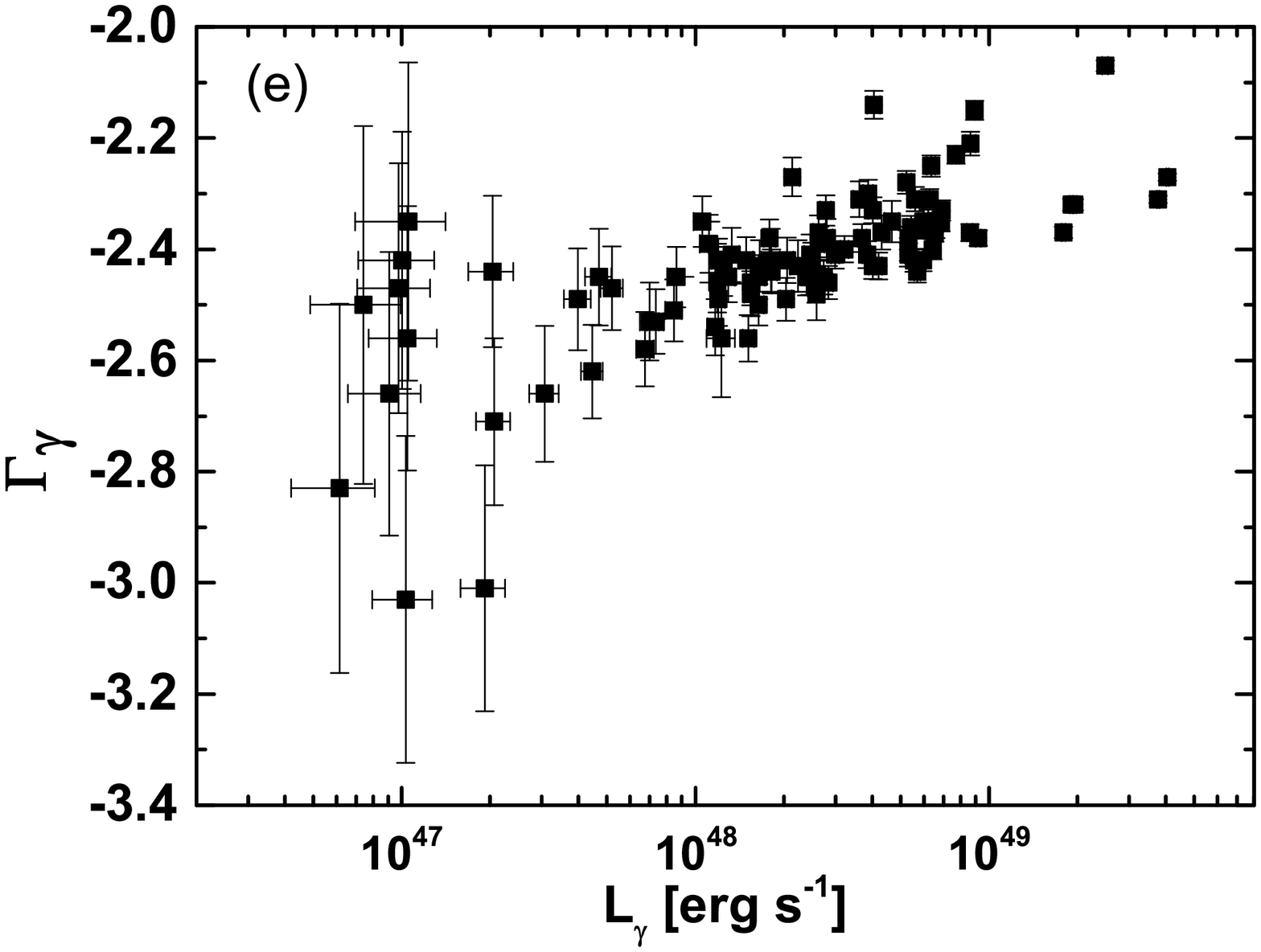}
   \includegraphics[angle=0,scale=0.30]{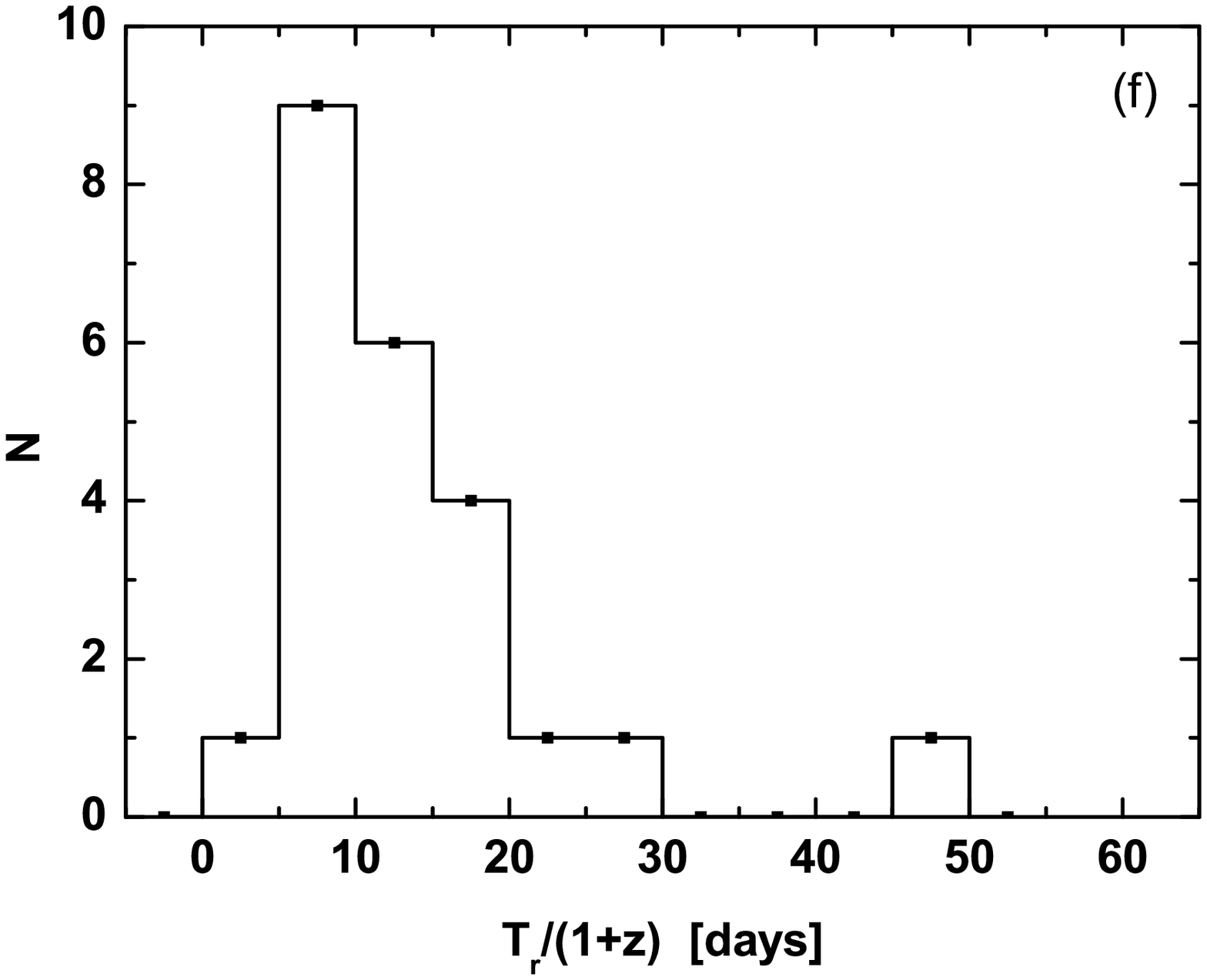}
   \includegraphics[angle=0,scale=0.30]{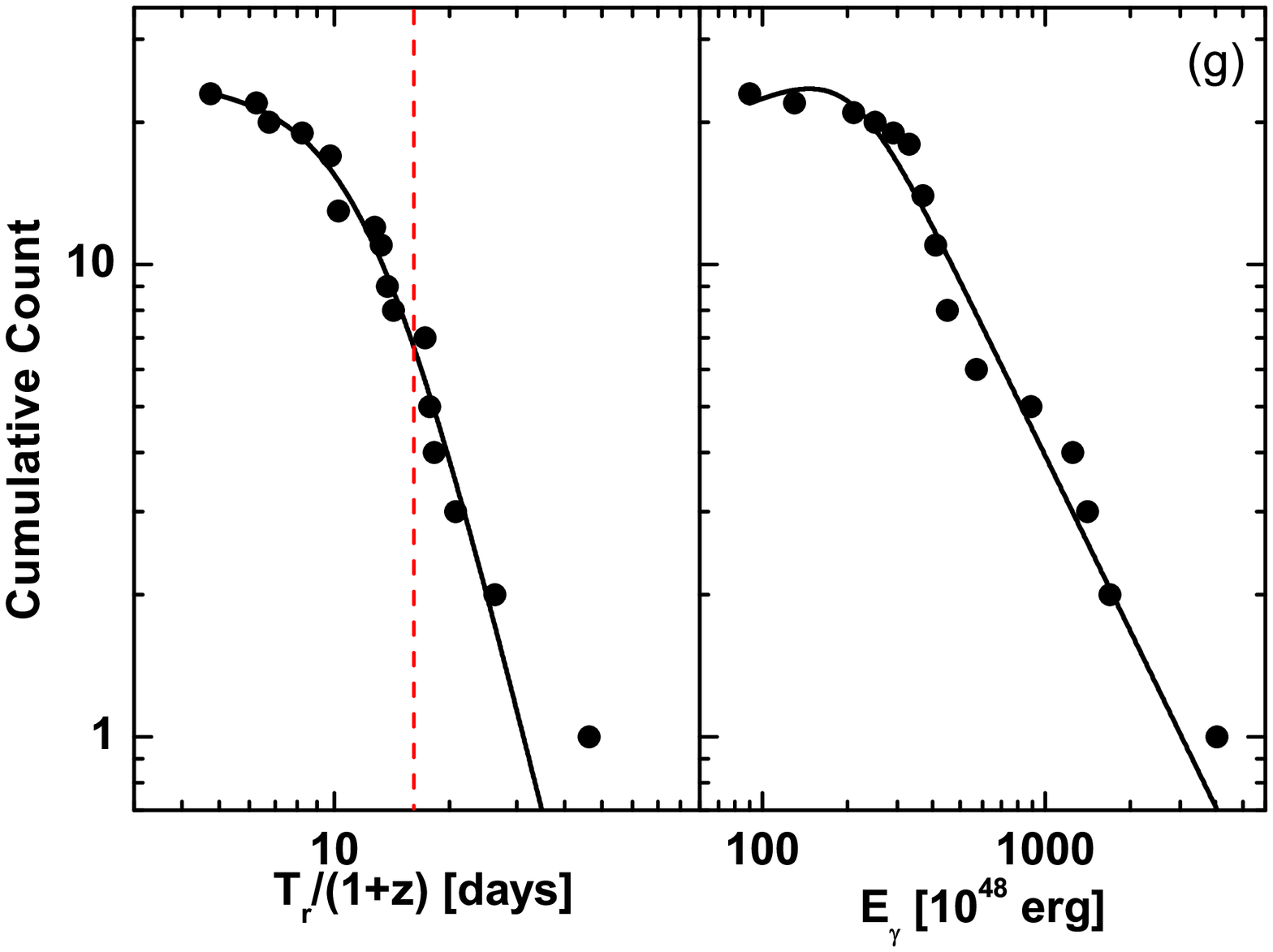}
\caption{Same as Figure 1, but for 3C 454.3.}\label{LC_454}
\end{figure}

\clearpage
\begin{figure}
 \centering
   \includegraphics[angle=0,scale=0.7]{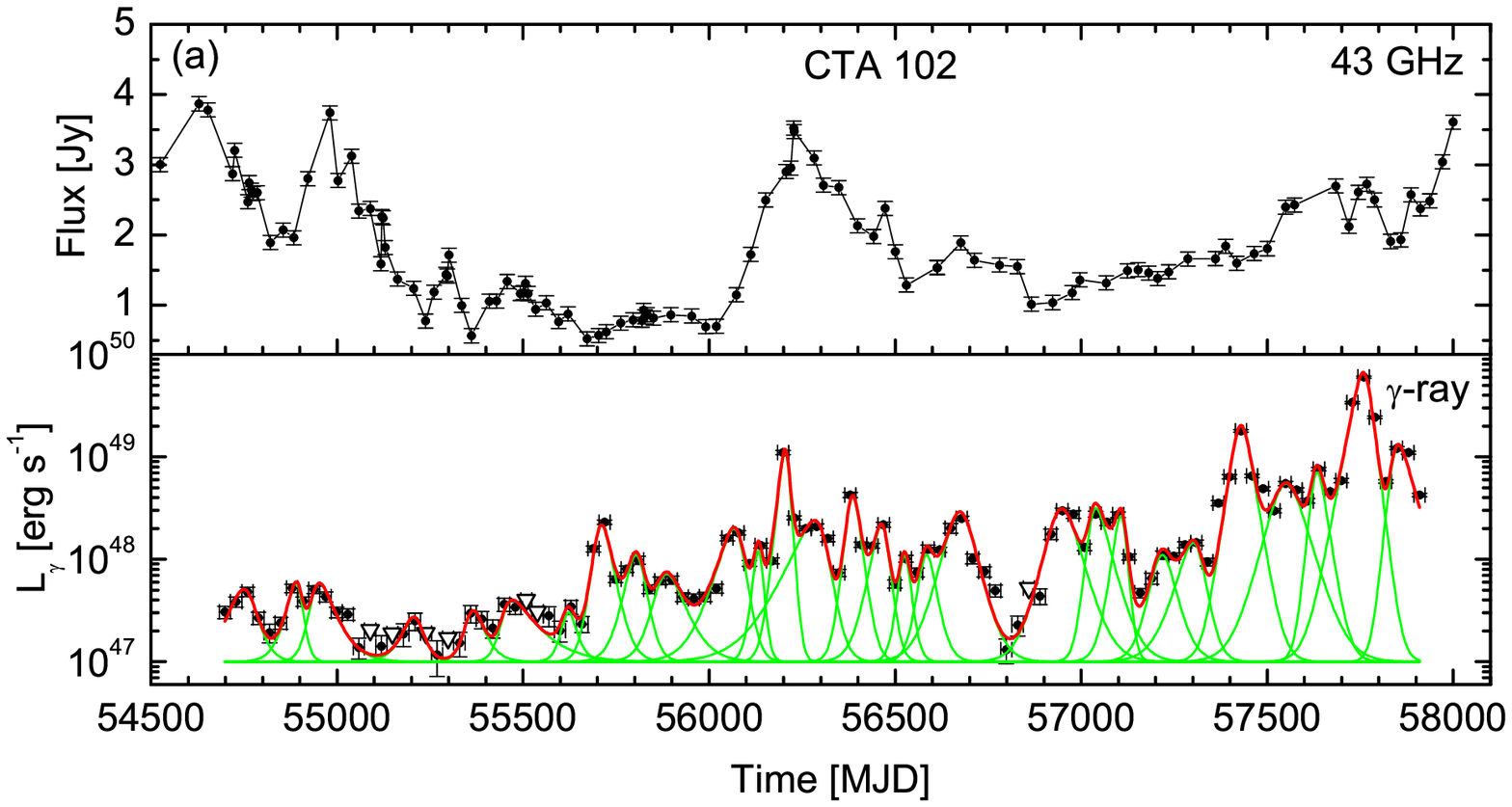}\\
   \includegraphics[angle=0,scale=0.26]{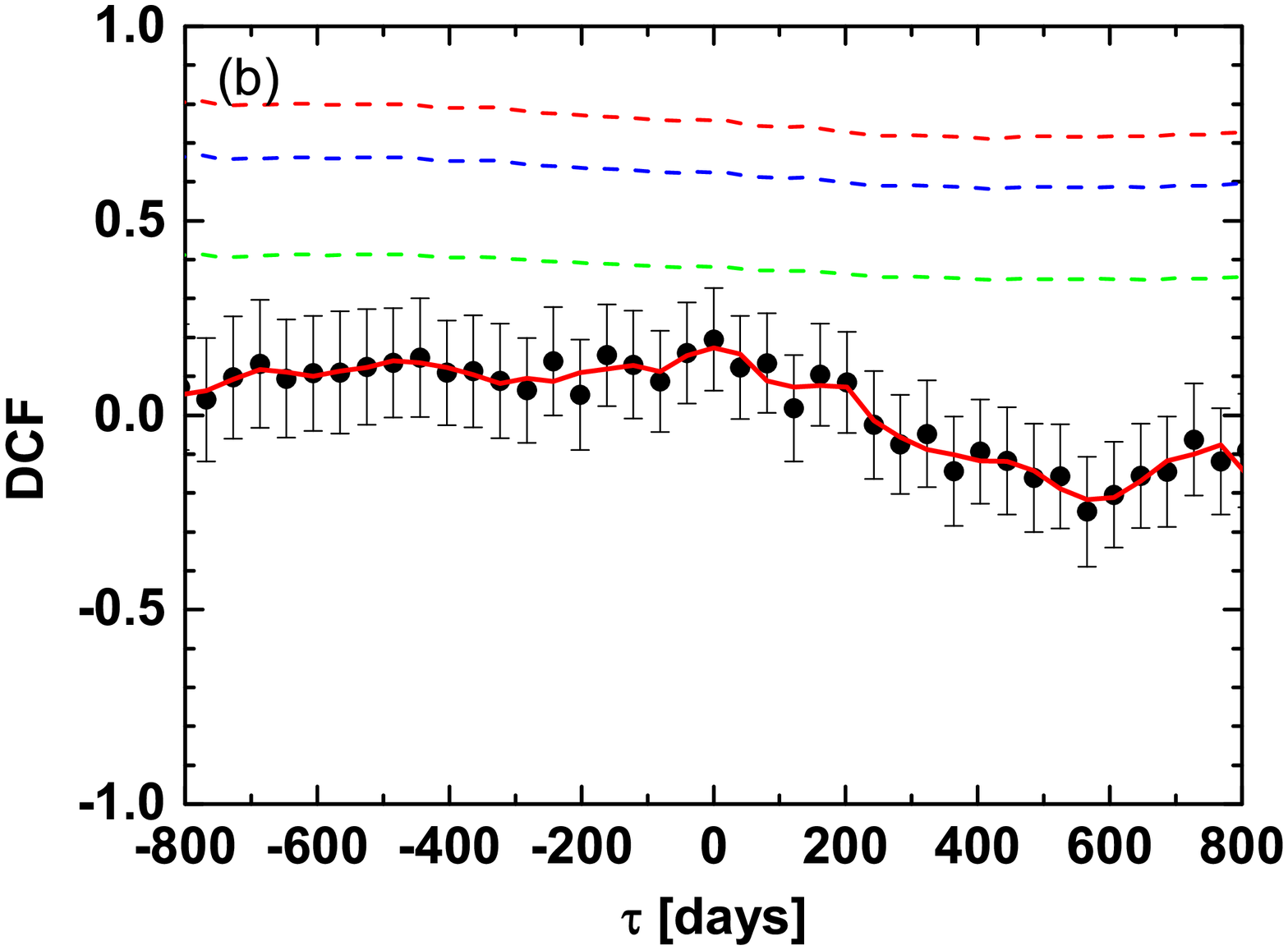}
   \includegraphics[angle=0,scale=0.24]{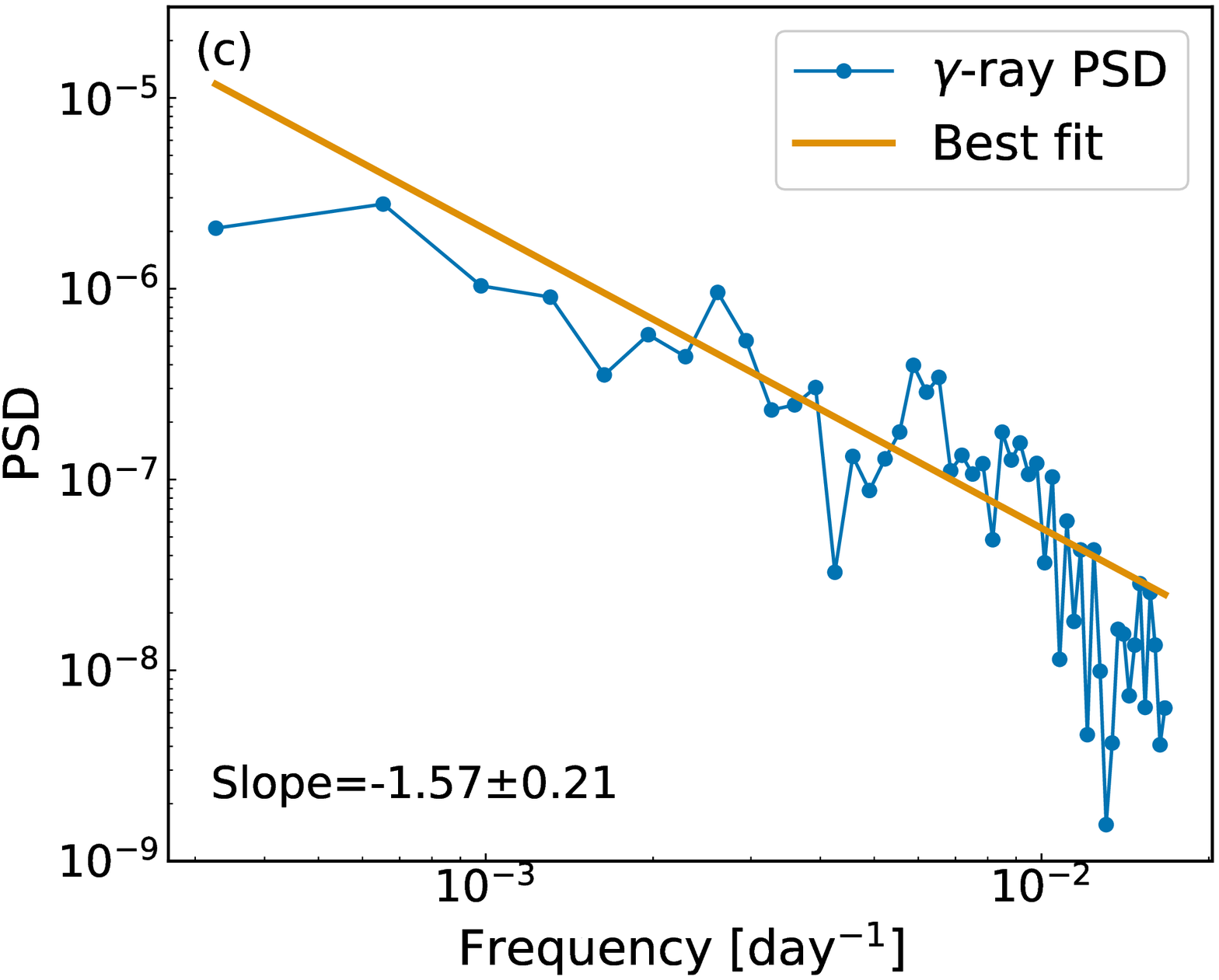}
   \includegraphics[angle=0,scale=0.24]{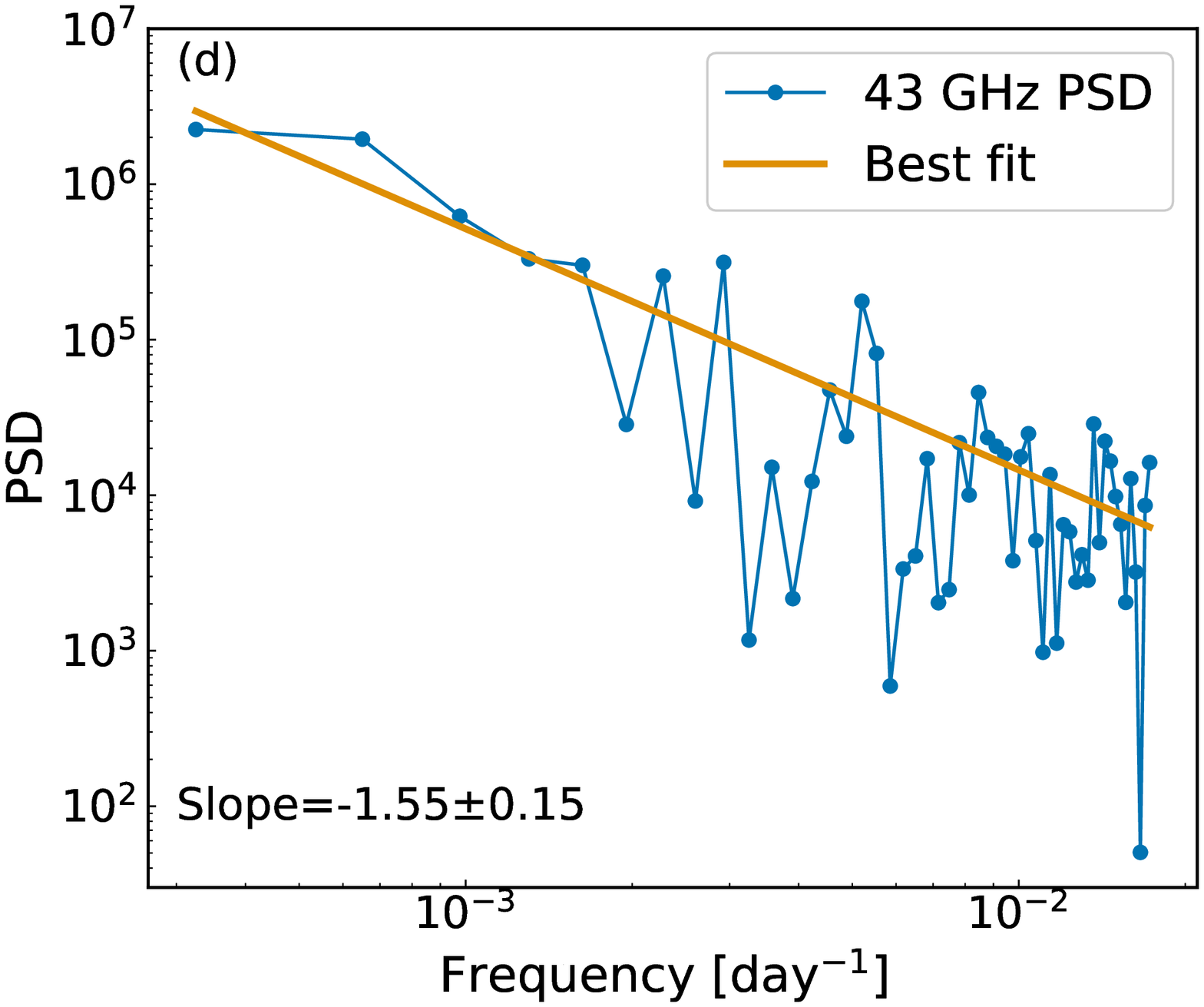}
   \includegraphics[angle=0,scale=0.25]{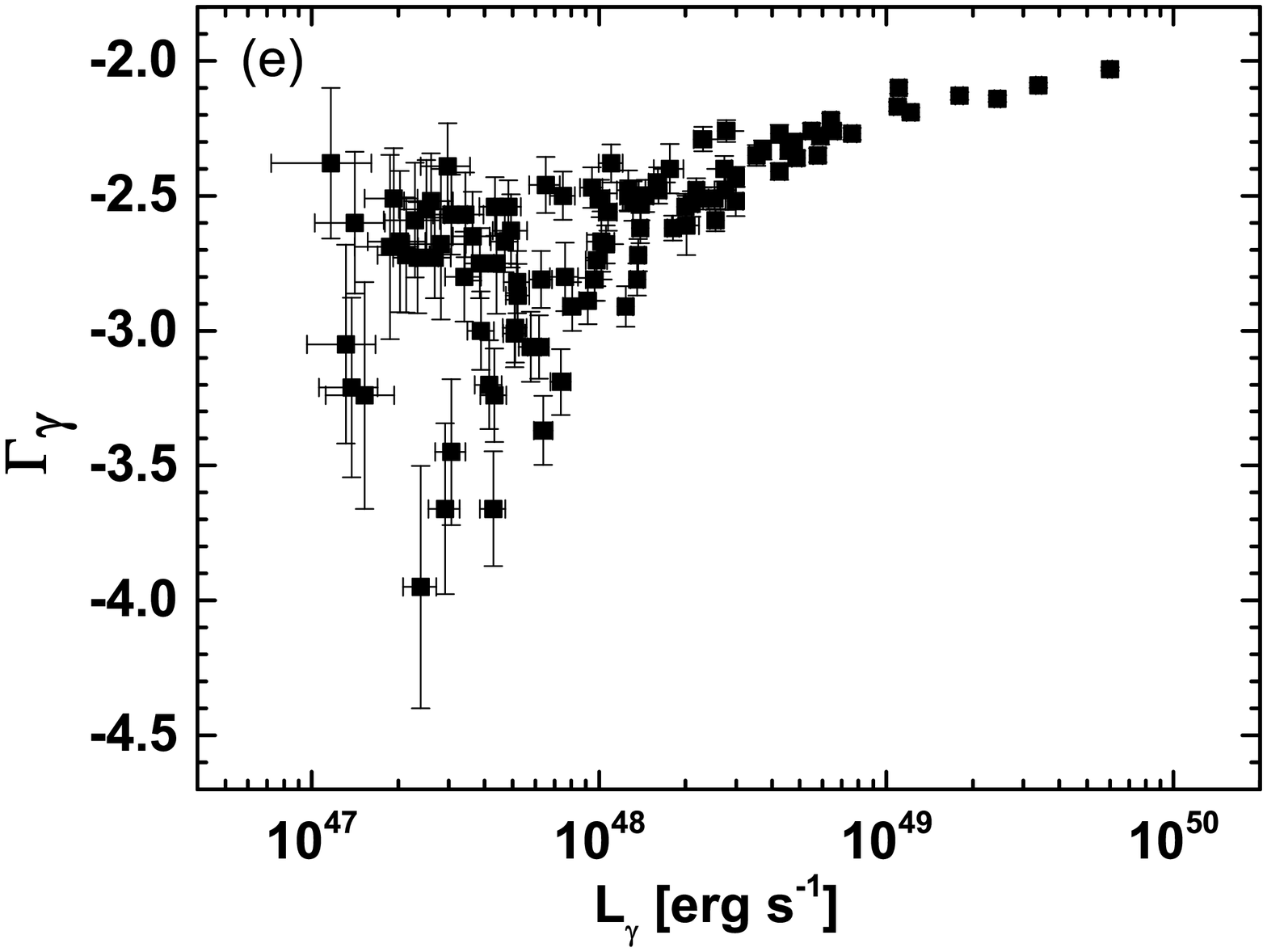}
   \includegraphics[angle=0,scale=0.30]{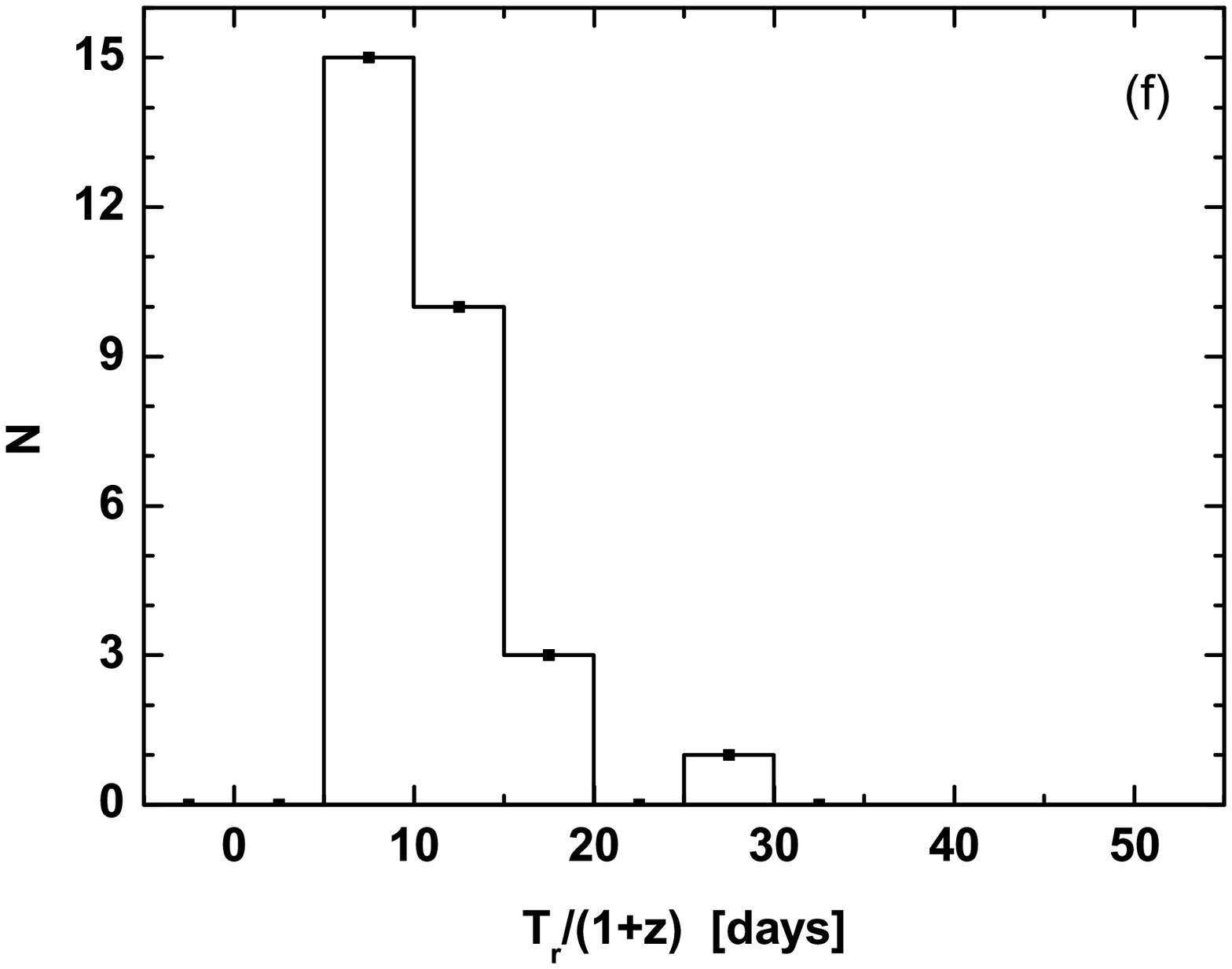}
   \includegraphics[angle=0,scale=0.30]{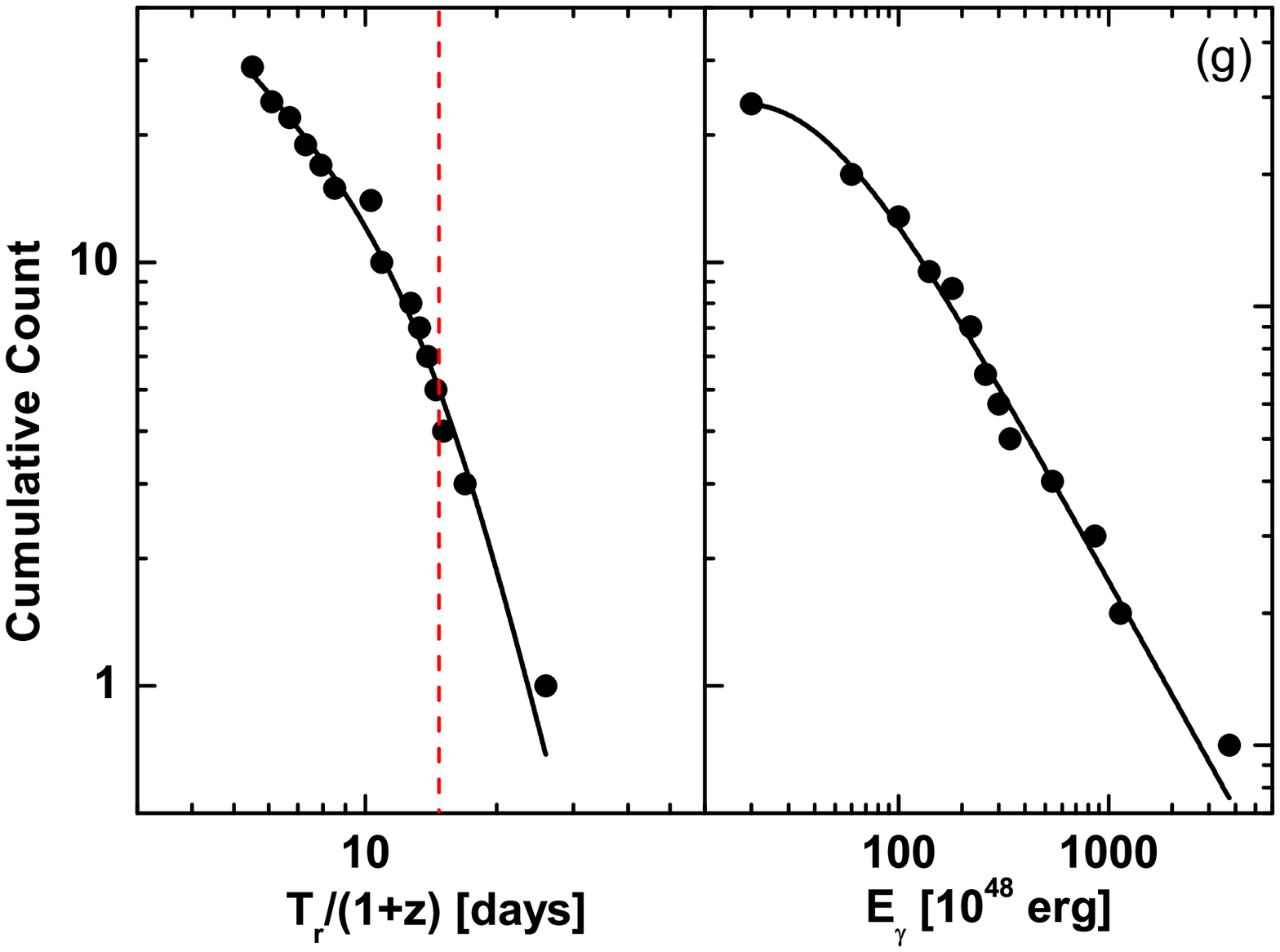}
\caption{Same as Figure 1, but for CTA 102.}\label{LC_102}
\end{figure}

\clearpage
\begin{figure}
 \centering
   \includegraphics[angle=0,scale=0.43]{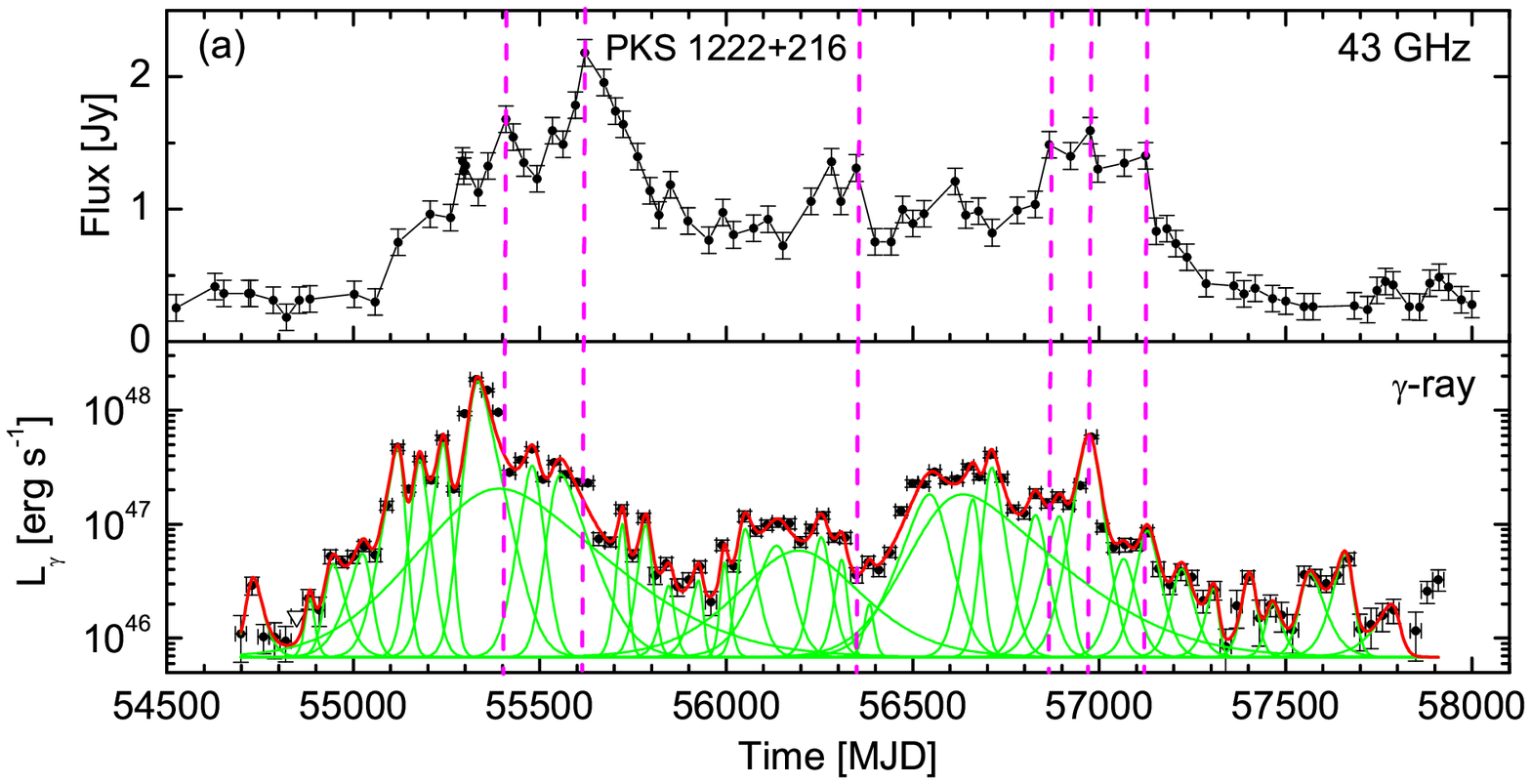}
   \includegraphics[angle=0,scale=0.43]{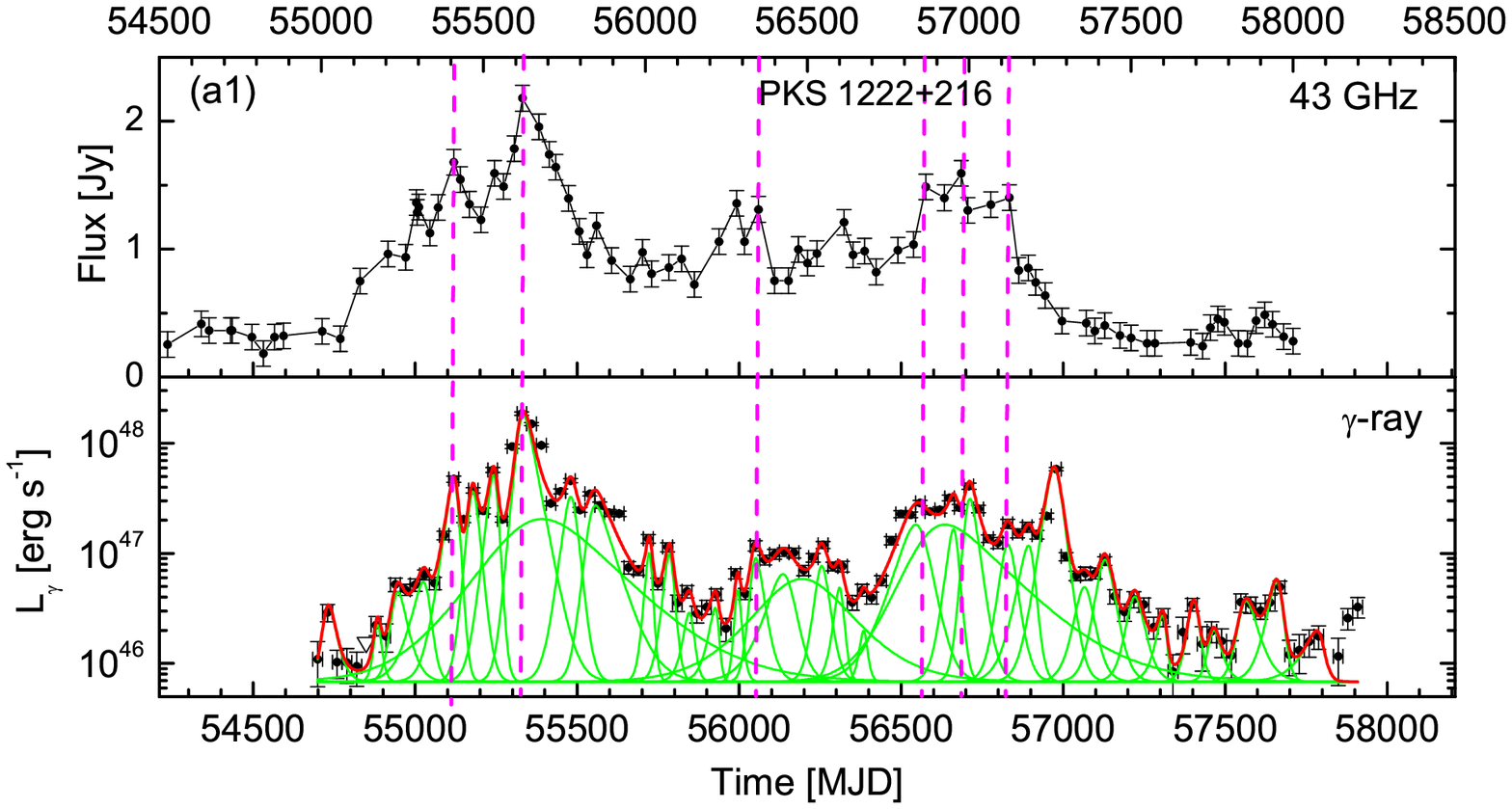}
   \includegraphics[angle=0,scale=0.26]{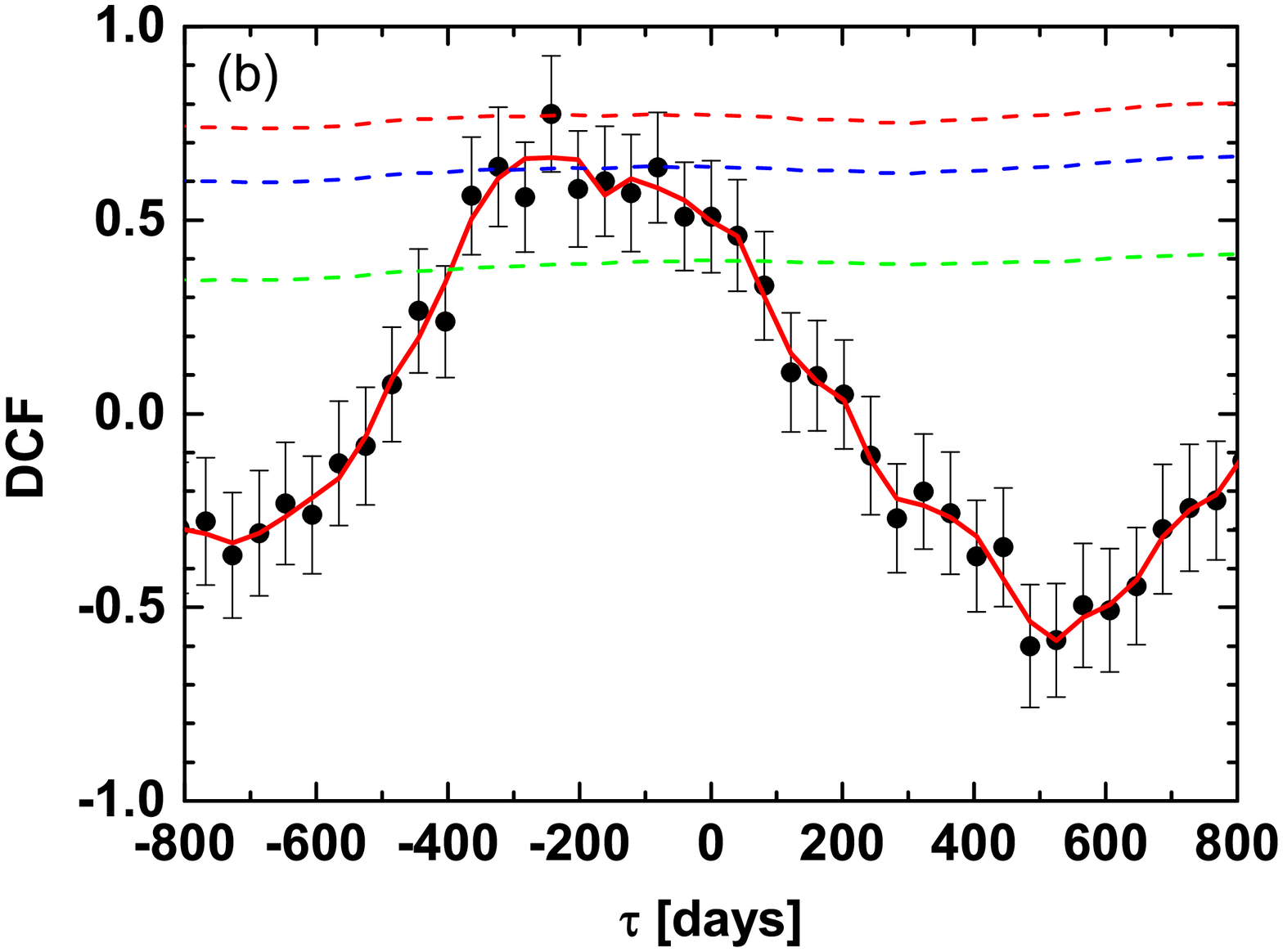}
   \includegraphics[angle=0,scale=0.24]{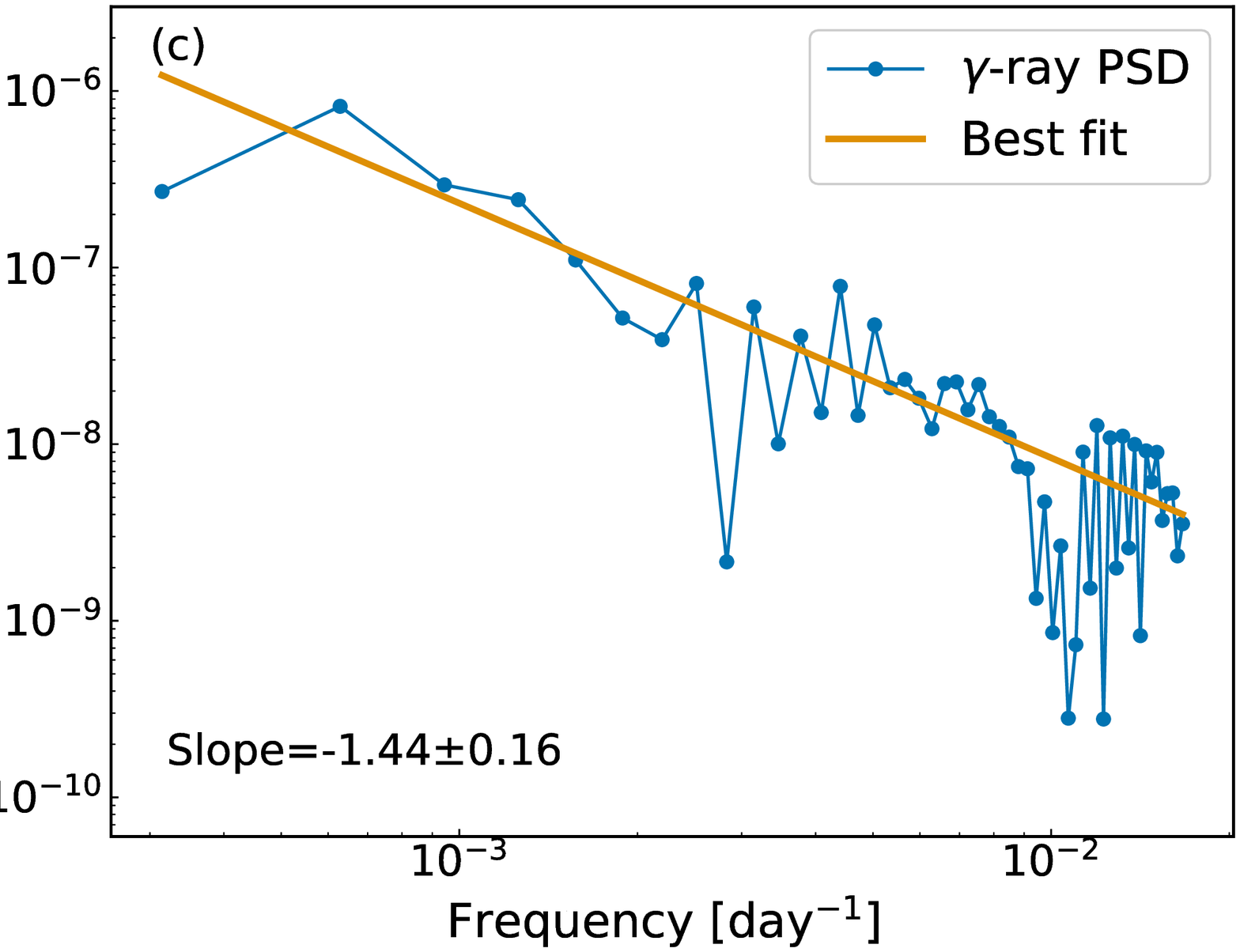}
   \includegraphics[angle=0,scale=0.24]{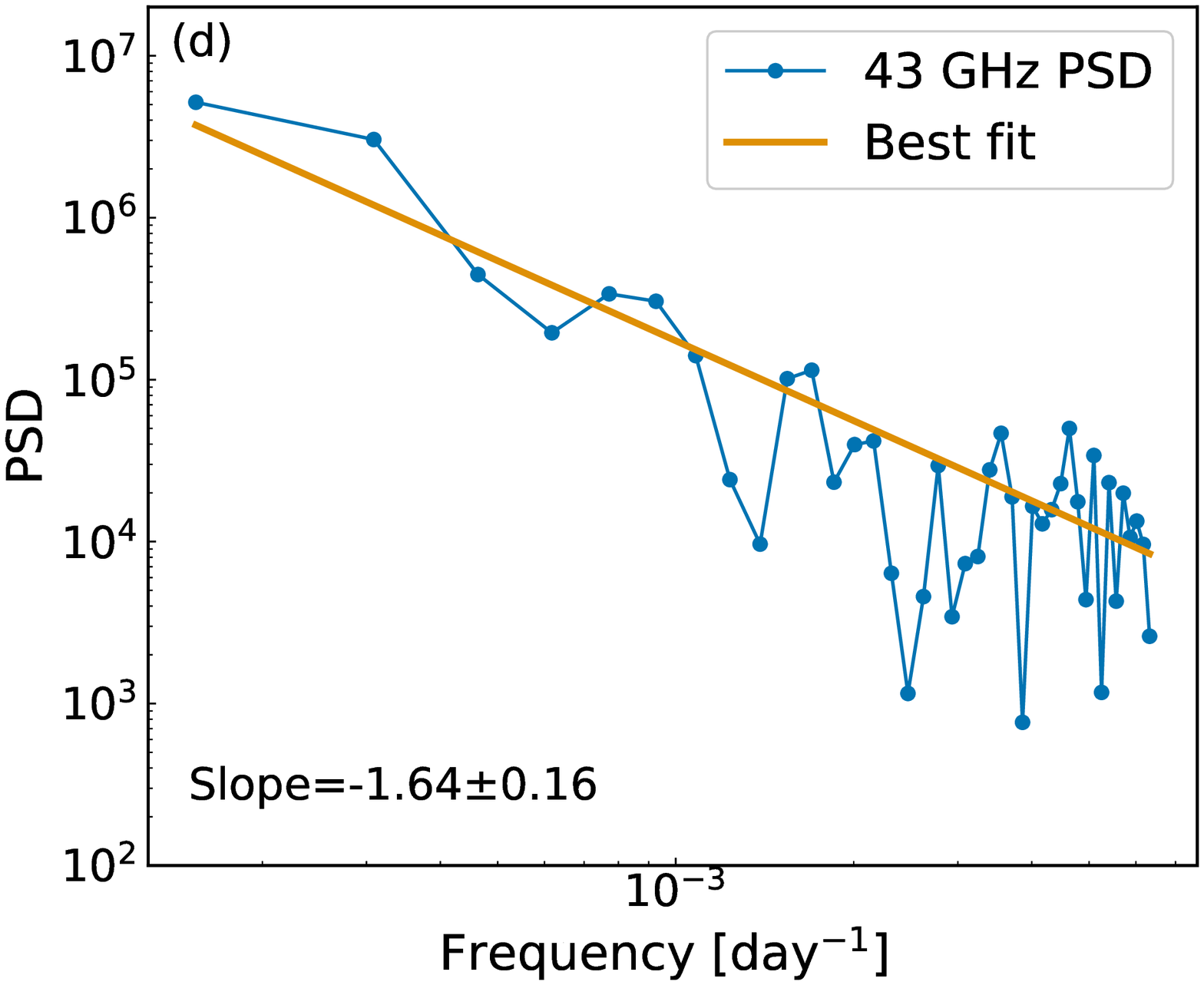}
   \includegraphics[angle=0,scale=0.25]{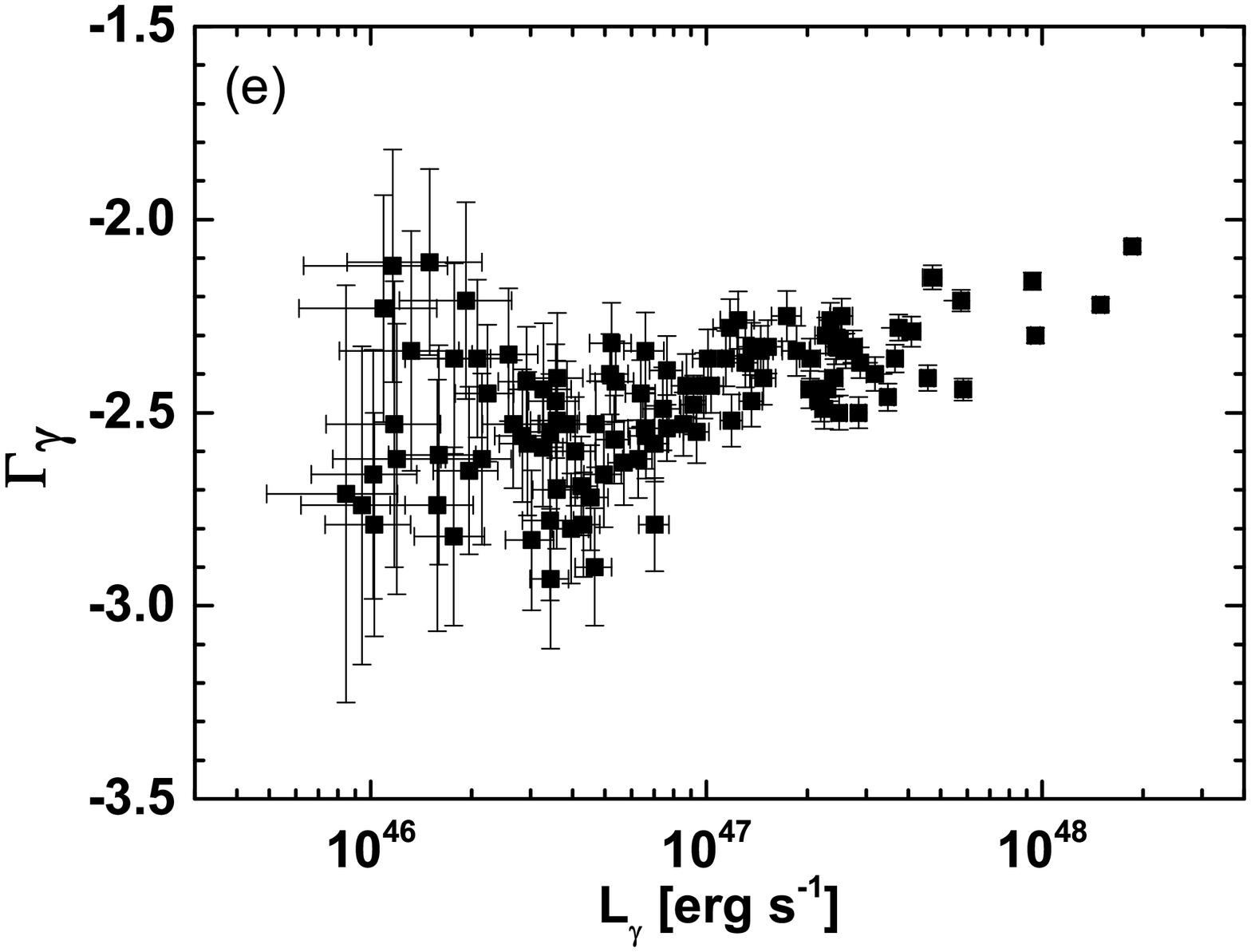}
   \includegraphics[angle=0,scale=0.30]{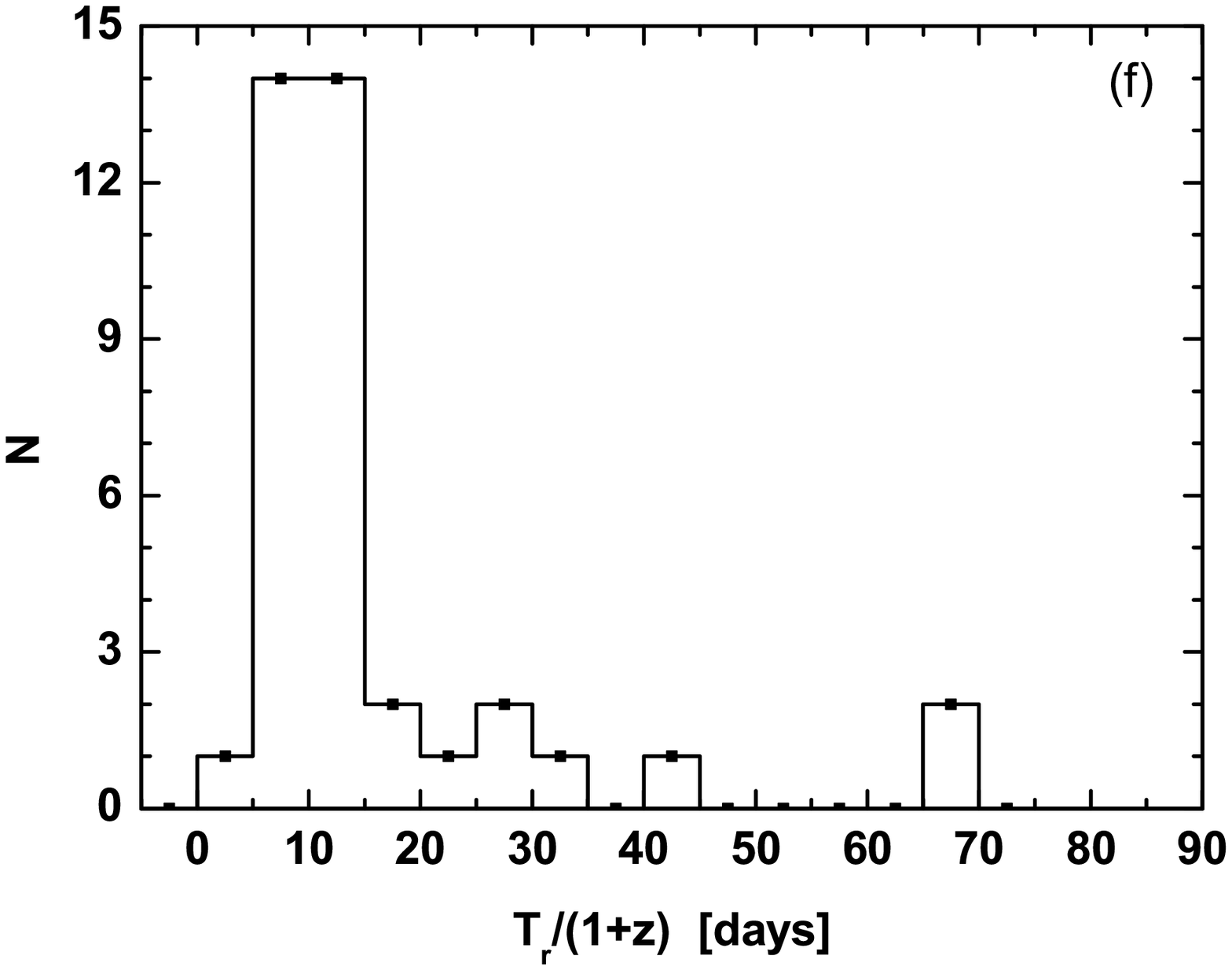}
   \includegraphics[angle=0,scale=0.30]{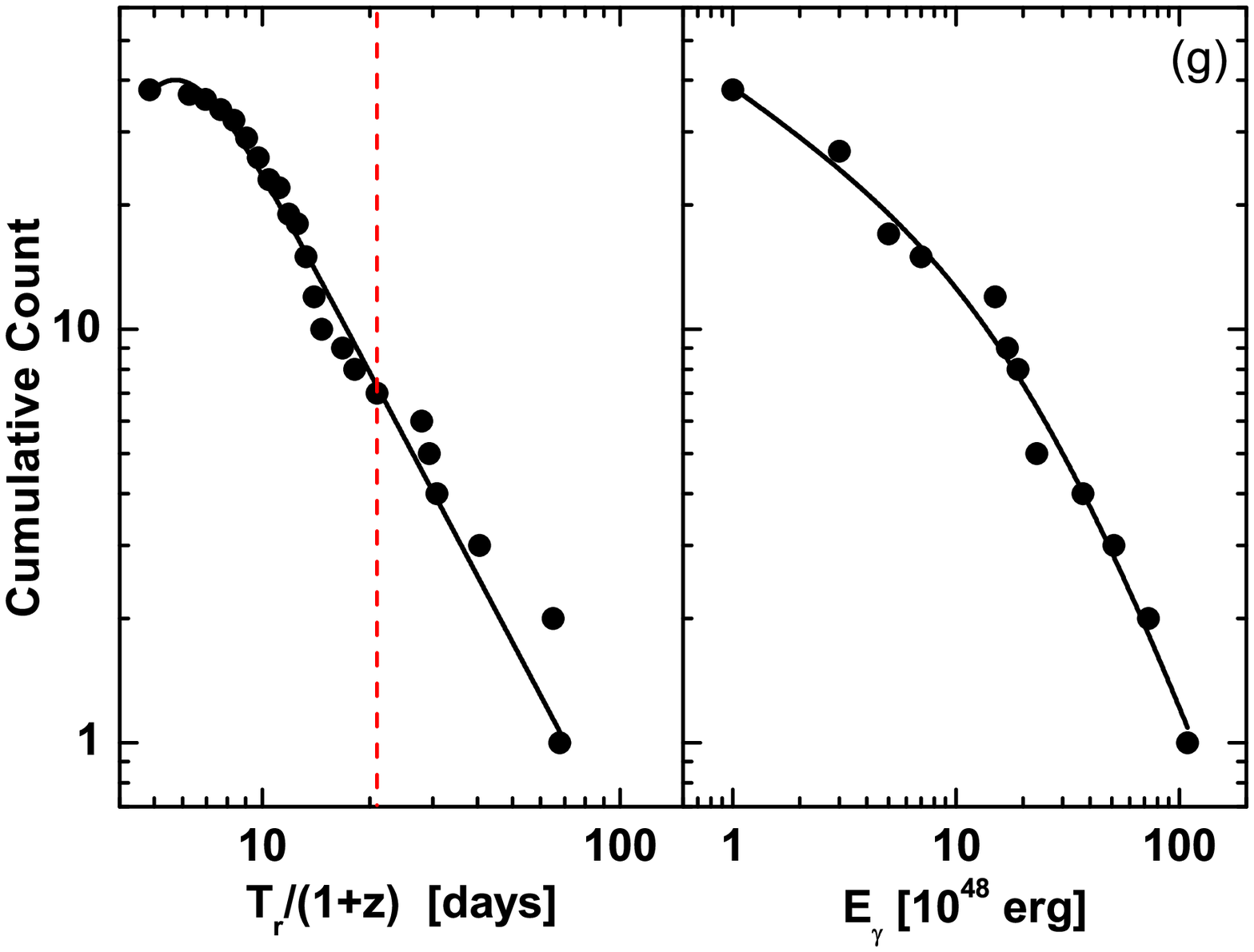}
\caption{Same as Figure 1, but for PKS 1222+216. Panel (a1): Same as Panel (a), but the time-axes of the light curves in the two energy bands are re-scaled according the time-lag of the highest fluxes in radio and $\gamma$-ray bands, i.e., 290 days, which is also consistent with the DCF result (panel (b)) between $\gamma$-ray and radio bands. }\label{LC_1222}
\end{figure}

\clearpage
\begin{figure}
 \centering
   \includegraphics[angle=0,scale=0.7]{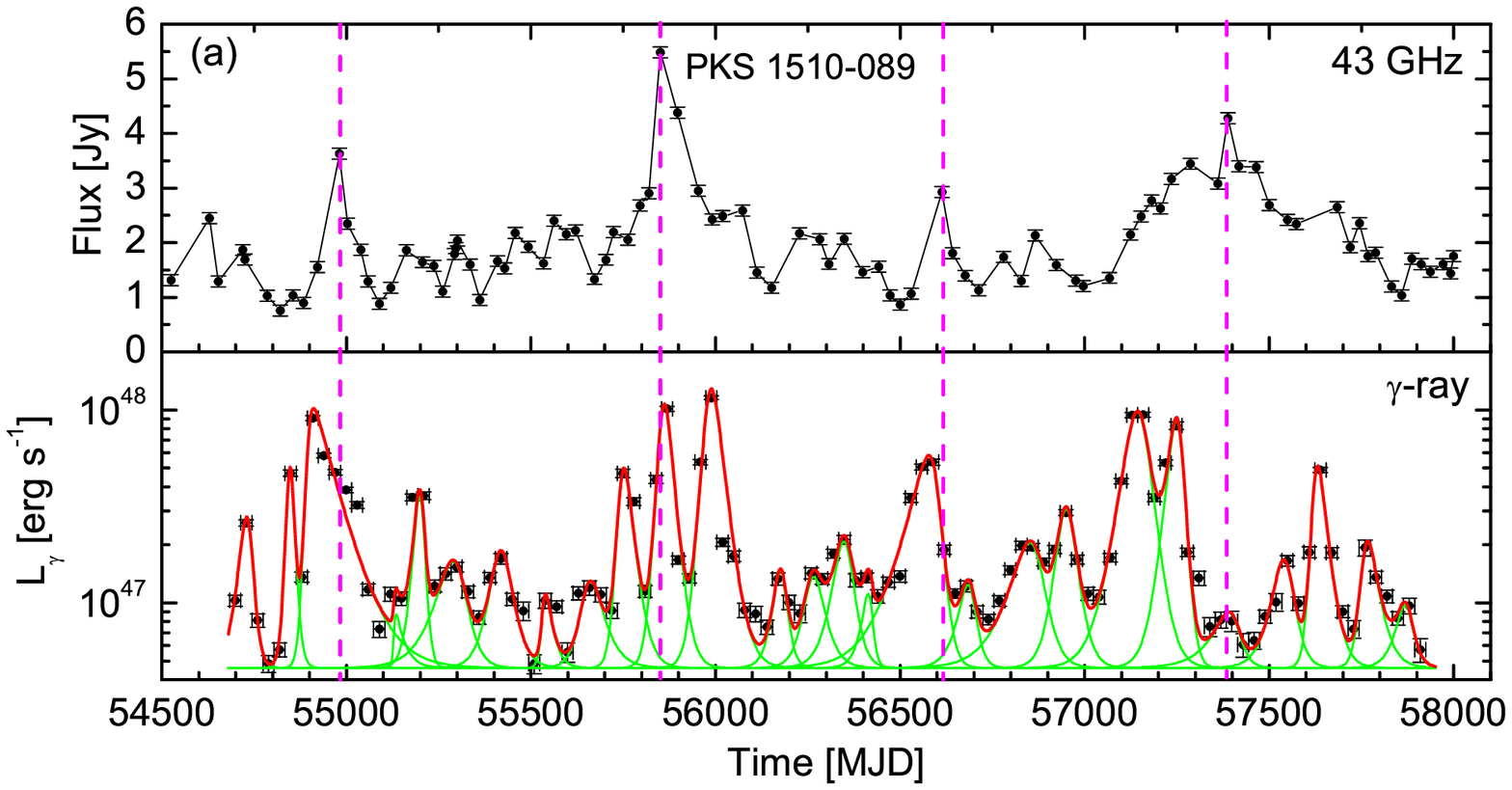}\\
   \includegraphics[angle=0,scale=0.26]{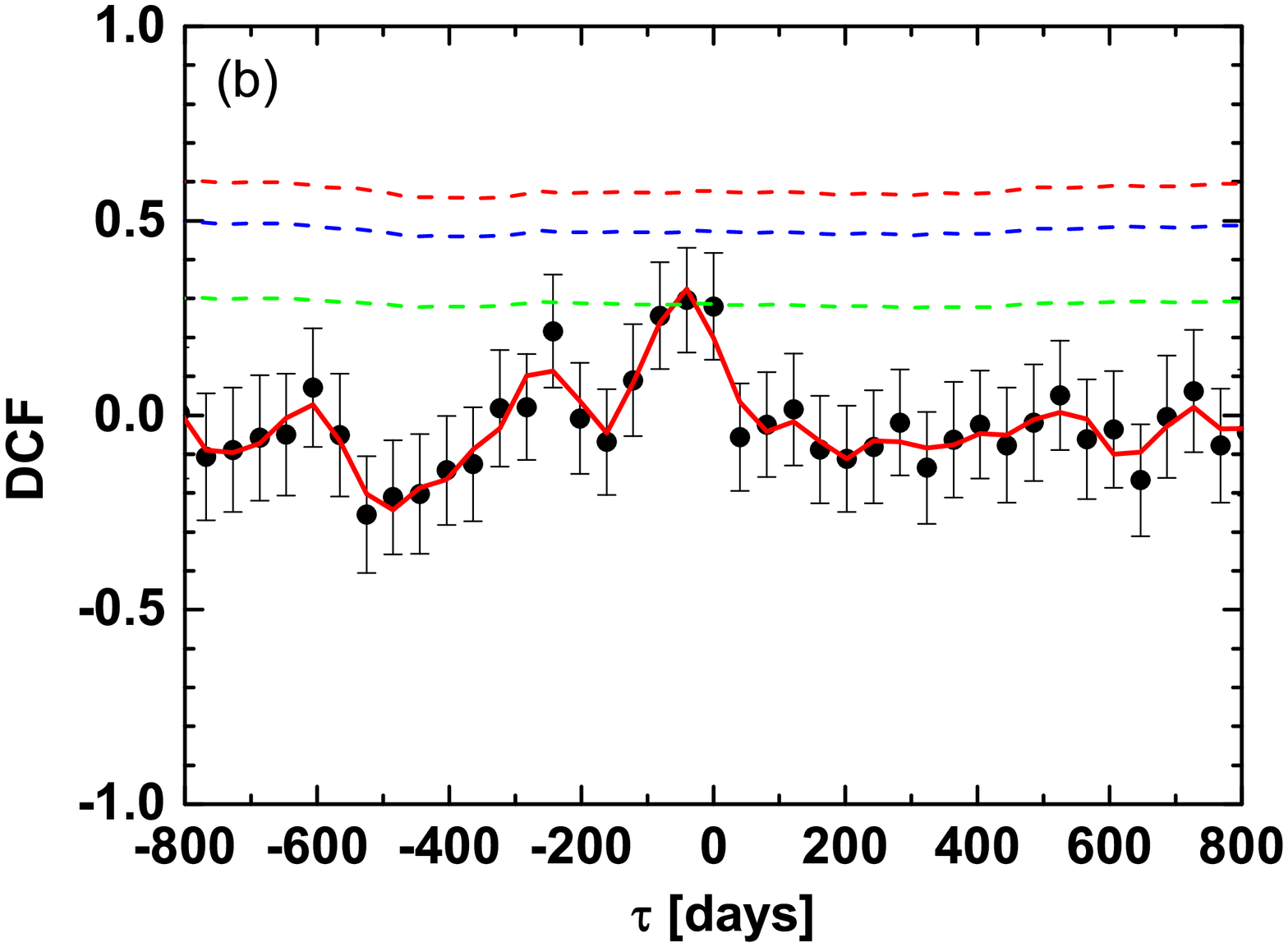}
   \includegraphics[angle=0,scale=0.24]{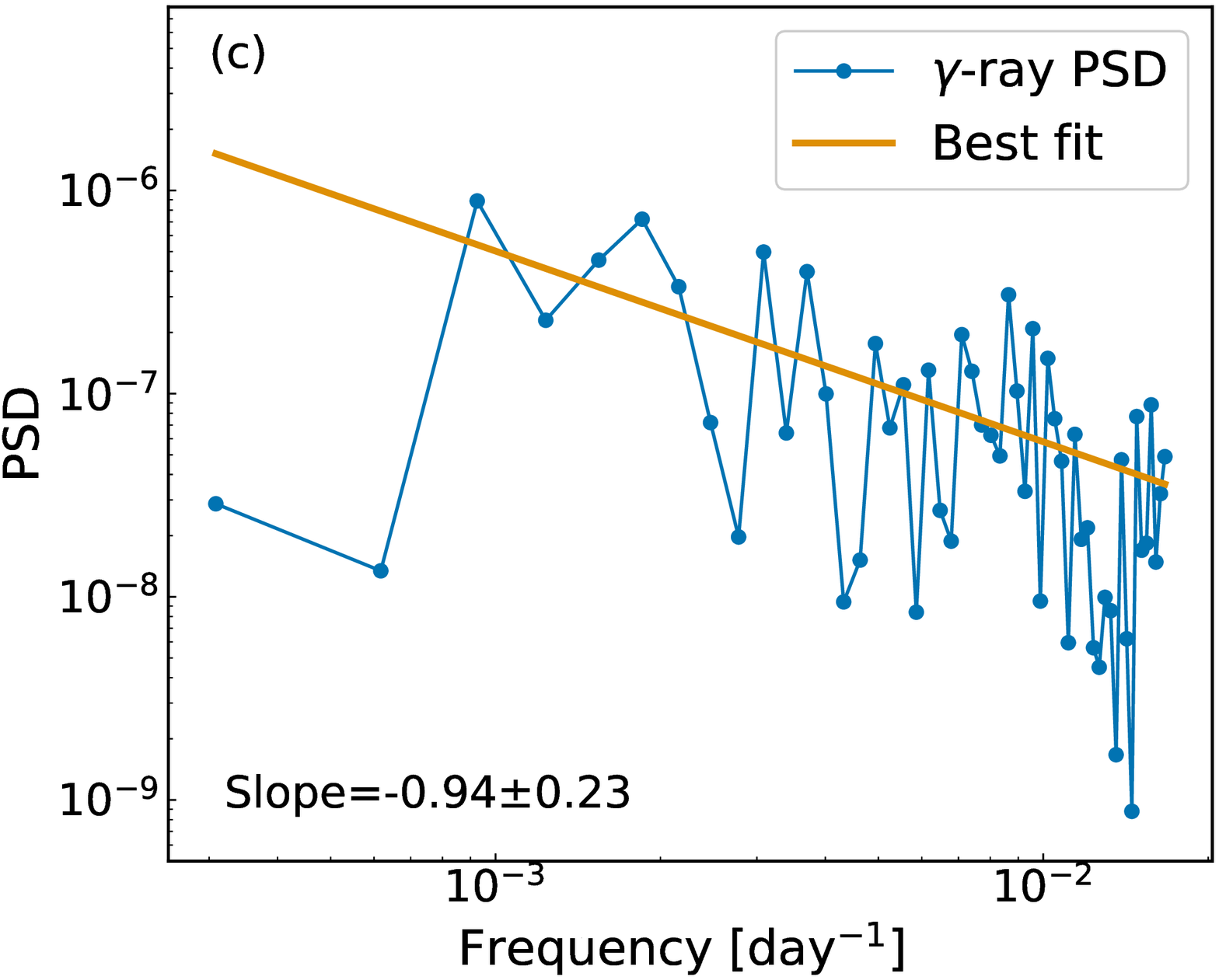}
   \includegraphics[angle=0,scale=0.24]{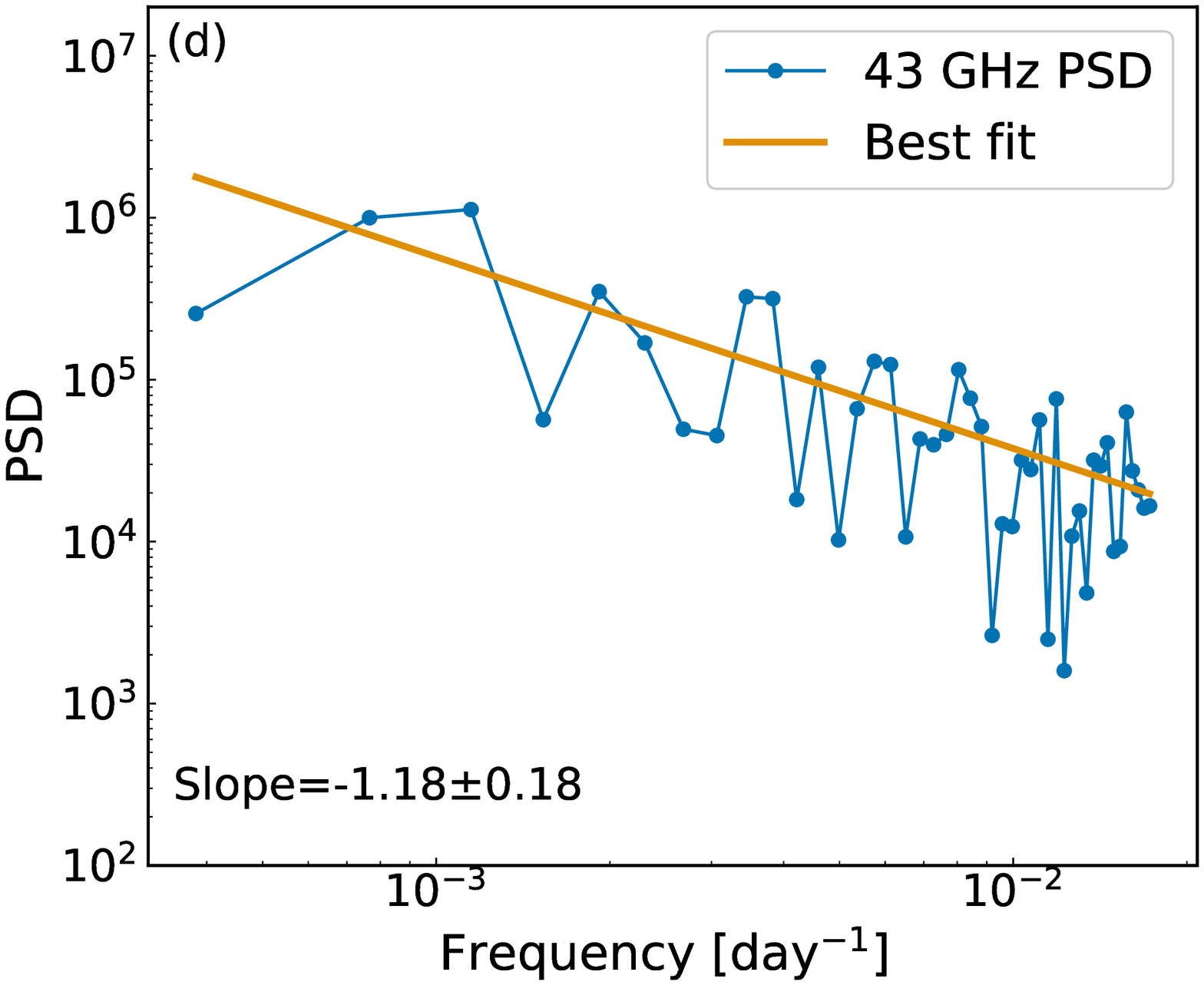}
   \includegraphics[angle=0,scale=0.25]{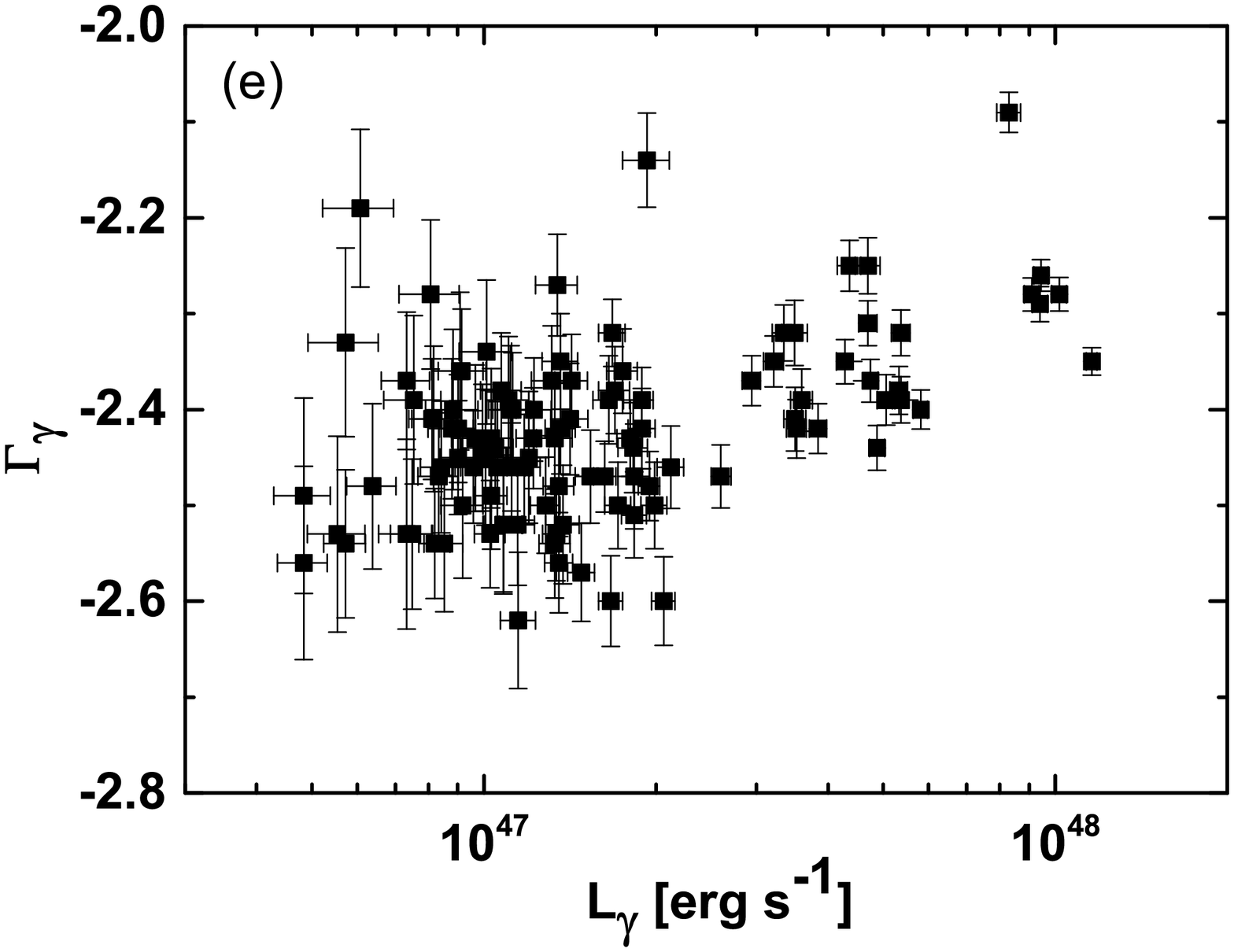}
   \includegraphics[angle=0,scale=0.30]{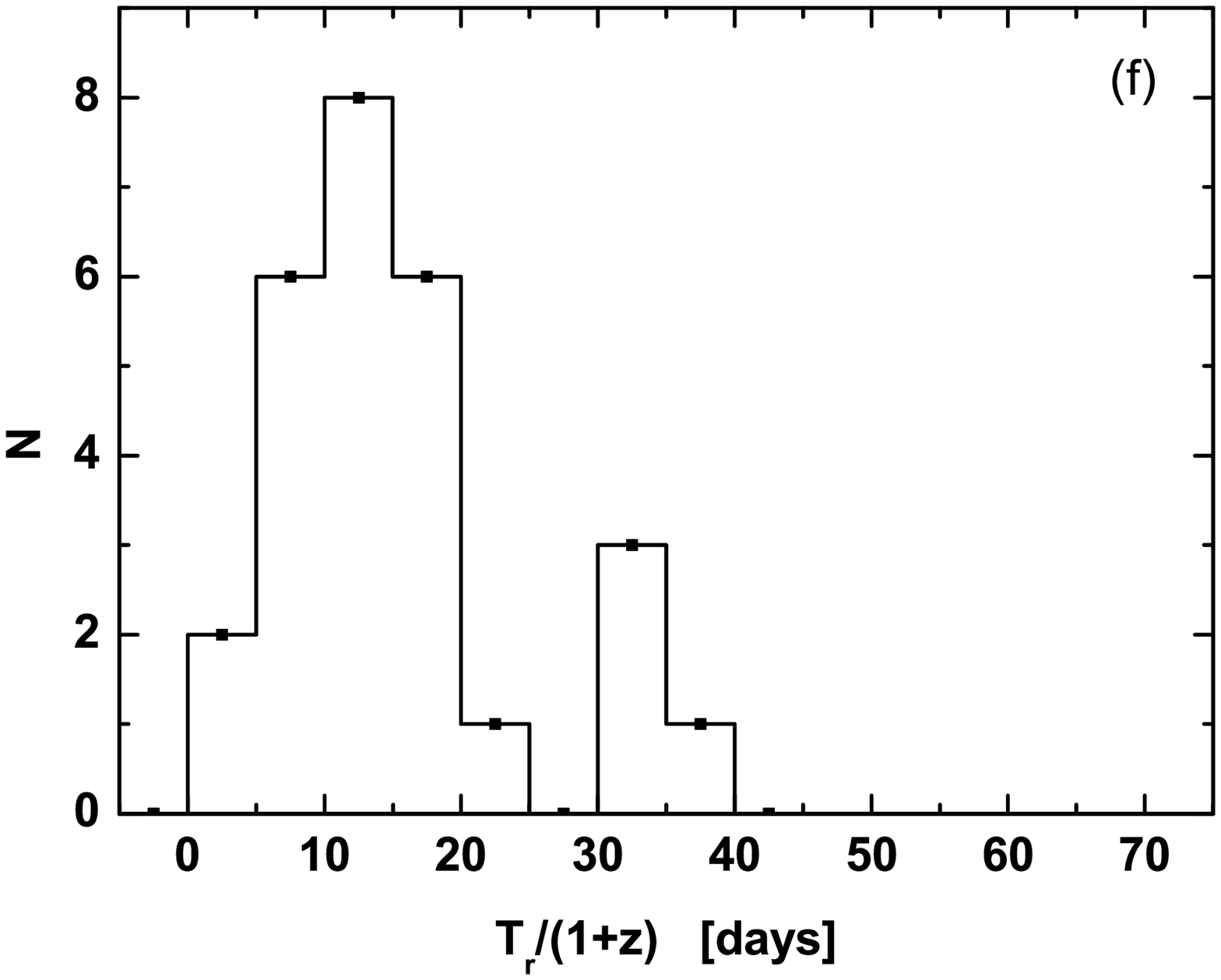}
   \includegraphics[angle=0,scale=0.30]{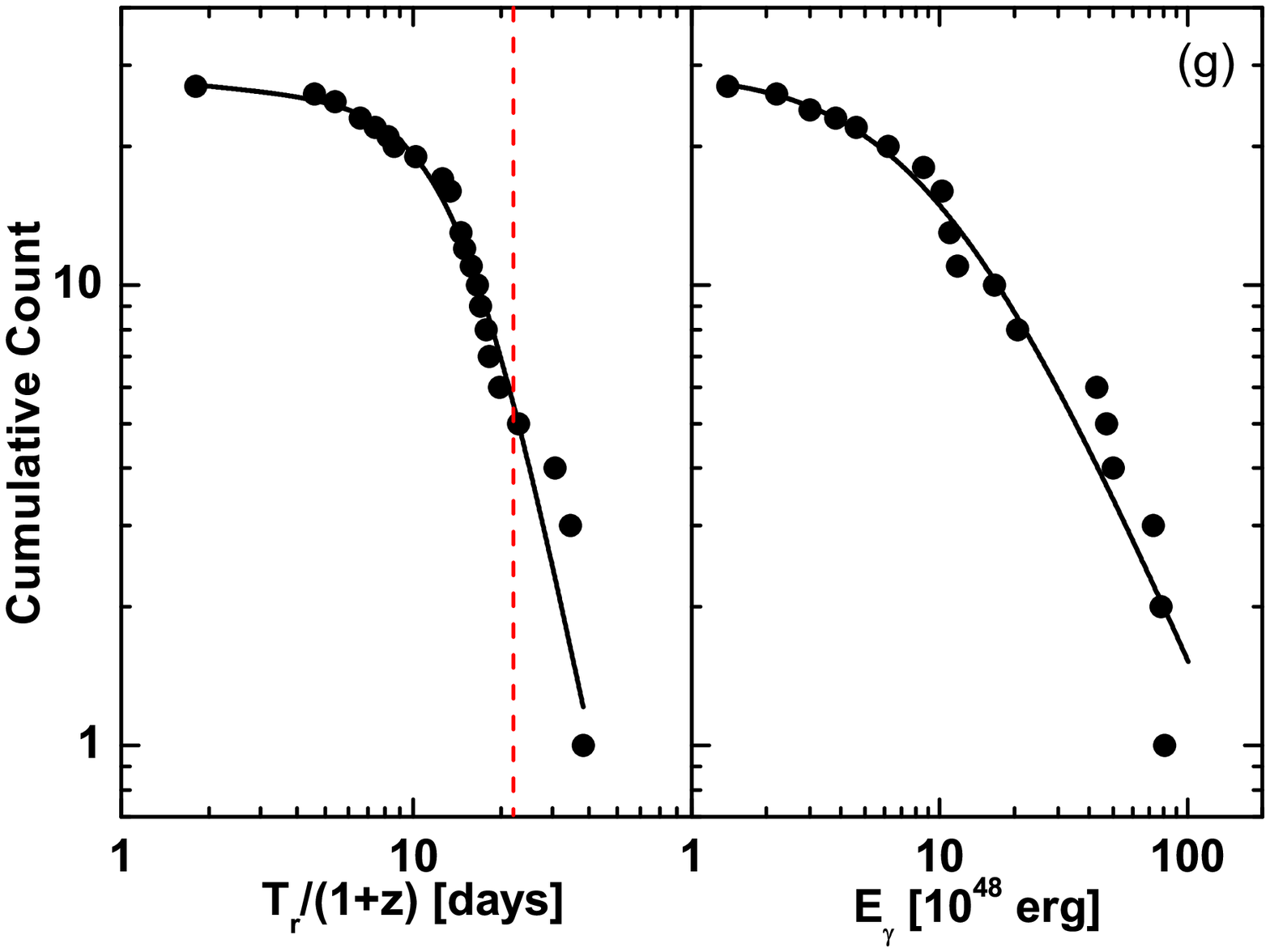}
\caption{Same as Figure 1, but for PKS 1510--089.}\label{LC_1510}
\end{figure}

\clearpage
\begin{figure}
 \centering
   \includegraphics[angle=0,scale=0.7]{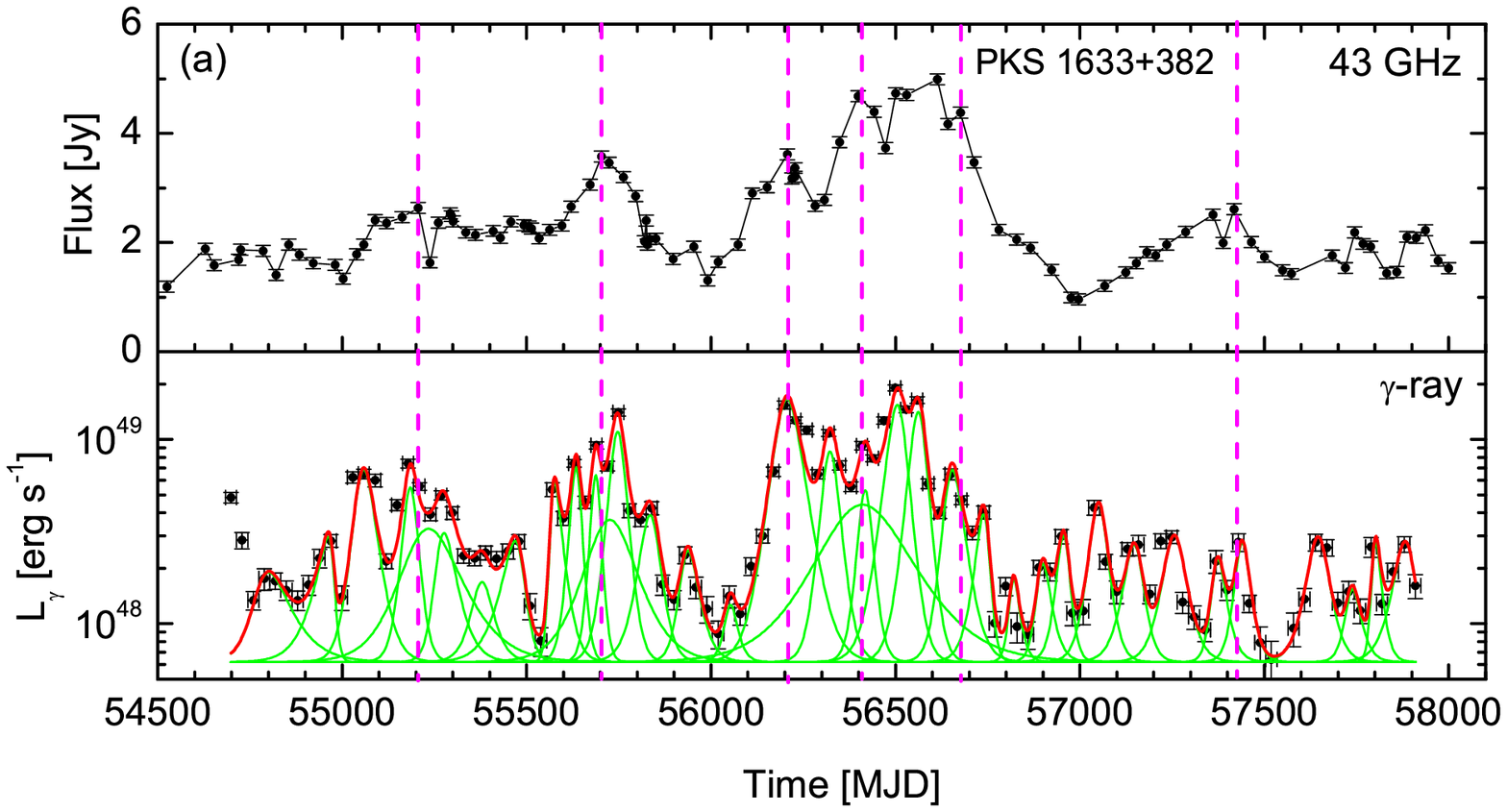}\\
   \includegraphics[angle=0,scale=0.26]{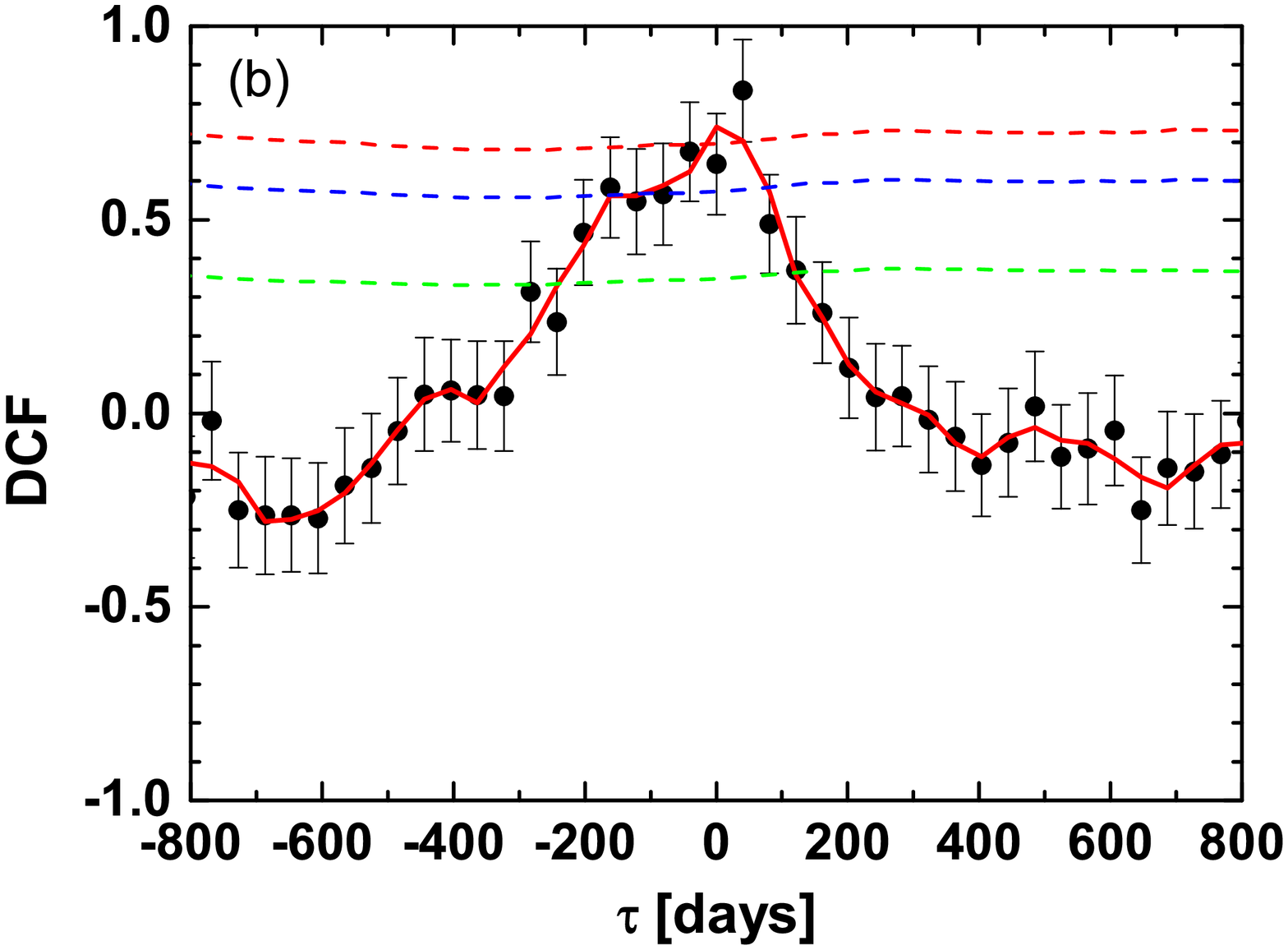}
   \includegraphics[angle=0,scale=0.24]{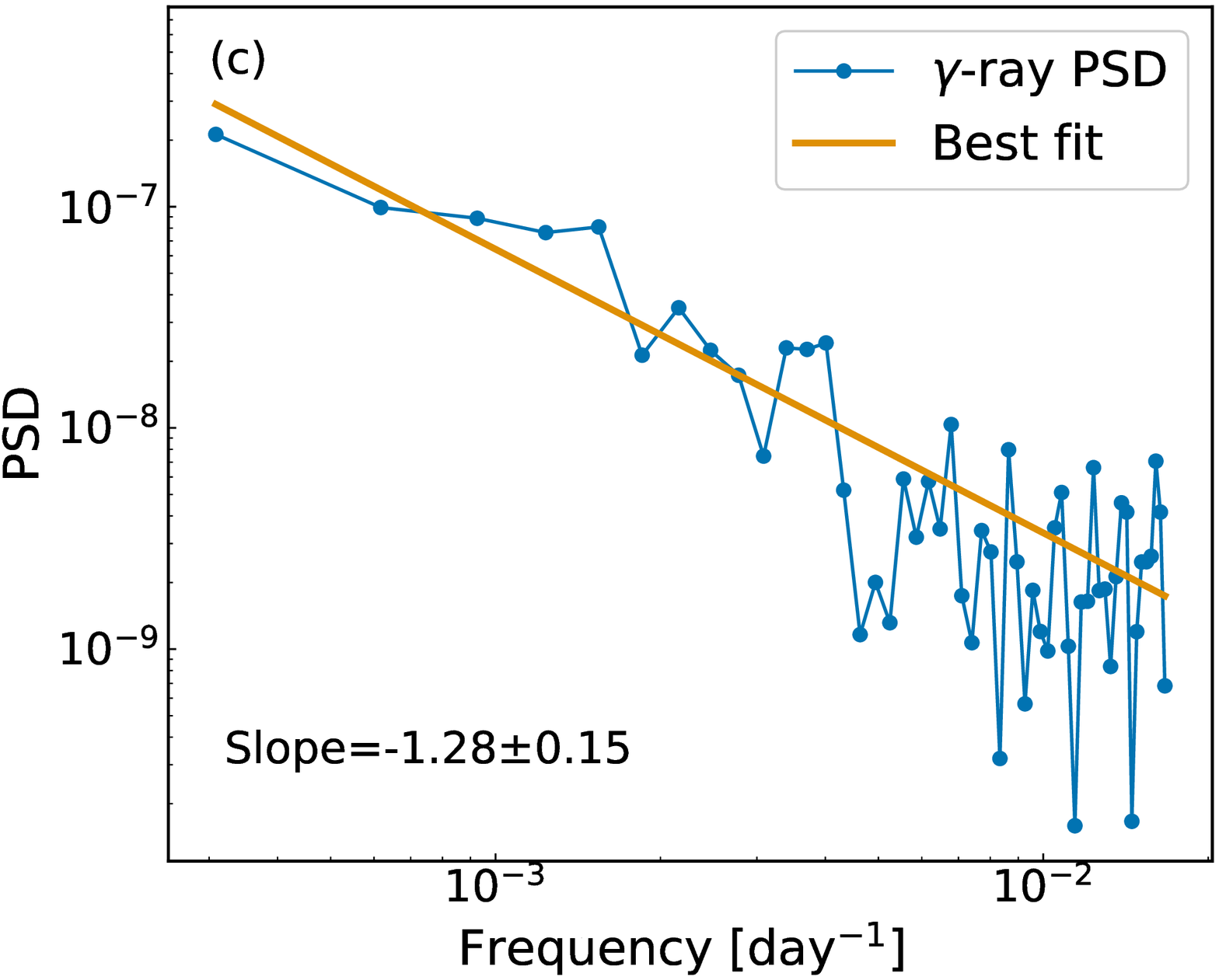}
   \includegraphics[angle=0,scale=0.24]{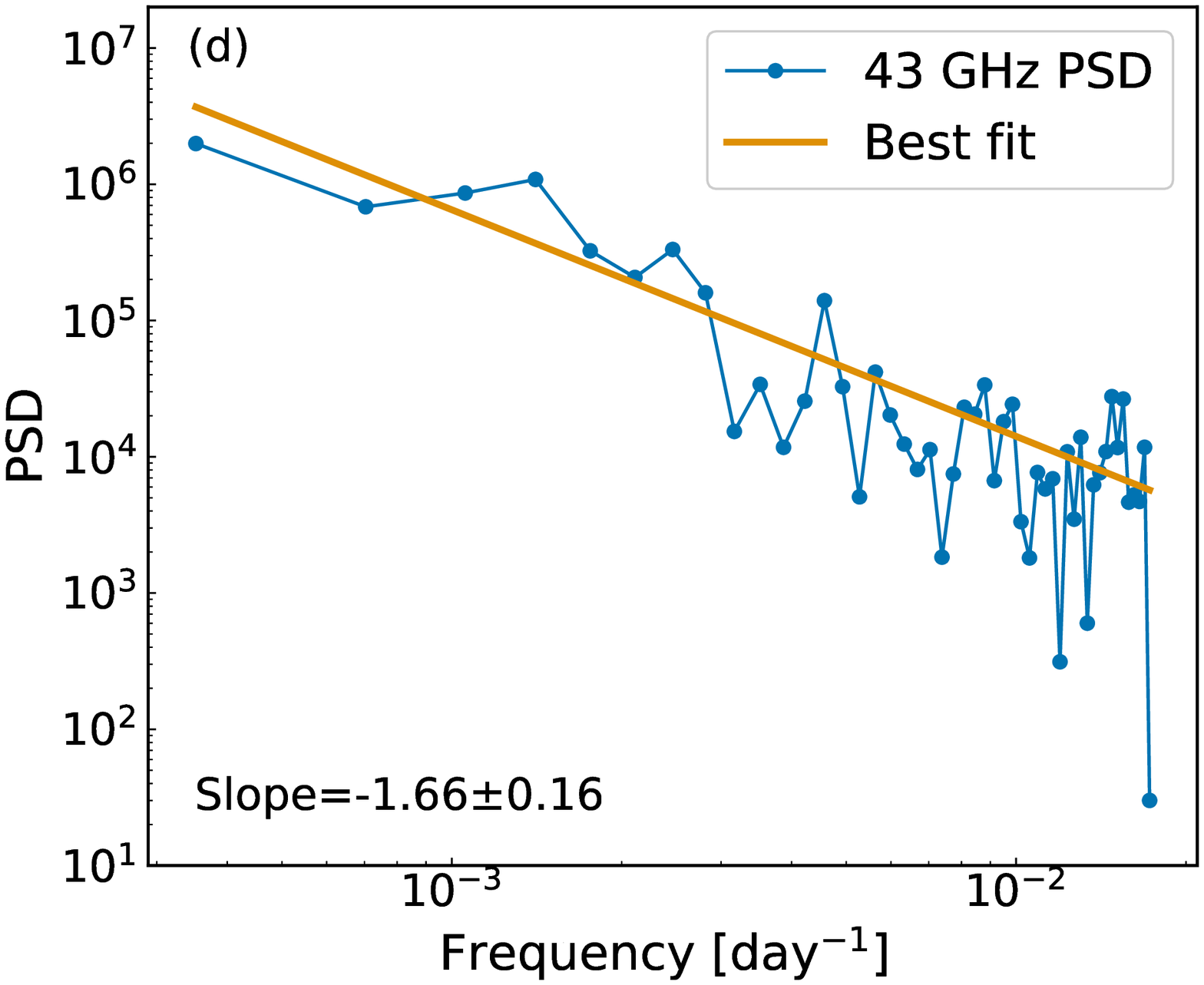}
   \includegraphics[angle=0,scale=0.25]{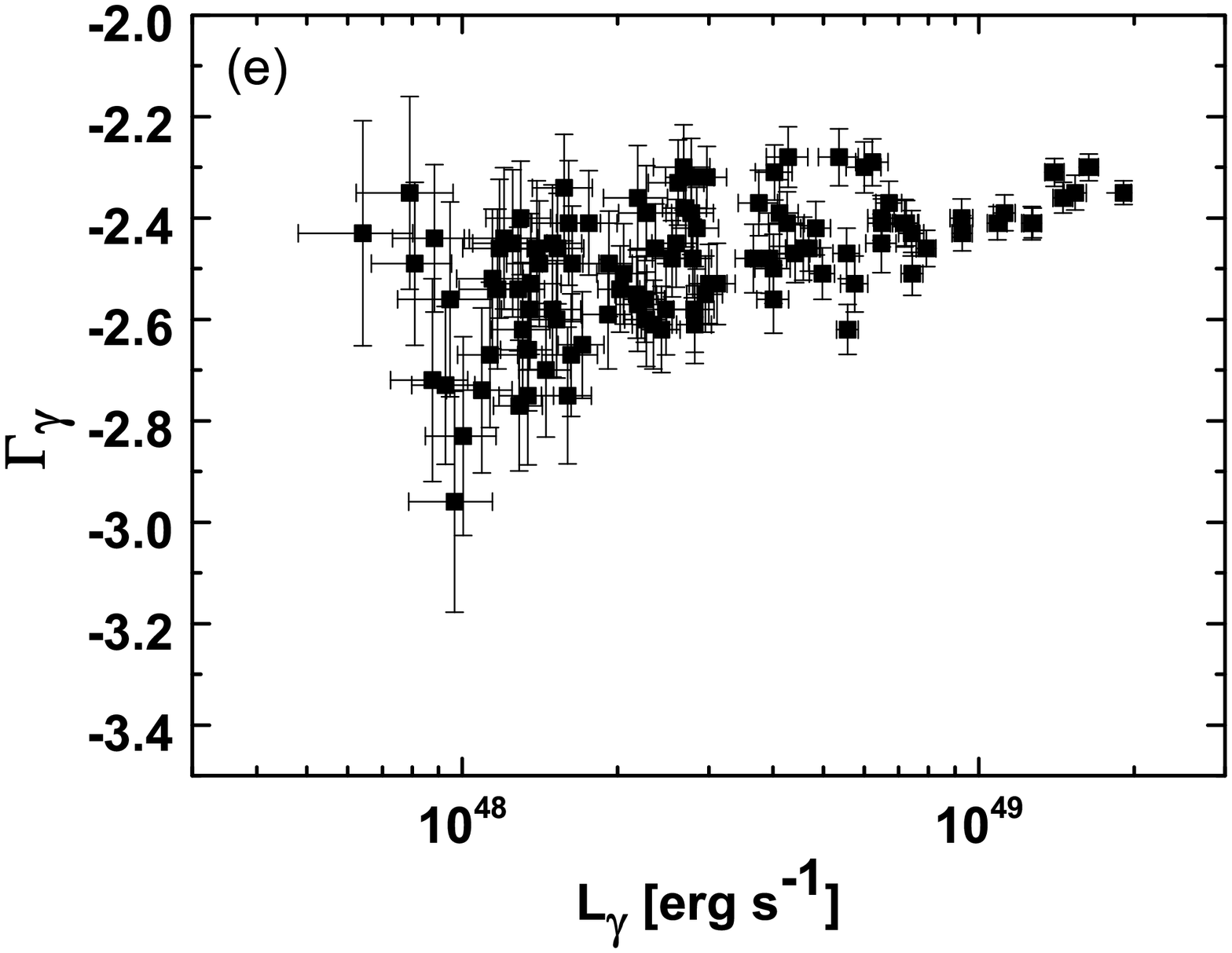}
   \includegraphics[angle=0,scale=0.30]{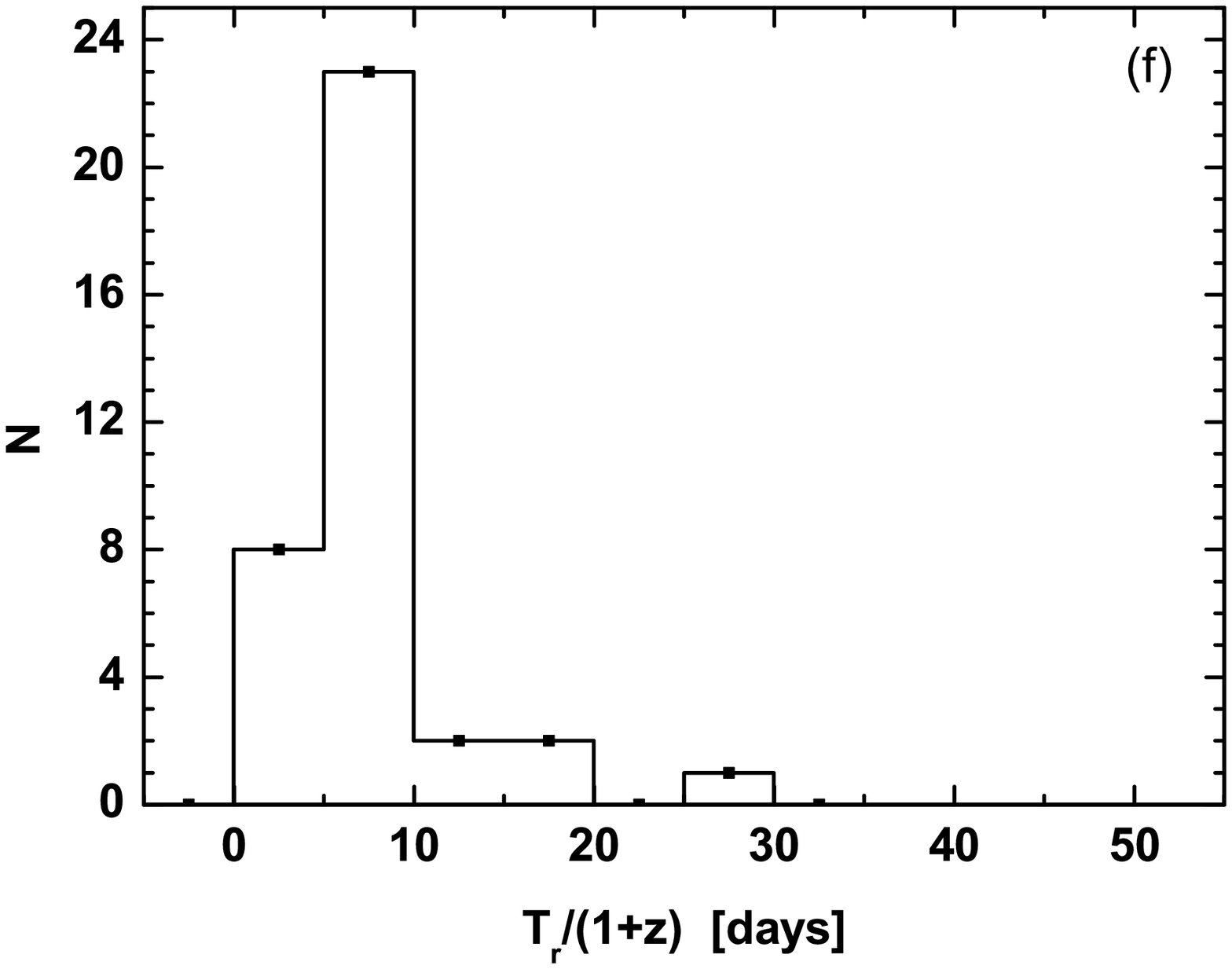}
   \includegraphics[angle=0,scale=0.30]{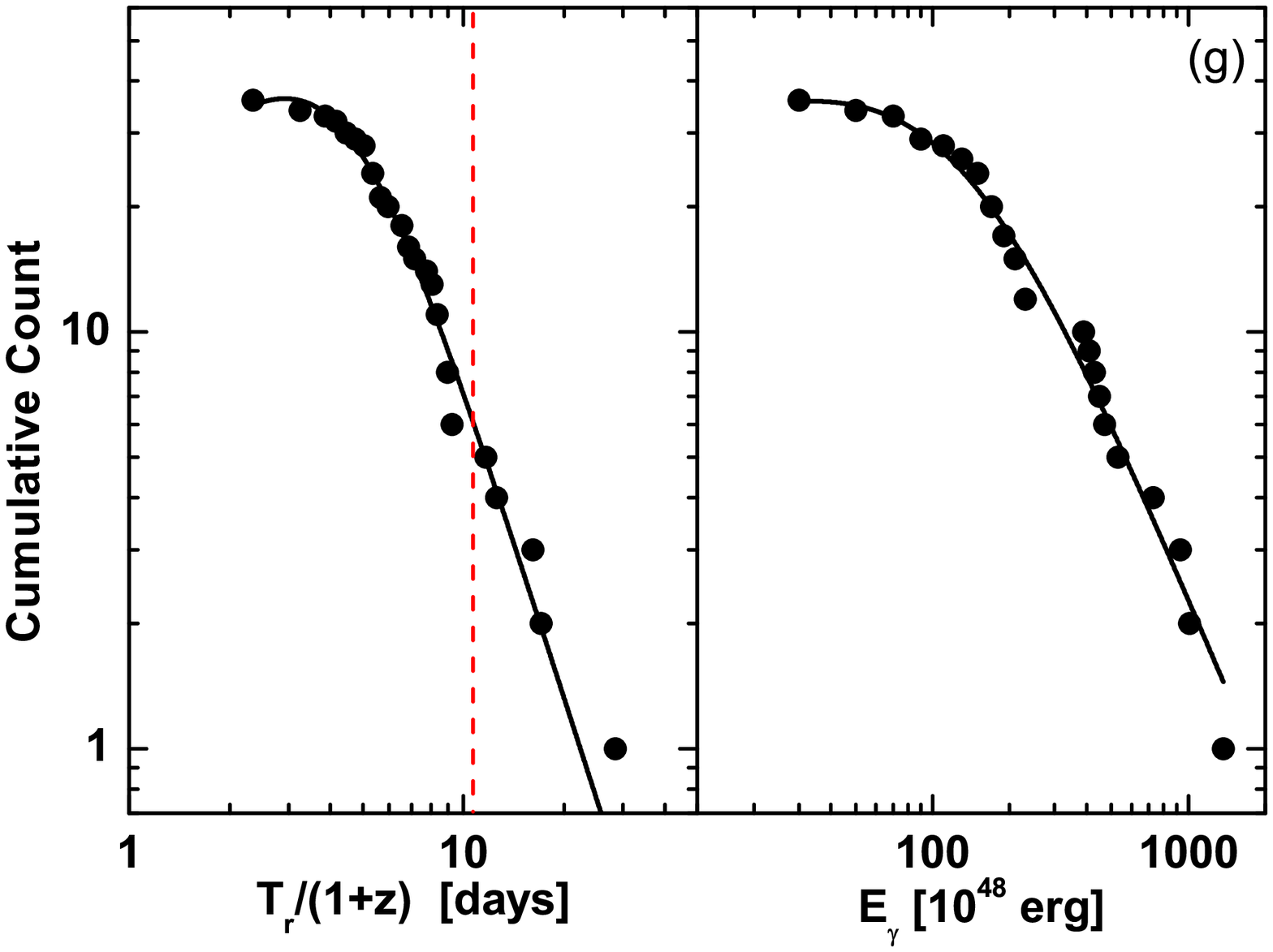}
\caption{Same as Figure 1, but for PKS 1633+382.}\label{LC_1633}
\end{figure}

\clearpage
\begin{figure}
 \centering
   \includegraphics[angle=0,scale=0.7]{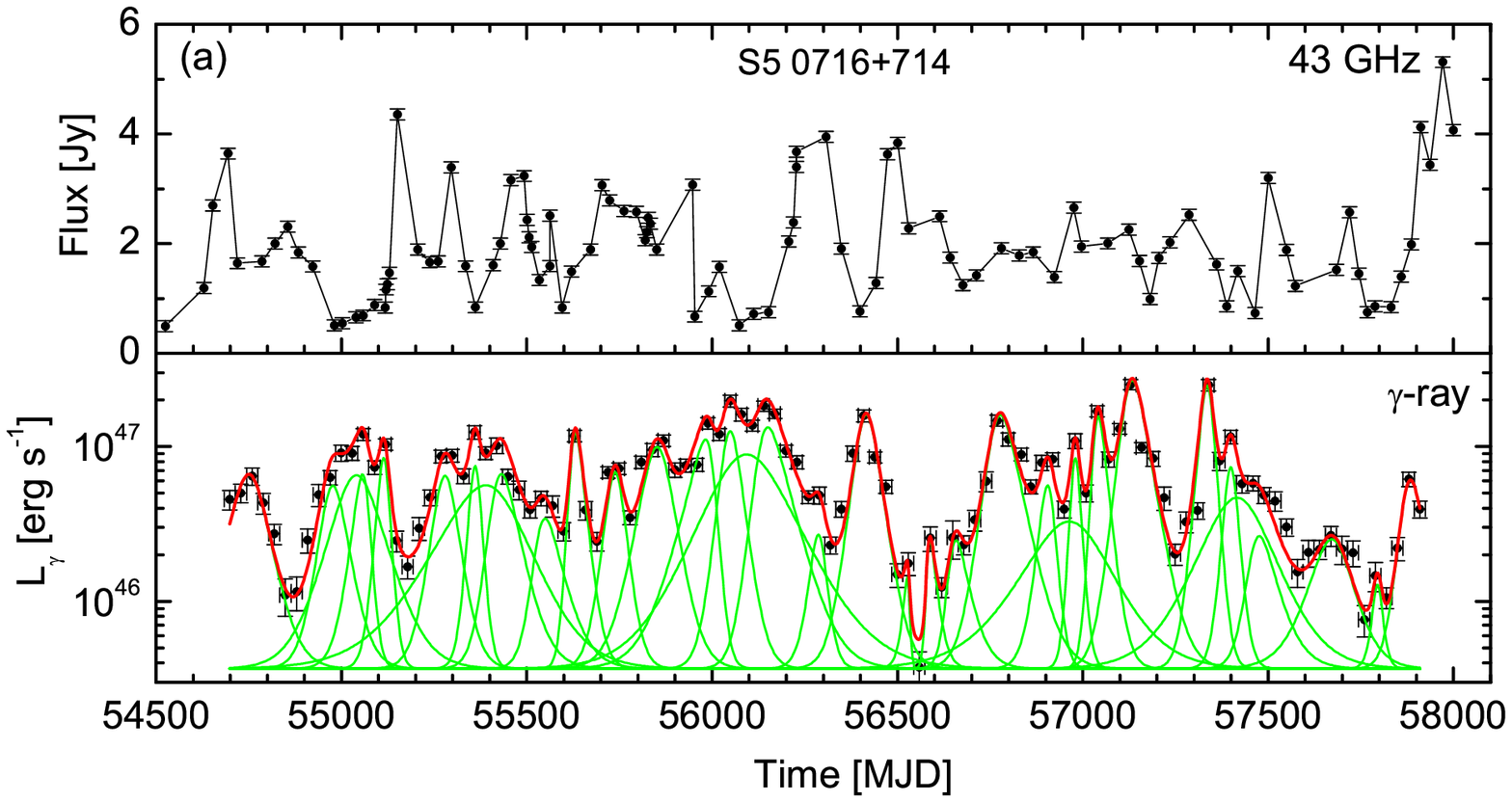}\\
   \includegraphics[angle=0,scale=0.26]{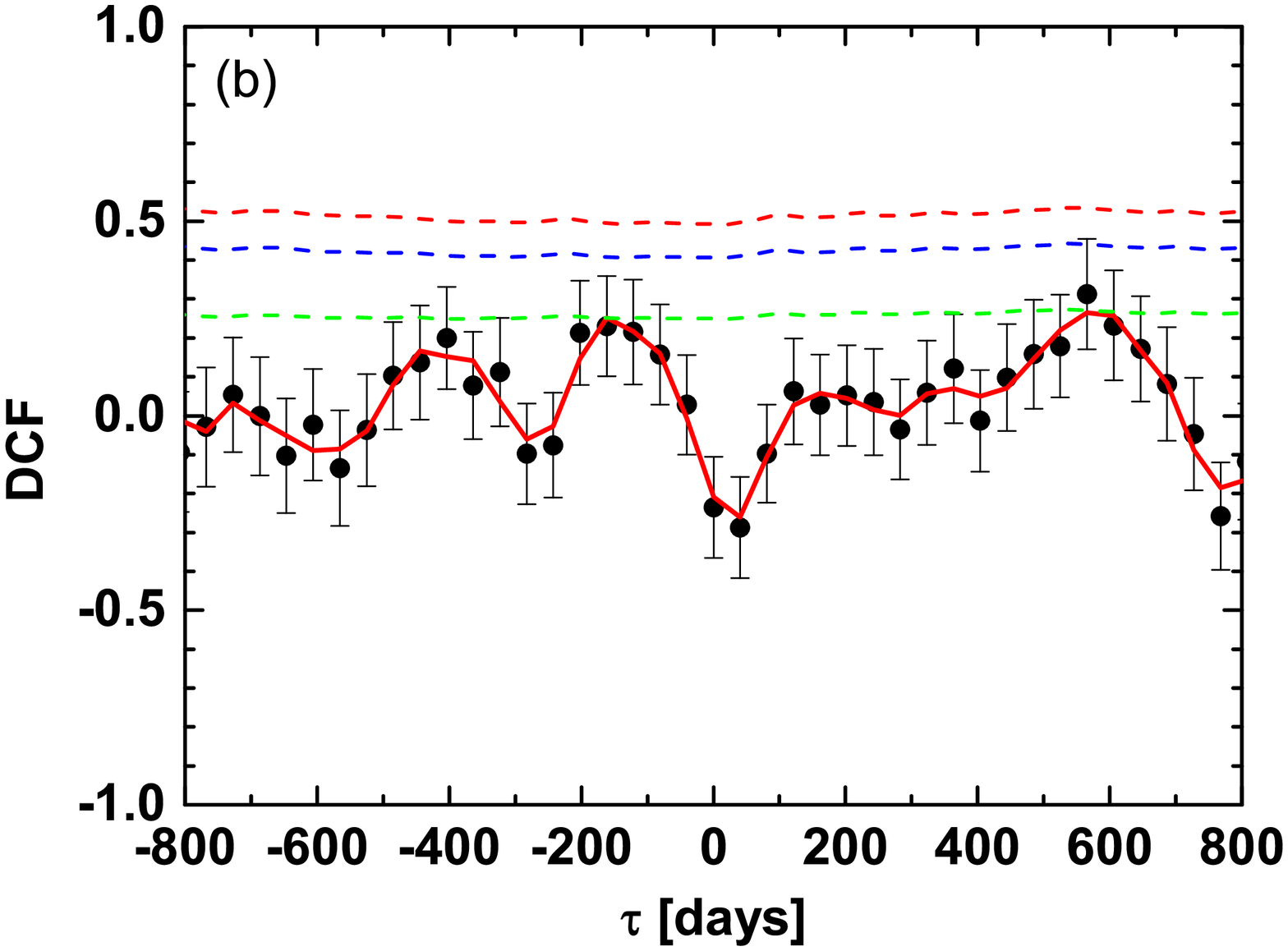}
   \includegraphics[angle=0,scale=0.24]{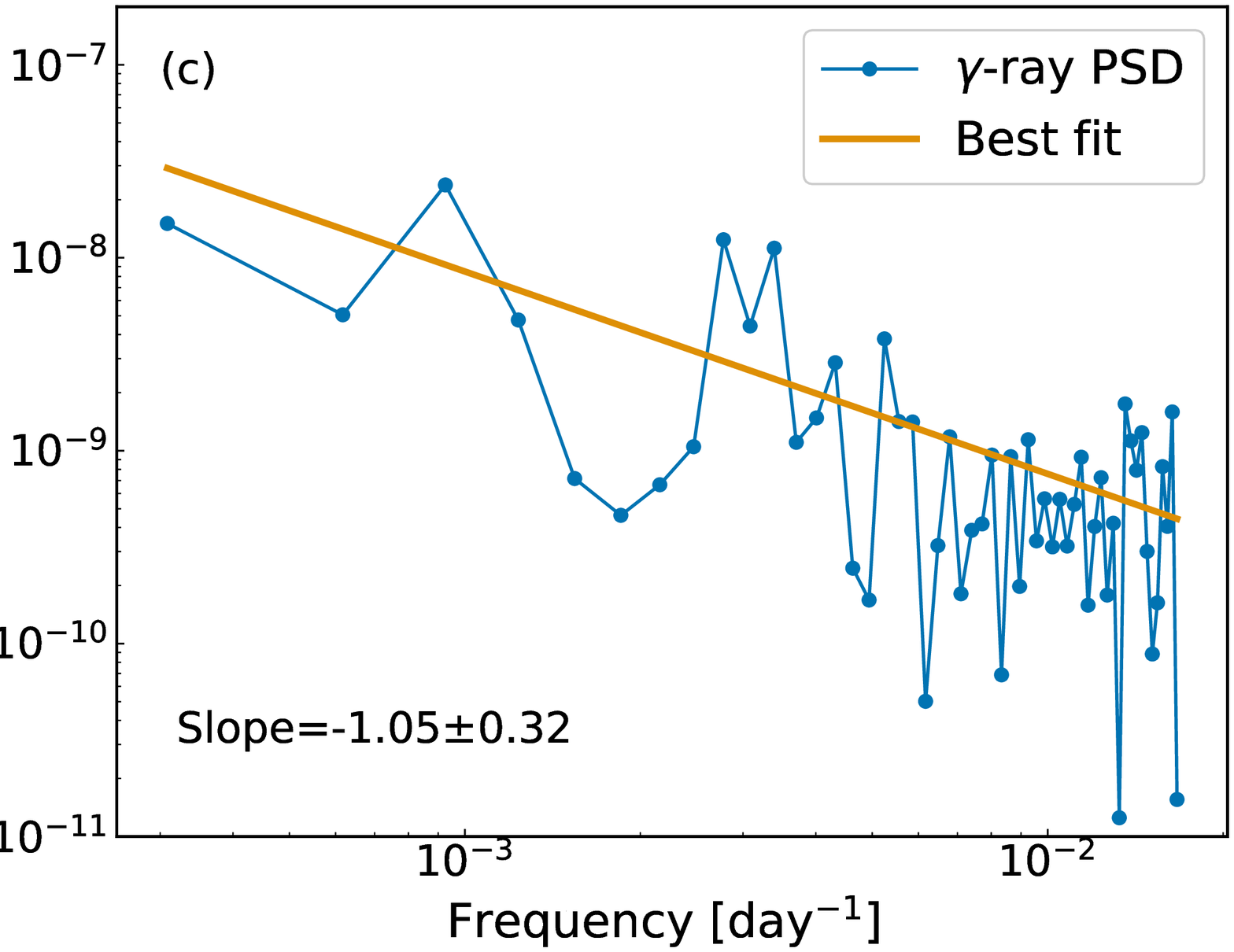}
   \includegraphics[angle=0,scale=0.24]{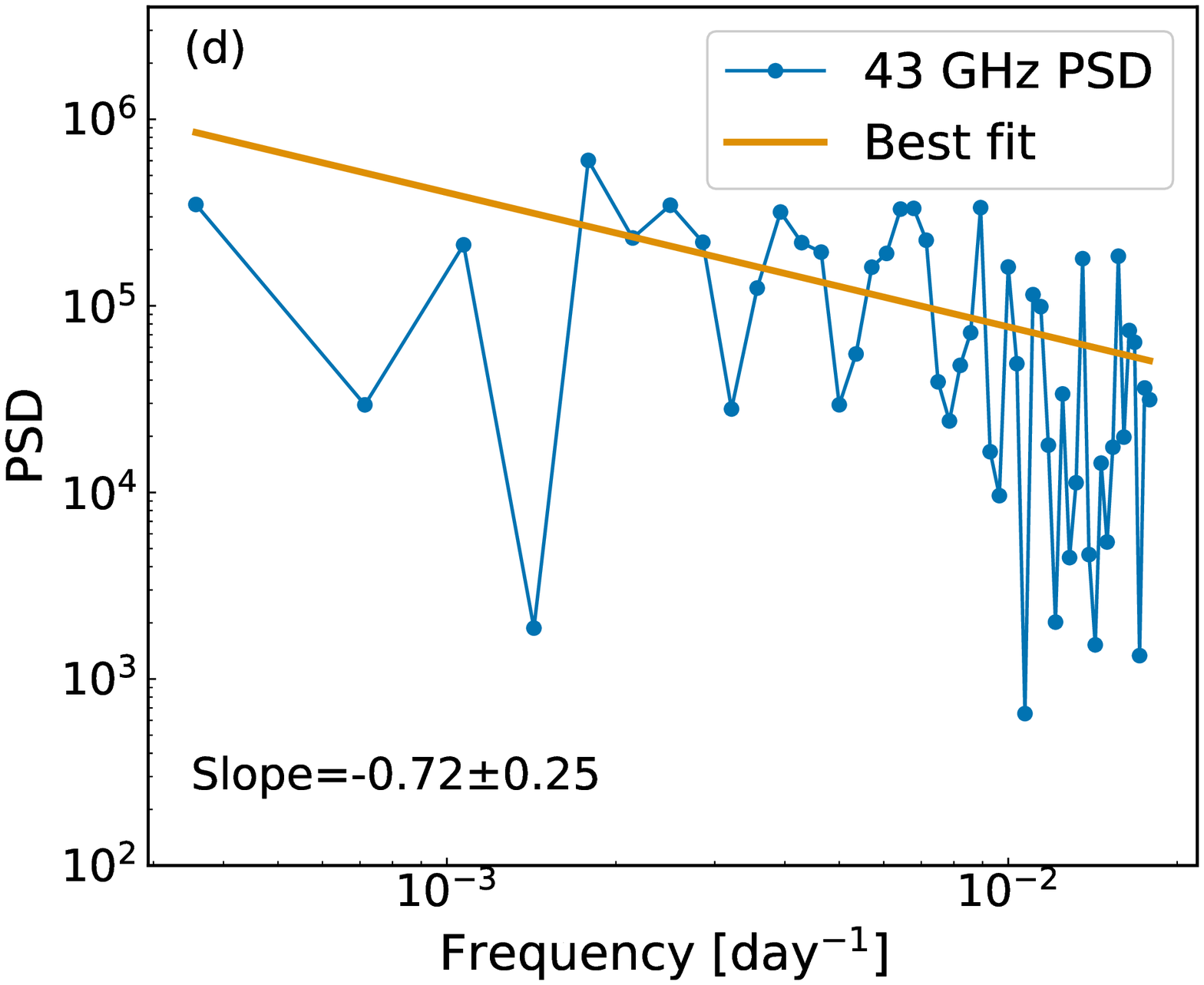}
   \includegraphics[angle=0,scale=0.25]{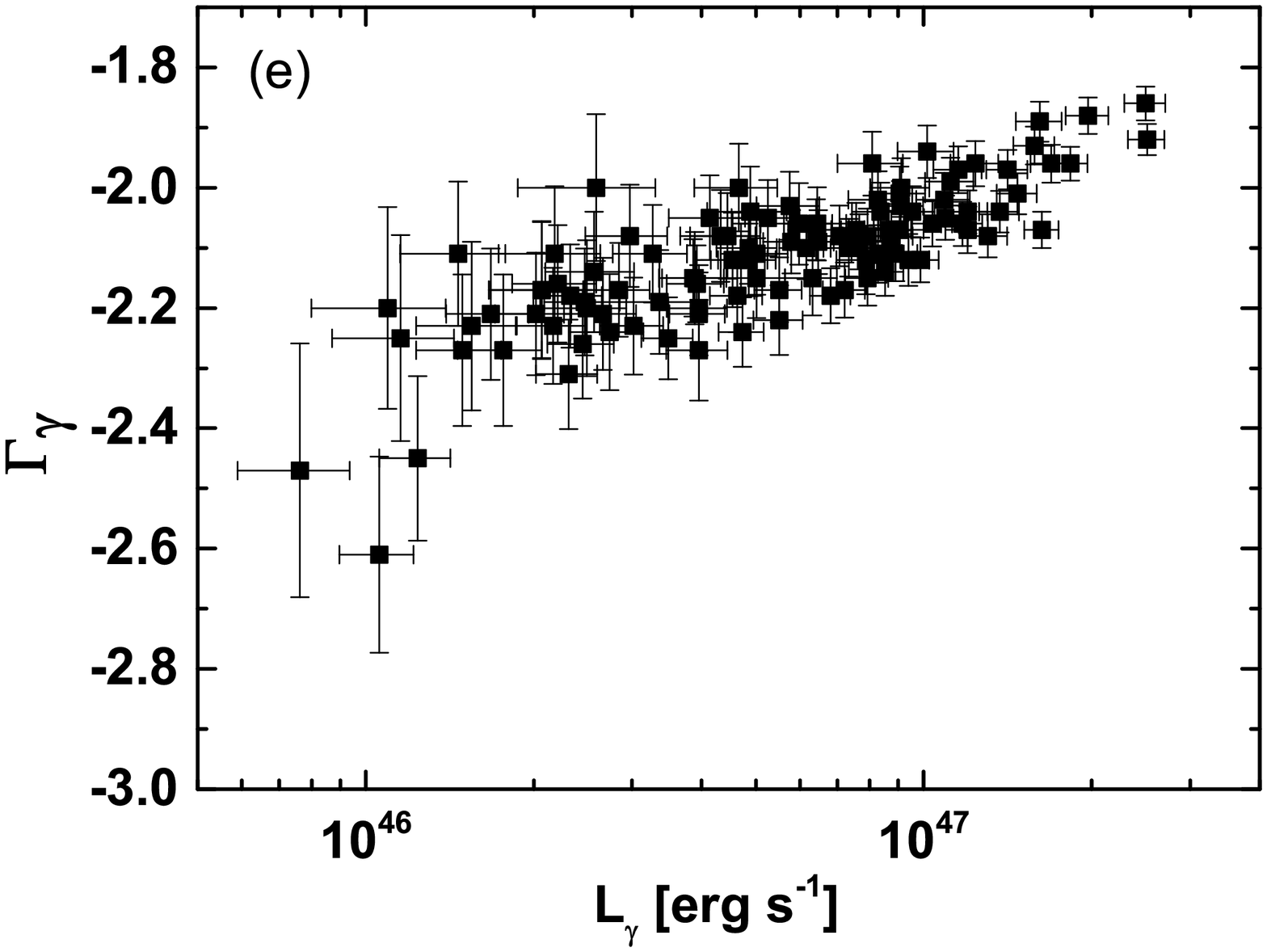}
   \includegraphics[angle=0,scale=0.30]{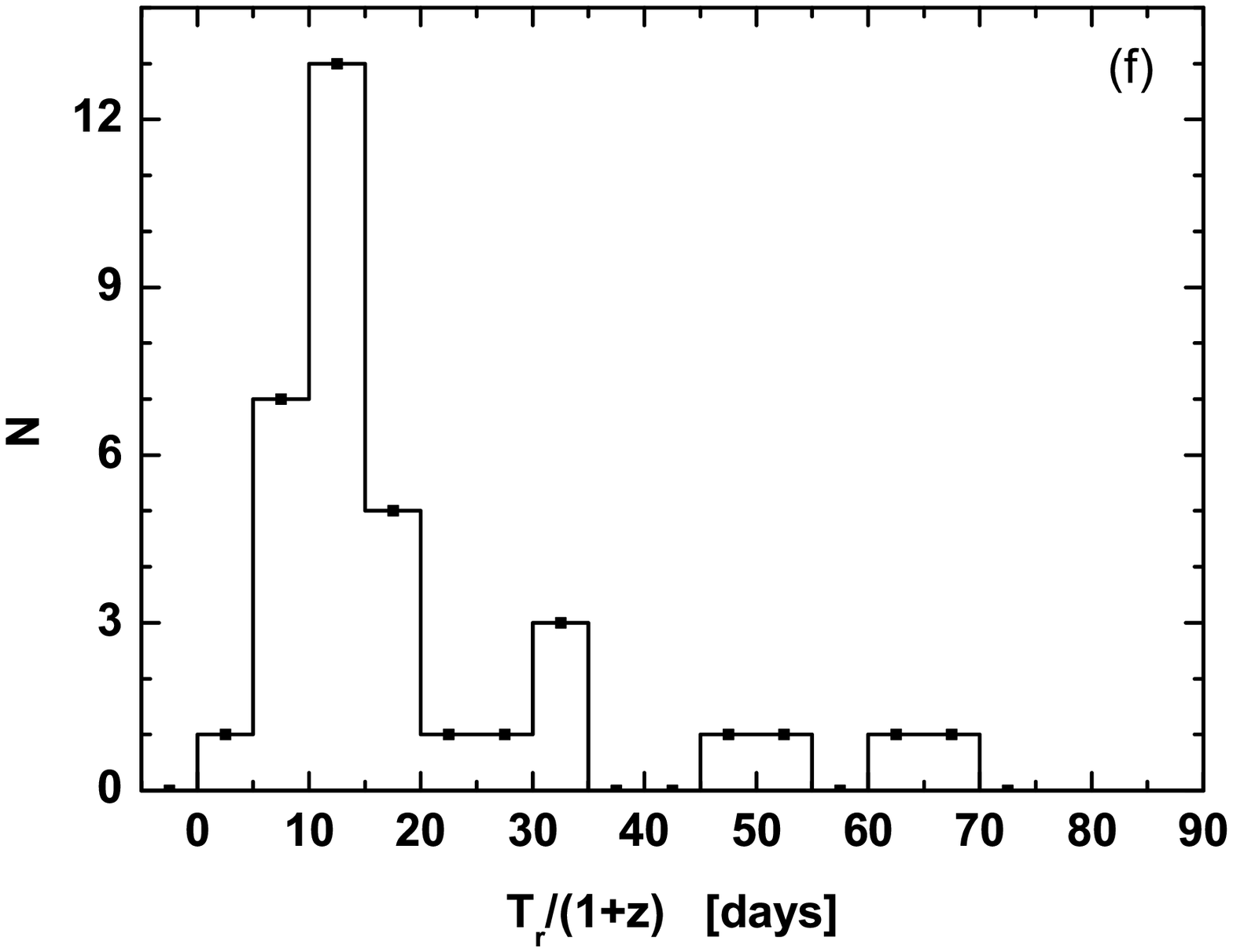}
   \includegraphics[angle=0,scale=0.30]{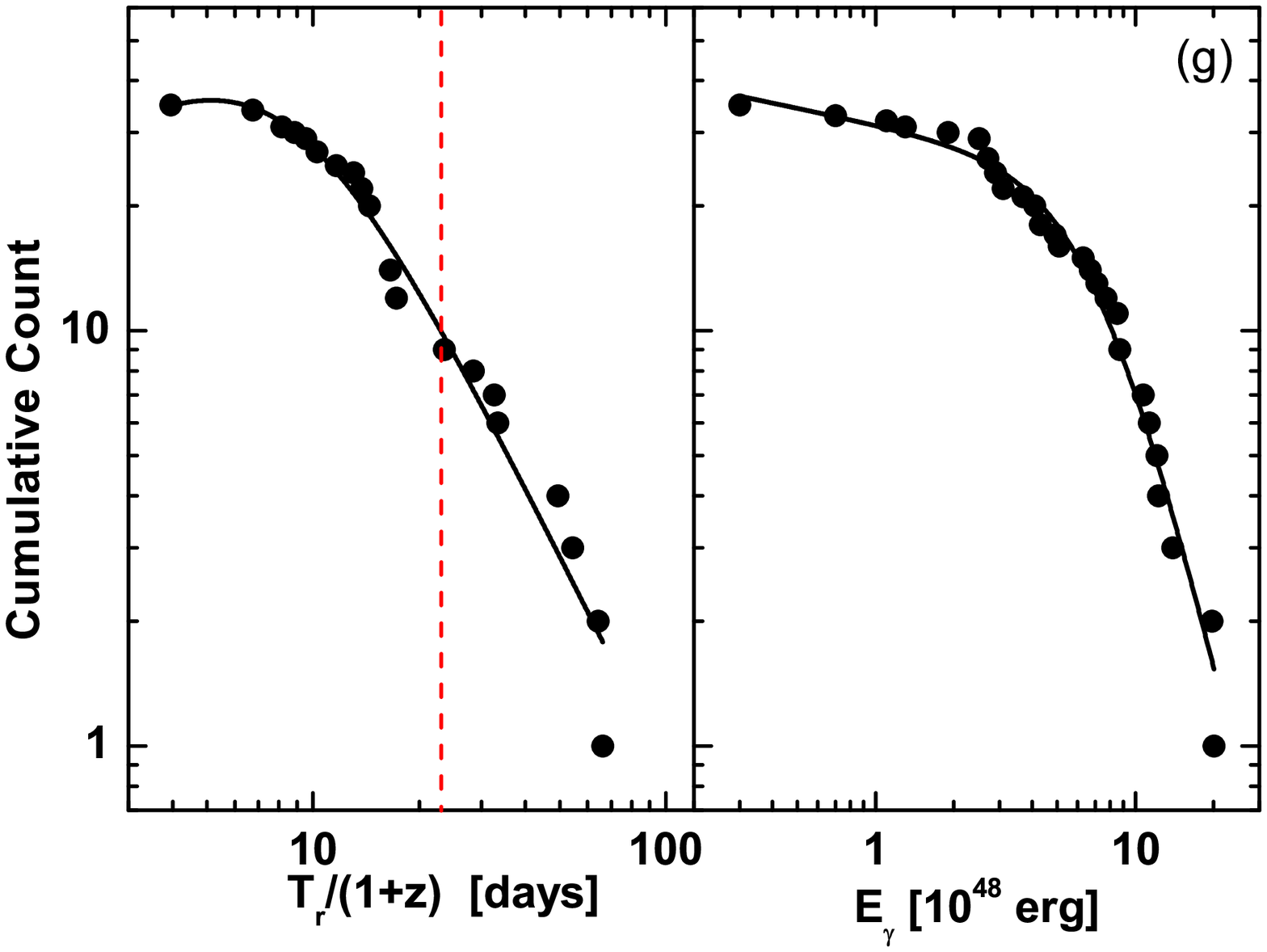}
\caption{Same as Figure 1, but for S5 0716+714.}\label{LC_0716}
\end{figure}

\begin{figure}
   \centering
   \includegraphics[angle=0,scale=0.5]{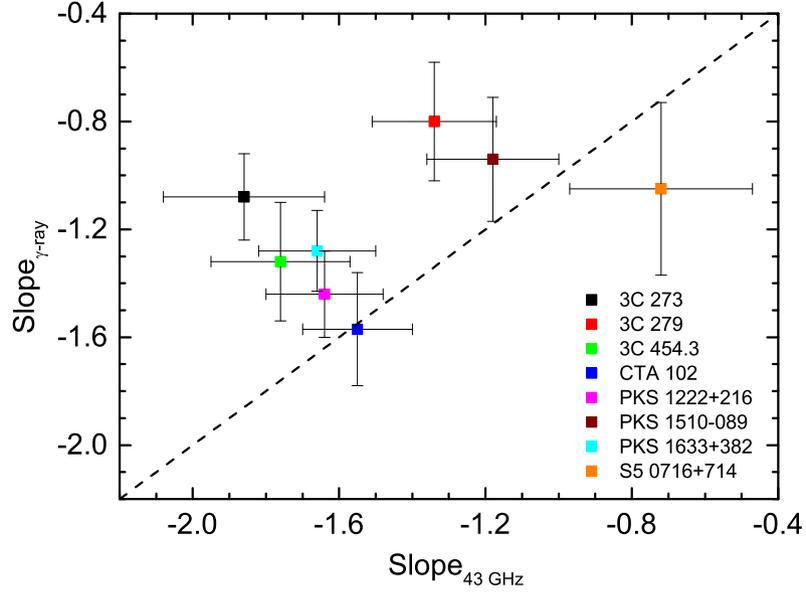}
\caption{PSD slopes in $\gamma$-ray band against that in 43 GHz for the eight blazars. The dashed line is the equality line. } \label{slope_PSD}
\end{figure}

\begin{figure}
   \includegraphics[angle=0,scale=0.45]{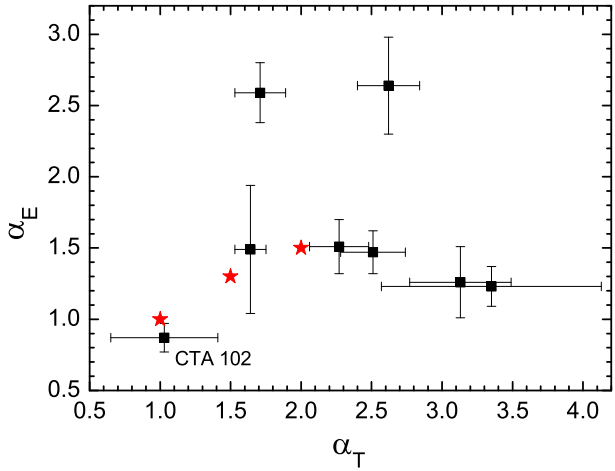}
\caption{Cumulative distribution slope of rise time ($\alpha_{\rm T}$) against that of energy ($\alpha_{\rm E}$) for flares in the $\gamma$-ray band. The three red stars indicate the values predicted by the SOC model, i.e., $\alpha_{\rm T}=(S+1)/2$ and $\alpha_{\rm E}=3(S+1)/(S+5)$, where $S=$1, 2, and 3 are the Euclidean dimensions (Aschwanden 2012). } \label{slope_TE}
\end{figure}

\clearpage
\begin{figure}
   \includegraphics[angle=0,scale=0.35]{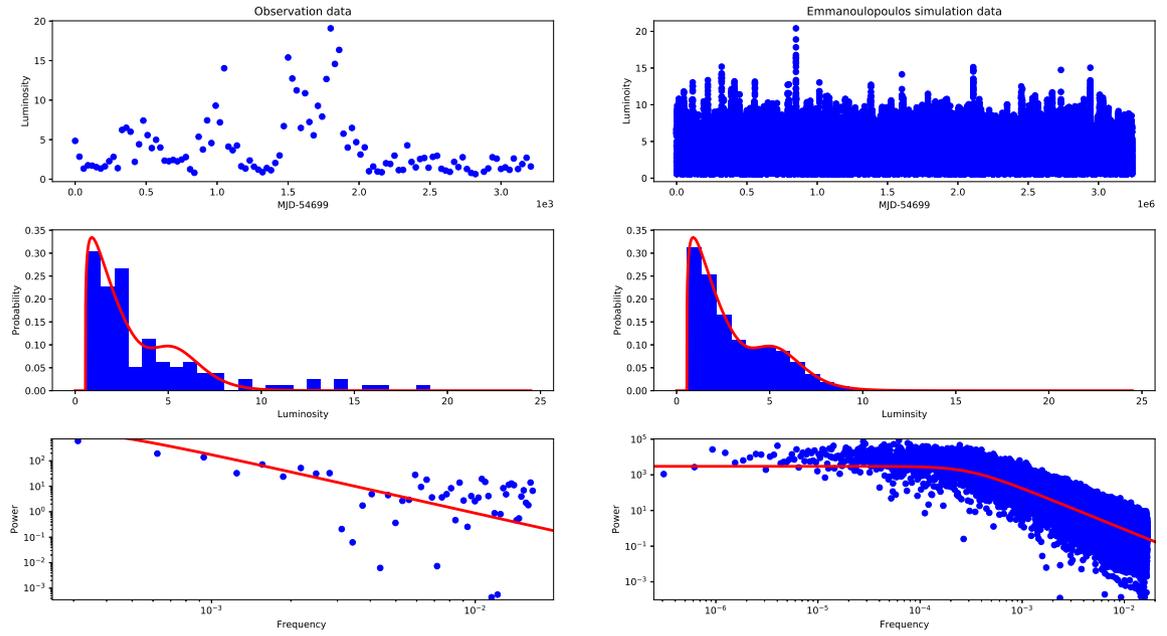}
\caption{The simulation results of the $\gamma$-ray light curve for PKS 1633+382. Left Panels: observed light curve (top), PDF histogram of the observed data (middle), and PSD of observed data (bottom). Right Panels: simulated light curve (top), PDF histogram of simulated data (middle), and PSD of simulated data (bottom). The red lines in the central panels represent the best-fitting mixture distribution model for the PDF histogram (details to see Emmanoulopoulos et. al 2013). The red line in the bottom right panel represents the PSD model (a broken power-law) for simulating the light curve, i.e., the slope$=0$ is set at low frequencies while at the high frequencies the slope is taken the corresponding best-fit value in Figure 11. The break frequency of the PSD model is at the lowest frequency ($\sim 1/3300$) sampled by the real data. The luminosity is in units of [$\times10^{48}$ erg s$^{-1}$]. } \label{simulation_LC}
\end{figure}
\clearpage


\end{document}